\documentclass[aps,prl,twocolumn]{revtex4}

\usepackage{graphicx}
\usepackage{docs}

\def\der#1#2{{\partial#1 \over \partial#2}}
\def\be{\begin{equation}}
\def\ee{\end{equation}}
\def\ba#1{\begin{array}{#1}}
\def\ea{\end{array}}
\def\bn{\begin{enumerate}}
\def\en{\end{enumerate}}
\def\r{\right}
\def\l{\left} 
\def\summ{\sum\limits}
\def\half{\frac{1}{2}}

\def\S{{\it S}~}
\def\N{{\it N}~}

\def\dx{D}
\def\at{\tau_0}

\def\B{B}
\def\scs{SC$^{\star}$~}
\def\oz{\overline{\zeta}}
\def\ax{a}
\def\vms{c}
\def\jx{\kappa}
\def\jt{J}
\def\j{K}
\def\xz{\lambda_Q}
\def\w{w}
\def\u{u}

\def\aeff{\alpha}
\def\beff{\beta}
\def\etas{\{\eta_s\}}
\def\xis{\{\xi_s\}}
\def\E{E}
\def\nuz{\mu}
\def\tzeta{\tilde{\zeta}}

\begin{document}
\title{Superconductor-to-Metal Transitions in Dissipative Chains of Mesoscopic Grains and Nanowires}
\author{{\sc Gil Refael${}^1$, Eugene Demler${}^2$, Yuval Oreg${}^3$, 
Daniel S. Fisher${}^2$}\\
 {\small ${}^1$\em Dept.\ of Physics, California Institute of Technology, Pasadena, CA 91125}\\
{\small ${}^2$\em Dept.\ of Physics, Harvard University, Cambridge, MA 02138}\\
{\small ${}^3$\em Department of Condensed Matter Physics, Weizmann Institute of Science, Rehovot, 76100, ISRAEL }} 

\date{\today}

\begin{abstract}

The interplay of quantum fluctuations and dissipation in chains of
mesoscopic superconducting grains is analyzed,  and the results are
also applied to  nanowires. It is shown that in   one dimensional
arrays of resistively shunted Josephson junctions,  the
superconducting-normal charge relaxation  within the grains plays an
important role.   At zero temperature, two superconducting phases can
exist, depending primarily on the strength of the dissipation.  In the
fully superconducting phase (FSC), each grain acts superconducting,
and the coupling to the dissipative conduction is important. In the
\scs phase, the dissipation is irrelevant at long wavelengths. The
transition between these two phases is driven by quantum phase slip
dipoles, and is primarily local, with continuously varying critical
exponents.  In contrast, the transition from the \scs phase to the
normal metallic phase is a Kosterlitz-Thouless transition with  global
character (i.e., determined by the field behavior at large
wavelengths). Most interesting, is the transition from the FSC phase
directly to the normal phase: this transition, which has mixed local
and global characteristics, can be one of three distinct types. The
corresponding segments of the phase boundary come together at
bicritical points. These behaviors are inferred from both weak and
strong coupling renormalization group analyses. At intermediate
temperatures, near either  superconductor-to-normal phase transition,
there are  regimes of super-metallic behavior, in which the
resistivity first decreases gradually with decreasing temperature
before eventually  increasing as temperature is lowered further.  The
results on chains of Josephson junctions are extended to  continuous
superconducting nanowires and the subtle issue of whether these can
exhibit an FSC phase is considered. Potential relevance  to
superconductor-metal transitions in other systems is also discussed.

\end{abstract}
\pacs{PACS numbers:}

\maketitle

\section{Introduction}

\label{intro_section}

\subsection{Motivation}

Quantum mechanical
systems that are  coupled to dissipative ``environments" arise in 
many areas of physics, including spin dynamics
in nuclear magnetic resonance \cite{NMR1, NMR2},
damping in atomic clocks and optical
interferometers \cite{Gardiner2000},
dephasing and decoherence in mesoscopic
systems \cite{mesoscopic}
and quantum computing \cite{clarke2004,Nielsen2000}.
Also in more conventional condensed matter contexts, dissipation has been
argued to play crucial roles. In particular: near to quantum magnetic phase transitions; \cite{Hertz,Millis, Sachdev}
in quantum Hall systems; \cite{Shimshoni1998}
and in various aspects of superconductivity, including Josephson junctions, and  thin superconducting films and wires 
\cite{Chakravarty2A,Wagenblast1997B,Larkin1999,Kapitulnik2000,Shahar, Dalidovich2000,Paalanen2002,Rimberg2002,W-H2003,Pekker,
  SachdevTroyerWerner,Tinkham2003,Bezryadin, Tinkham}. The last of
these is the primary focus of the present paper.

For theoretical studies of  dissipative effects in 
macroscopic  --- and some mesoscopic --- quantum systems, the degrees of freedom that cause the dissipation are often modeled as a 
heat bath following  Caldeira and Leggett.
\cite{c-leggettA, c-leggettB,leggett2} The best studied example is a single resistively-shunted Josephson junction (RSJJ)
\cite{Chakravarty,schmid,bulga,Korshunov1987,simanek,weiss,INGOLD1992}. 
Recent
experiments by Pentill\"a et al. \cite{Haviland2003,paalanen} have shown good agreement
with the theoretical predictions.  
Extensions to {\it arrays} of  RSJJs have been analyzed by several groups 
focusing on the existence and location of phase boundaries between 
superconducting and insulating regimes of the set of junctions.
\cite{Schoen-Zaikin,pz,MPAFisher1987,Chakravarty2A, Chakravarty2B,Bobbert,KorshunovA, KorshunovB,Zwerger1989,Schoen,Fazio}
Experimental studies of one and two dimensional arrays of large superconducting grains 
coupled by dissipative Josephson junctions, agree qualitatively with  
results of the theoretical analyses 
\cite{Fazio,HavilandA, HavilandB, HavilandC, HavilandD, HavilandE,Rimberg,japan,Haviland2A, Haviland2B,NEWROCK2000,Miyazaki2002}. 

The understanding developed from
studying the destruction of superconductivity by quantum fluctuations in
arrays of Josephson junctions has became a useful paradigm for
more general considerations of  quantum phase transitions in dissipative environments.
In general, dissipation suppresses certain types of  quantum fluctuations and thus can favor
states with spontaneously broken symmetries, such as superconductivity. 
Considerable interest in further theoretical analysis
of RSJJ arrays thus  stems not only from direct experimental relevance in the context of superconductivity, 
but also from expectations that the concepts and approaches will be useful far more generally, especially for understanding
universal aspects of quantum phase transitions that can occur at zero-temperature in the presence of dissipation.
But to do this, it is crucial to take into account the small size of the components involved, whether they are small grains or individual atoms in a crystal.

Because the models on which they are based were initially introduced
to understand macroscopic quantum phenomena, \cite{c-leggettA,
  c-leggettB,leggett2} most theoretical analyses of  RSJJ arrays have
assumed that the effective charges associated with the superconducting
and normal  currents are perfectly mixed within the superconducting
grains while passing separately between grains via the Josephson
junctions and normal ``shunts", respectively,.  Such an approximation
is  reasonable for macroscopic superconductors but will break down in mesoscopic or microscopic systems
\cite{Refael2003}.  As we showed in  Ref. \cite{Refael2003},  a
consequence of the breakdown of the macroscopic paradigm is that the
superconducting and normal fluids can effectively decouple at low
energies. For the  simple case of two junctions in series through a
mesoscopic grain, we showed that this leads to changes in both the nature and location
of the superconductor-normal transition that occurs. In the present
paper,  this analysis is extended to show that decoupling of the two
fluids has equally important consequences for chains of mesoscopic grains and for superconducting
nanowires.

The  goal of this paper is to provide a detailed analysis of  superconductor-to-normal
transitions in one dimensional
mesoscopic systems for which dissipation plays a role.
The main emphasis  is on  chains of mesoscopic
grains that are connected both by Josephson
junctions and some form of shunt resistance, although we also consider  continuous superconducting nanowires.   We will not 
discuss the possible origins of the assumed Ohmic dissipation in such systems, but
rather assume that it is present and study its consequences. Furthermore, because we are primarily interested in dissipative effects that arise simply only if the diameters of the grains or wires are substantially larger than atomic sizes, we will also ignore the limits in which the discreteness of the electrons or Cooper pairs becomes most essential, such as in Giamarchi and Schulz's treatment  of superfluid-to-normal transitions in Luttinger liquids, \cite{Giamarchi-Schulz1988} and phenomena associated with localization by randomness.
Within a simple but relatively general mesoscopic model, we analyze
the nature of the phases that can exist and the locations and
character of the several types of quantum phase transitions that
occur.  In particular, we study the universal scaling behavior of the 
resistivity in the vicinity of the superconductor-to-normal
transition(s). To do so, we develop both  strong- and  weak-coupling
renormalization group approaches which are tailored to deal with both the
 local resistive and the  long-wavelength superconducting  degrees of freedom.

\subsection{Outline}

\begin{figure*}
\includegraphics[width=11cm]{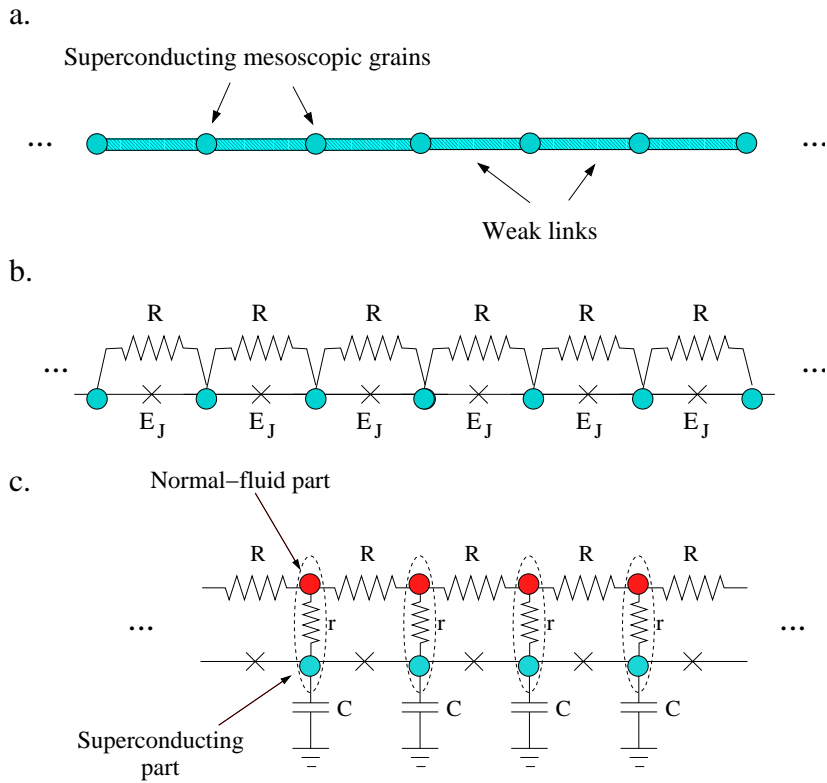}
\caption{ (a) A system of superconducting grains connected
by weak links. The grains are coupled by flows of both superconducting 
Cooper-pairs and normal electrons. (b) We model the system in (a) by a
chain of superconducting grains connected  by Josephson
junctions and shunt resistors. (c) Two fluid model: each mesoscopic
grain is represented as a combination of a superconducting
and a normal parts, depicted here as separate grains. Charge relaxation between the \S and \N grains
is via a conversion resistance $r$. 
\label{Fig1intro}}
\end{figure*}

This paper is organized as follows. 
\begin{itemize}
\item Sec. \ref{intro_section} provides a general introduction. Sec. \ref{2fintro} introduces the two-fluid approach to
  mesoscopic superconducting grains, as first given in
Ref. \onlinecite{Refael2003}. Sec. \ref{resintro} gives a
  summary of the main results of the paper, omitting technical
  details. 
\item Sec. \ref{secmm} derives the quantum two fluid 
 model that describes an infinite chain of mesoscopic
two-fluid grains as shown in Fig. \ref{Fig1intro}. Sec. \ref{regimes}
discusses the various possible regimes and  Sec. \ref{ctm} gives
an analysis of the linear electrodynamics of the model which 
provides intuition for the location and nature of the various
transitions.

\item In Sec. \ref{sec2} we discuss the strong coupling limit of the
  chain, first deriving the quantum phase-slip
  representation of the chain (Sec. \ref{QPSsec}), from which we
  construct a sine-Gordon action (Sec. \ref{sgsec}). Using the sine-Gordon
  action for the chain, we discuss the possible phases of the system
  (Sec. \ref{phases1}). Finally, in Sec. \ref{sec3} we derive the
  strong coupling RG flow equations for the system. These 
  are constructed by an anisotropic scaling procedure, suited to the
  dissipative environment. 

\item In Sec. \ref{sec5} we use the RG flow equations to determine the
phase diagram of the system. Although the system exhibits three
phases, the transitions between them have a variety of types. Each
transition is discussed separately in
Secs. \ref{scs11}-\ref{ssF}. Special aspects of the phase diagram,
such as bicritical and multicritical points are discussed in Secs. \ref{bicrit}-\ref{multi}. 

\item Sec. \ref{wcsec} analyzes the
weak-Josephson coupling limit. First, we cast the action in terms of
pair-tunnel events (Sec. \ref{ptesec}). Then we construct the RG flow
equations (Sec. \ref{wrg}), from which we obtain the weak coupling
flow diagram (Sec. \ref{wpd}). 

\item Sec. \ref{ContinuumLimitSection} considers
superconducting nanowires by considering them as the continuum
limit of the JJ chains. 

\item Sec. \ref{res} presents scaling
forms for the resistivity of the chain and discusses various interesting
parameter regimes.

\item We conclude in Sec. \ref{discuss}
by  reviewing the implications of our results for various experimental
systems and raising open questions.  

\end{itemize}

Some technical details are relegated to the Appendix.

\subsection{Quantum Two Fluid Description\label{2fintro}}

\subsubsection{Mesoscopic Grains and Shunted Josephson Junctions}

The primary system we will study is shown in 
Fig. \ref{Fig1intro}(a): a chain of identical mesoscopic
superconducting grains which are connected by weak links that allow the
flow of both Cooper pairs and normal electrons. 
Following Refs. \onlinecite{MPAFisher1987,Chakravarty2A, Chakravarty2B} 
we describe this system as a chain of resistively
shunted Josephson junctions (Fig. \ref{Fig1intro}(b)).
Such an RSJJ chain has a natural interpretation in terms of  a ``two fluid"
model: Cooper pairs that can tunnel between the superconducting
grains comprise the superfluid, and electrons that can flow in the shunting
resistors  the normal fluid. The presence of both fluids  suggests considering each grain as consisting of two (physically overlapping) parts --- a superconducting
grain ({\it S}) and a normal grain ({\it N})--- as shown in Fig. \ref{Fig1intro}(c).

Changes in the super or normal charge on a grain will induce  changes
in the corresponding electrochemical potentials with the coefficients
depending on both the {\it capacitance} of the grain and the {\it
  compressibilities} of the normal and superfluid components. A
difference between the normal and the superfluid electrochemical
potentials  will lead to charge relaxation between them; we model this
as a {\it conversion current}, $I_{NS}$, through a phenomenological
Ohmic {\it conversion resistance}, $r$.

This classical model of a chain of grains can be made into a quantum model straightforwardly by analogy with previous work, e.g., ref. \cite{Refael2003}.
The length scale is set by the spacing, $a$, between the grain centers, and various  time scales by the normal--superfluid relaxation rate within a grain,  the plasma frequency of the Josephson junctions, and the high frequency cutoffs of the superconducting degrees of freedom (typically of order the energy gap) and of the dissipative processes.

We will show that the behavior of the quantum chain depends crucially on the normal-superfluid conversion resistance, $r$.
In the limit $r \rightarrow 0$ the relaxation
between the \S and \N fluids is infinitely fast, and our model reduces
to that of an RSJJ chain composed of macroscopic superconducting grains each with a {\it single} electrochemical potential . This is the
limit studied previously 
\cite{KorshunovA, KorshunovB,Schoen}. The
opposite limit, $r \rightarrow \infty$, describes a system with purely
capacitative couplings between the {\it S} and the {\it N}
fluids. This case has been discussed  in Refs. 
\onlinecite{WAGENBLAST1997,Vishwanath2002} and a two dimensional version  
realized experimentally. \cite{Rimberg,MASON2002} The intermediate case of finite $r$
involves new behavior, which, to our knowledge, has not been analyzed previously:  it should  be relevant for experiments on arrays of 
mesoscopic grains. The new physics associated with the interplay between  the \S and \N fluids dominates below a temperature $T^*$, roughly proportional to the electron-energy-level spacing within a grain.  \cite{Refael2003}

\subsubsection{Nanowires}

The two fluid model can be readily generalized to a continuous wire that is thin enough to ignore dependences on the transverse coordinates. Classically, the corresponding two-fluid model is defined by the
generalized Josephson equation for the superfluid current, 
Ohm's law for the normal current, and a constitutive equation for the conversion current, 
\be
\ba{c}
\frac{\partial I_S}{\partial t}=-\Upsilon\nabla V_{S},
\label{SuperfluidEOM}\vspace{2mm}\\
I_{N}= -\sigma \nabla V_{N}
\label{NormalEOM} \\
{\cal I}_{NS}=\gamma (V_N-V_S)  \label{ConvCurrWire}
\ea
\label{eq1}
\ee
 with $V_{S}$ and $V_{N}$  the electrochemical
potentials for the superfluid and normal electrons respectively.  
 Note that for a wire, the conversion current  ${\cal I}_{\rm NS}$
is a current per unit length, so $\gamma$ has dimensions
conductance per unit  length. $\gamma$, the
 conductivity between the normal and superfluid parts, reflects the
 relaxation rate for a population imbalance between the two fluids, as
 investigated, e.g., by Clarke {\it et al.} \cite{Clarke1979}.
The current equations must be supplemented, as for the chain of grains, by current conservation laws and constitutive relations between the excess normal and superfluid charge densities and the electrochemical potentials; again these will involve compressibilities and capacitances.

The continuum two fluid equations, Eqs. (\ref{SuperfluidEOM}), can be made into
a quantum model by defining the superfluid velocity as the gradient of
the superconducting phase, imposing phase-charge conjugation, and introducing appropriate degrees of freedom to mimic the dissipation associated with the resistive processes. But, in addition, the superconducting phase should be allowed to undergo {\it quantum phase-slips} (QPS); these are  implicit in the chain of grains but {\it not} included in the linear continuum model (for a
discussion of phase-slips see Sec. 
\ref{StrongJosephsonCouplingSection}). To introduce phase slips, a short distance cutoff must be imposed. as, in contrast to the chain, there is no intrinsic length scale. Care must be
exercised,however, when imposing a  cut-off on this nanowire model. For example, a simple
lattice regularization implicitly assumes that the size of  quantum phase slips  is also
the shortest possible distance between them. In reality
the core size of a QPS is non-zero, but,
if two phase slips occur at different times, their {\it centers} can be 
arbitrarily close to one other. We discuss this important subtlety in detail in Sec.
\ref{ContinuumLimitSection}.

\subsection{Overview of Results \label{resintro}}
\label{intro_overview}


\begin{figure*}
\includegraphics[width=15cm]{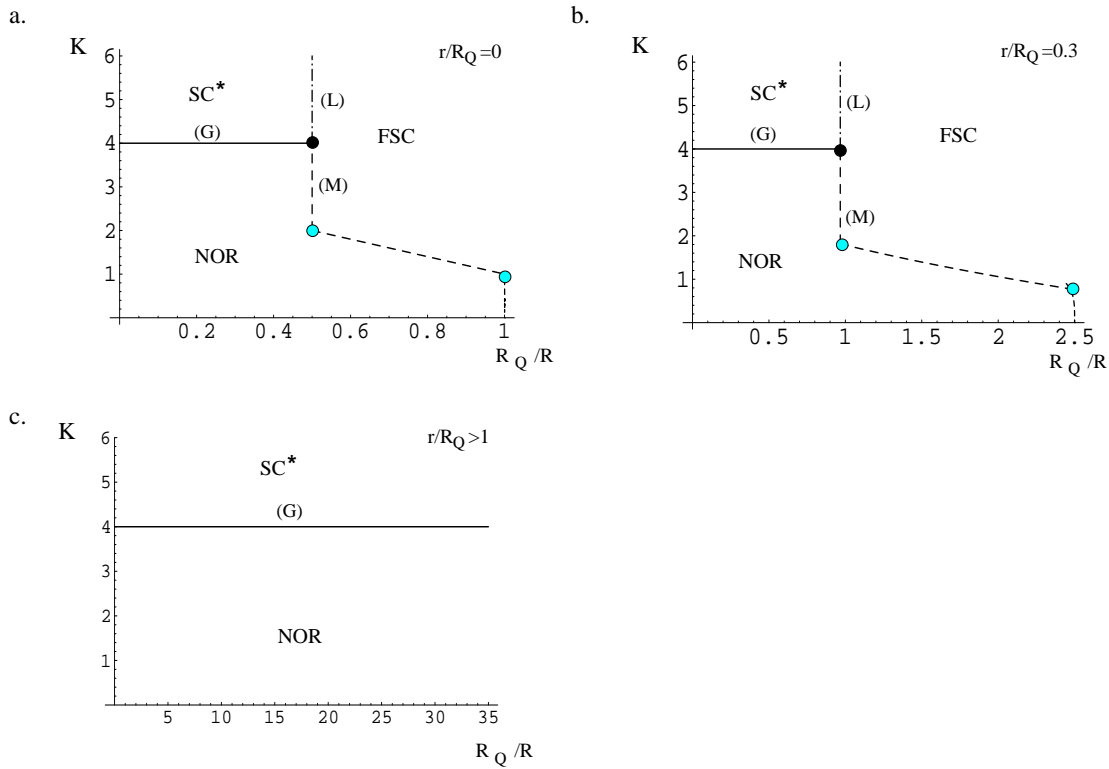}
\caption{Schematic phase diagram of the quantum two fluid model of
  Fig. \ref{Fig1intro}(c) as a function of the shunt resistance, $R$, and the quantum superconducting stiffness,  $\j =2\pi\sqrt{\frac{E_J}{E_C}}$, for various values of the conversion resistance, $r$. The phase boundaries between the FSC, \scs, and NOR
  phases are of different nature: (G) indicates global (solid line), (L) 
  local (dashed-dotted line), and (M) mixed (dashed lines).  These come together at a multicritical point (black dot). The FSC-NOR phase boundary has sections with three different characters, separated by bicritical points (gray dots)  Most of the phase boundaries are derived from  the strong
Josephson limit (Sec. \ref{sec5}), but their positions will depend on the fugacity, $\zeta$, of quantum phase slips, which should be another axis. For small $K$, the 
  weak coupling expansion (Sec. \ref{wcsec}) is needed: that portion of the FSC-NOR boundary is inferred from it. 
  \label{wphases_intro}}
\end{figure*}


Before formulating the quantum model of a chain of grains and analyzing it in detail, we give a
brief overview of the results obtained in this paper.

\subsubsection{Phases}

The JJ chain has two superfluid phases, which we call {\it fully
  superconducting} (FSC) and {\it \scs},  plus a normal metallic phase (NOR). [The two types of superconducting  phases were
discussed earlier in Refs. \onlinecite{KorshunovA, KorshunovB, Schoen} and
referred to as SC-2 and SC-1. ] In the FSC phase, which occurs when dissipation is strong, quantum fluctuations are suppressed enough that 
conventional superconducting tunneling into a grain is
  possible, and the Josephson junctions behaves completely classically at zero temperature. Nevertheless the superconducting correlations are not
  truly long range, but decay algebraically with distance.  In the
  opposite limit, the normal phase,  phase fluctuations are large, and
  number fluctuations are small, so that the charge of a grain is well defined and superconducting correlations decay exponentially. 

 In between the FSC and NOR phases is the \scs phase, in which  {\it both} the phases and the charges of
  individual grains exhibit large fluctuations.
  Phase differences between grains, however, do exhibit quasi-long
  range order. In this remarkable phase, tunneling into a grain will result in
  singular low energy behavior, and will be suppressed, which is an indication of the
  strong fluctuation of the phase variables on each grain in this
  state.  Nevertheless a uniform supercurrent can flow {\it through} the
  chain unimpeded at low energies. 
The \scs phase tends to occur when the dissipation is weak: at low
energies in the \scs phase, the dissipation is irrelevant with the
quantitative values of the  shunting, $R$, and conversion, $r$,
resistances playing little role, in the determination of global
  transport properties.

 The three phase structure is somewhat analogous to the behavior of two-dimensional solids describable  by disclinations: at intermediate temperatures the disclinations bind into pairs forming dislocations --- the hexatic phase with quasi-long range orientational order --- and at lower temperatures, these dislocations bind into pairs --- disclination quadrupoles --- forming the solid phase which has true orientational long range order.

\subsubsection{Transitions}

Three types of quantum phase transitions occur between the three phases.  We denote the critical values of parameters for these transitions, $G,\ L$ and $M$ for global, local and mixed respectively.

Separating the \scs phase and the normal phase, is a transition of
Kosterlitz-Thouless (KT) type driven by unbinding of pairs of quantum
phase slips. It is thus controlled by  an effective space-time phase
``stiffness" parameter that we denote $K$, proportional to the square
root of the ratio of the Josephson and capacitative energies: the
former favors superconductivity, while the latter favors localization
of charge, and hence normal behavior. We refer to the \scs-NOR transition, as it
is intrinsically controlled by long-wavelength physics, we refer to as
{\it global} ($G$). Interestingly, in this transition, the resistive
shunts play almost no role: they are screened by the fluctuations of
{\it phase slip dipoles}, i.e., a pair of phase-slip and anti-phase-slip
occurring simultaneously on neighboring junctions.  The dynamical exponent, $z$, that relates time or
inverse-energy scales to length scales,  is, for this transition,
equal to unity: $z_G=1$. Deviations from criticality are only
marginally relevant, in the RG sense, and give rise to characteristic
energy scales going to zero at the critical point with the exponential
form typical of KT transitions.

Between the two superconducting phases, the transition is driven by
dissipation. It is essentially {\it local} (L), being related to the
superconducting-normal transition of a single Josephson junction. The
important excitations that control this local transition are phase
slips between one grain and the rest of the chain.  In terms of
individual quantum phase slips of the chain, these are equivalent to
{\it quantum phase slip dipoles}, which, as mentioned above, consist of a pair of opposite
sign phase slips, one on each side of the grain.  Associated with the
locality of this physics, there is no simple diverging length scale
and the dynamical exponent is thus $z_L=\infty$.  How the
characteristic energy scale goes to zero at this transition depends
on values of resistances: the corresponding critical exponents vary
continuously.

Perhaps the most interesting transition is that which can occur from
the FSC phase {\it directly} to the normal phase: this is driven by
changes in the dissipation, yet because it also involves destruction
of superconductivity, it has {\it mixed} (M) character with  both
local and global aspects and involves the interplay between individual
quantum phase slips, and phase slip dipoles. Related to this more
complicated mechanism, there  is more than one type of critical
behavior --- probably three types --- for the mixed transition.

\subsubsection{Phase diagram}

A schematic zero-temperature phase diagram for the chain   is shown in
Fig. \ref{wphases_intro}.   It is convenient to show this as a function of
the shunt resistance, $R_Q/R$, and the stiffness superconducting
stiffness, $K$, at various values of the conversion resistance,
$r$. [Although a fourth parameter, related to QPS fugacities, is
  really needed as well to exhibit the range of possible behaviors.]
Here $R_Q$ is the quantum resistance for Cooper pairs:
\be
R_Q=\frac{h}{4e^2}\approx 6.5k\Omega.
\ee
For large $r$, $r>R_Q$, the phase fluctuations on each grain are
large, and only the normal and \scs phases can exist.  In this regime,
the phase diagram is simple (Fig. \ref{wphases_intro}) with the global transition between these phases at $\j=\j_G\approx 4$. This transition is driven by proliferation of quantum phase slips, is analogous to the classical Kosterlitz-Thouless transition in two dimensions, and has  characteristic energy scale going to zero exponentially rapidly as the transition is approached, and a correlation length that diverges with the inverse of this energy scale.

For intermediate $r$, $R_Q>r>r_c\approx \half R_Q$, all three phases
exist, but the mixed transition
between the FSC and normal phases is always driven by dipoles. The
critical $R$, $R_M$, varies with $r$ and weakly with $K$: it  is close
to when a particular combination of $R$ and $r$ is $R_Q$. The energy
scale goes to zero as a power of $|R-R_M|$ with an exponent, $\nuz$ that varies continuously with $r$.
The characteristic length scale on the normal side --- associated with
proliferation of individual dipoles ---  diverges as a power of
$R-R_M$  with a continuously variable exponent that is not simply
related to the that of the  inverse energy scale. The FSC-\scs local
transition is similarly driven by dipoles, and  occurs when the same
combination of $R$ and $r$, is close to $R_Q$.  The energy scale
similarly goes to zero with a continuously variable exponent.  

For $r<r_c\approx \half R_Q$, the phase diagram is far richer (Fig. \ref{wphases_intro}), and qualitatively similar to the previously studied $r=0$ case of  infinitely fast relaxation between the normal and superconducting electrons.  The \scs-normal and FSC-\scs transitions are similar to the intermediate $r$ regime discussed above.
But the mixed-character phase transition between the FSC and normal phases   is more complicated.  For large-intermediate Josephson coupling, $\j_B<\j<4$ with $\j_B\le 2$ depending on $r$ and $R$, the mixed transition is similar to that in the intermediate $r$ regime (above).   In the limit of weak Josephson coupling --- small $\j$ --- its character is different.   In this regime, the mixed transition is most easily thought of as being driven --- from the normal phase --- by Cooper-pair tunneling. It 
occurs when 
$R=R_M\approx R_Q-2r$ with $R_M$ decreasing to this value as $\j\rightarrow 0$. There is power law scaling of energy  in the vicinity of this transition, with continuously variable exponents. 

In some regime of parameters, the small coupling and large-intermediate coupling phase boundaries may join together: if they do so, it will be in an intermediate coupling regime that is far from all three phases and thus for which we have 
no controlled methods to study. At this point, it is not clear whether or not these two regimes can join continuously. Naively , their character, particularly the behavior of length scales, seems rather different.  But it is possible that they are related and then could join continuously.

In the regime in which we can understand the behavior in terms of
quantum phase slips, there is a section of the FSC-normal phase
boundary which has different character than those discussed above. For
$\j<\j_B$ but not too small --- small-intermediate -- direct
proliferation of individual QPS (rather than their proliferation
caused by dipole proliferation) drives the transition of the the FSC
phase.  Although the behavior is more subtle and there are complicated
crossovers, the asymptotic critical behavior is similar to the
Kosterlitz-Thouless transition between the \scs and NOR phases with
exponentially vanishing energy scale and length scale diverging as the
inverse of this: i.e. isotropic behavior.  Surprisingly, the location
of the transition is determined by a combination of the low energy
properties of the superconducting and normal --- i.e. dissipative ---
degrees of freedom. But how this changes measurable asymptotic
properties we have not worked out.

In the low range of $r$ in which the mixed transition can have more
than one character, the different sections of the phase diagram join
together at two bicritical points. The  weak-coupling power-law regime
joins up to the small-intermediate coupling  KT-like regime at a point
that is neither near the $\j=0$ nor the large $\j$ regime and thus not amenable to 
study by the methods we use.  But the other bicritical point at which the dipole driven and
individual QPS driven segments of the phase boundary come together can
be analyzed: we discuss it briefly in Sec. \ref{bicrit}.

The three phases come together at a multicritical point as shown in
the figures.  The behavior near this point involves crossover from the
FSC to either the \scs or normal phase to the left of the
nearly-vertical phase boundary. The asymptotic critical behavior,
however, is probably controlled by the \scs phase. We have not investigated this in
detail.

\subsubsection{Renormalization group analysis}

The primary methods we use are renormalization-group  (RG) analyses of 
the effective low energy action for the
quantum two-fluid JJ chain model;  both Coulomb gas and sine-Gordon
representations of this action are derived and used in the various regimes. The zero-temperature phase
diagram in various limits, and the nature of the quantum phase transitions more generally,  are derived from these.  
The RG method naturally leads to  detailed
understanding of both the qualitative and the quantitative roles of various aspects of the model, for example, that the dissipation
is irrelevant in the \scs phase and near  the ``global" normal-to-\scs phase transition. For the normal-to-FSC
transition, on the other hand,  dissipation plays a key role by 
suppressing local fluctuations of the phase, and the characteristic length scale is the relaxation length  between the superconducting and the normal
fluids;  as $r$ varies from zero and infinity, this length-scale changes
from the inter-grain spacing, $a$, to infinity, although the low energy properties of the transition from the FSC to the \scs phase remains  ``local".

The RG flow equations  can also be used to obtain 
the temperature dependence of various quantities, notably the resistivity close to the superconductor-to-normal
transitions.  In the vicinity of the global normal to \scs transition
the dissipation plays little role and the temperature dependence of the measured total resistivity near this quantum KT transition
has been analyzed by other authors \cite{ZAIKIN1997, Duan1995}. 
But in the vicinity of the mixed normal to FSC
transition, the behavior is strikingly different. 

The RG analysis also gives information about crossovers and regimes of validity of the asymptotic behaviors. It is important to emphasize that over much of the parameter ranges, 
the superconductor-to-normal transitions are likely to be  characterized
by wide crossover regions (see Fig. \ref{wphases_intro}) in which, for example, the resistivity can be almost temperature independent over extended temperature ranges. This may correspond to the
``supermetallic'' behavior in the vicinity of superconductor-to-normal transitions,
that has been observed in several experiments in one dimensional
\cite{GiordanoA,GiordanoB,GiordanoC,GiordanoD,HavilandE}
and two
dimensional systems \cite{MASON2002,HavilandA}.

The RG can also be used
to study finite size properties and effects of boundary conditions. 
For example, we show that
when the RSJJ chain is connected to superconducting
electrodes, at asymptotically low temperatures  in the  \scs regime,
the appropriate measure of the dissipation that controls the location
of the macroscopic  superconductor-to-normal transition is the {\it
  total}  shunting resistance, in contrast to the traditional picture
that the {\it local} inter-grain shunting resistance will control the
macroscopic behavior (see also discussion, Sec. \ref{discuss}). \cite{finitechain}

\subsubsection{Nanowires}

Late in the paper we go briefly  from the realm
of discrete grains and Josephson junctions to that of continuous superconducting wires.
For these  there is a subtle question about whether the FSC phase exists.  The approximate model we use, leads to the conclusion that it does not. If this is correct, then at sufficiently low temperatures,  the wires will always approach the \scs or the normal phase, but there can be wide regimes of crossover in which FSC-like behavior may be observable.   But aspects that have been left out of the model, in particular aspects of charge discreteness and interference between quantum phase slips, may invalidate this conclusion.  Preliminary indications are that these can stabilize the FSC phase.  Its existence is thus left as an open question.


\section{Mesoscopic Model of Chain of Grains \label{secmm}}

The system we study consists of 
identical mesoscopic superconducting grains that are connected by 
weak links which allow the
flow of both Cooper pairs and normal electrons (Fig. \ref{Fig1intro}(a)). The flow of Cooper pairs is via 
Josephson coupling between the grains and the flow of normal electrons via shunting resistors. 
As discussed above, it is convenient to describe this system in terms of a two fluid
model. The superfluid is transported by Cooper pairs tunneling between superconducting
grains, and the normal fluid is transported by electrons through the shunting
resistors.   Each grain is thus considered as a double grain with superconducting
and normal parts as shown in Fig. \ref{Fig1intro}(c).  The
{\it S} (superconducting) and
{\it N} (normal) parts of each grain experience the {\it same} electrostatic
potential, $\varphi$, but they will generally have different chemical potentials. [An analogous
decoupling of the chemical potentials for the {\it S} and the {\it N}
fluids  near phase-slip centers in superconducting wires
in the presence of a transport current was discussed in Refs. \onlinecite{beasley,tinkham2,ivlev}.] 

The
possibility of having different chemical potentials for the normal and superfluid components is a consequence of the
mesoscopic size of the grains. The conversion resistance between the
two fluids within a grain is proportional to some power of the size of
a grain, and therefore it will provide significant dissipation only
for small grain sizes (see discussion in Appendix \ref{appA}
and Ref. \cite{Refael2003}). The dependence of $r$ on the size of a
grain should be obtained from an appropriate microscopic model, which
is beyond the scope of this paper. 
The sum of the electrostatic potential, $\varphi_i$ and the chemical potentials, $\mu_{N\,i},\ \mu_{S\,i}$, of the $i$'th grain  yield the total normal and superfluid electrochemical potentials, $V_{N\,i}$ and $V_{S\,i}$, which will drive the currents. 

Changes in the total charge on a grain  modify its electrostatic potential via the capacitance, $C$, while changes in the normal or superfluid charges on a grain  correspondingly modify the chemical potentials with the coefficients the  inverse  compressibilities $D_{N}$ or  $D_{S}$ of the
 {\it N} and {\it S} fluids on an individual grain; these compressibilities thus have the character of ``quantum capacitances" \cite{Buttiker93}.

When the electrochemical potentials of the {\it N} and {\it S} fluids on a grain
differ, relaxation processes will occur to equilibrate the two components. The simplest form for such relaxation is an {\it Ohmic}  conversion current, $I_{NS\,i}$: \cite{Refael2003}
\begin{eqnarray}
I_{NS\,i} = \frac{V_{N\,i}-V_{S\,i}}{r},
\label{ConversionCurrent}
\end{eqnarray}
where  $r$ is the conversion resistance within a grain. We assume that $r$ remains finite even 
in the limit of zero temperature.

The charge
relaxation time between superconducting and normal fluids on a grain is set
by the  $RC$ time of the effective circuit, 
\be
\tau_{NS}=r (D_S
+D_N)^{-1} \ .
\ee
Therefore our assumption of an Ohmic conversion resistance is equivalent to assuming a form for the relaxation rate between the
normal fluid and the superfluid.

The effective low-energy model of the system shown
in Fig. \ref{Fig1intro}(c) should include the charging energy
for the grains, the Josephson
coupling energy, and appropriate heat bath Hamiltonians
for the shunting and conversion resistors. 
We now construct the appropriate Hamiltonian and
thence obtain the corresponding quantum action. 

We start with the charging energies. 
The electrochemical potentials for the superconducting and the normal
fluids include contributions from both electrostatic and
electrochemical ``capacitances" (see Fig. \ref{Fig1intro}(c))
\be
\ba{rcl}
V_{S\,i} &=& \varphi_i + D_S Q_{S\,i},\\
V_{N\,i} &=& \varphi_i + D_N Q_{N\,i}
\label{electrochemical}
\ea
\ee
where $\varphi_i$ is the electric potential of grain $i$, and $Q_{S\,i},\,Q_{N\,i}$ are
the superconducting and the normal
parts of the charge on grain $i$; and $D_{S}$ and $D_{N}$ are
the inverses of the compressibilities of the
{\it S} and {\it N} grains respectively. In this paper we consider a simplified
model in which only the self-capacitance of each grain, $C$, is
included. We can then write the electric potential as
\be
\varphi_i=\frac{1}{C}\l(Q_{S\,i}+Q_{N\,i}\r).
\ee
We expect that including mutual capacitances between the grains
\cite{Bradley1984,Gurarie2003} will not change the qualitative
conclusions, although it may modify the energy
scales involved.  

By integrating the
electrochemical potentials in (\ref{electrochemical}) with respect to
the charge, we obtain the charging part of the Hamiltonian. The resulting term is
\be
\ba{c}
{\cal H}_Q =
 \frac{1}{2} (C^{-1}+D_S) \summ_i
Q_{S\, i}^2 + \\
 \frac{1}{2} (C^{-1}+D_N) \summ_i
Q_{N\, i}^2 + C^{-1} \summ_i Q_{S\, i} Q_{N\, i}.
\label{HamiltonianCharging}
\ea
\ee

In order to write down the action for the dissipative  and 
Josephson terms, we define phase-angles conjugate to the
charges $Q_S$ and $Q_N$. For each grain $i$ the superconducting phase
$\phi_i$ and the ``normal phase'' $\chi_i$ are defined via 
\cite{Schoen-Zaikin,Refael2003}
\be
\begin{array}{ccc}
\l[ Q_{Ni},\chi_j \r]=-ie \delta_{ij}	&&	\l[Q_{Si},\phi_i \r]=
-2ie \delta_{ij} \vspace{2mm}\\
\l[Q_{Ni},\phi_j \r]=0	&&	\l[Q_{Si},\chi_j \r]=0.
\label{QPhiCommutation_relations}
\end{array}
\ee
The physical interpretation
of $\chi_i$ follows from the observation
that its time derivative gives the 
electro-chemical potential of the normal fluid, by analogy with the Josephson relation for the superfluid 
\cite{Schoen-Zaikin,Kamenev1999,INGOLD1992}.

The Hamiltonian term arising from the Josephson tunneling between
grains can be readily written in terms of $\phi_i$:
\be
{\cal H}_J = -E_J \summ_{\langle i j \rangle} \cos (\phi_i - \phi_j ),
\ee
where the summation is over nearest-neighbor grains, $i$ and
$j$. $E_J$ is the Josephson coupling energy of the Josephson junctions, given by $E_J=\frac{\hbar}{2e}I_J$ in terms of $I_J$, the
critical current of the junctions.

The dissipative parts of the system can be described  
by heat bath Hamiltonians with appropriately
chosen spectral functions. These are written as follows,
\be
{\cal H}_{\rm dis} = \summ_{\langle i j \rangle}{\cal H}_{\rm bath} (R, 2 \chi_i - 2 \chi_j)+ \summ_i {\cal H}_{\rm bath} (r, \phi_i - 2 \chi_i)
\label{HamiltonianDissipation}
\ee
We do not give the explicit form of ${\cal H}_{\rm bath}$
here, but below provide the corresponding effective actions
obtained after integrating out the heat-bath degrees of freedom. The
crucial requirement for Ohmic heat baths is that their density of states
is linear at low frequencies. Note that  this is the case for
particle-hole excitations of a Fermi liquid, one likely source of dissipation in dirty gapless superconductors especially near transitions to normal metallic behavior.

Combining Eqs.(\ref{HamiltonianCharging}) - (\ref{HamiltonianDissipation})
we construct the imaginary time
action and partition function for the  RSJJ array of
Fig. \ref{Fig1intro}(c):
\begin{widetext}
\be
\ba{c}
Z= \int {\cal D} Q_{N} {\cal D} Q_{S} {\cal D} \phi 
{\cal D} \chi \exp \left( -S \right)
\vspace{2mm}\\
S=
-\frac{i}{2e} \summ_i \int_0^\beta d\tau\, Q_{Si}\, \dot{\phi}_i 
- \frac{i}{e} \summ_i \int_0^\beta d\tau\, Q_{Ni}\, \dot{\chi}_i 
+ \int_0^\beta d\tau {\cal H}(Q_{Ni}, Q_{Si}, \phi_{i}, \chi_{i})
\vspace{2mm}\\
{\cal H}(Q_{Ni}, Q_{Si}, \phi_{i}, \chi_{i}) 
= {\cal H}_Q + {\cal H}_J + {\cal H}_{dis}.
\label{ZwithQ} 
\ea
\ee 
\end{widetext}
In the presence of Ohmic
dissipation or an external current source, the phase variables $\phi_i$ and $\chi_i$ should be
periodic at $\tau=0$ and $\tau=\beta$ with {\it no phase twists} by
multiples of $2\pi$ allowed: in contrast to simple non-dissipative Hamiltonians, there is a physical distinction between $\phi_i - \phi_j=0$ and $2\pi$ (see e.g. the discussion in Sec. IIB  of Ref. \onlinecite{Refael2003}).

We can integrate out the  $\{Q_{N\,i}\}$ and $\{Q_{S\,i}\}$, as these appear quadratically in (\ref{ZwithQ}), and obtain
\be
\ba{rcl}
Z &=&\int D \phi D \chi \exp\l(-S_{chain}\r)\vspace{2mm}\\
S_{chain} &=& S_Q + S_{\rm dis}^r + S_{\rm dis}^R +S_J\vspace{2mm}\\
S_{Q} &=& \int d\tau \summ_i \frac{1}{( D_N + D_S + C D_N D_S)}
\l[\frac{1}{2}\l(\dot{\phi}_{i}{(\tau)}-
2 \dot{\chi}_{i}{(\tau)}\r)^2\vspace{2mm}\r.\\
&+&\l.\frac{C D_S}{2}\, (2 \dot{\chi}_{i})^2 +\frac{C D_{N}}{2}\, \dot{\phi}_{i}^2\r]\vspace{2mm}\\
S_{\rm dis}^r&=&\beta \summ_{\omega_n}
\l(\frac{|\omega_n|}{r}
|\phi_{i}{(\omega_n)}-2 \chi_{i}{(\omega_n)}|^2\r)\vspace{2mm}\\
S_{\rm dis}^R&=&\beta \summ_{\omega_n} 
\sum _{\langle i j \rangle} \frac{|\omega_n|}{R}
| 2\chi_{i\, (\omega_n)}
-2\chi_{j}{(\omega_n)} |^2\vspace{2mm}\\
S_J &=& - E_J \int_0^\beta d \tau 
\summ_{\langle i j \rangle} \cos\, \left( \phi_i - \phi_j \right)
\label{allaction}
\ea
\ee
where Matsubara frequencies $\omega_n = 2 \pi n T$
and Fourier transforms of the periodic
functions $\phi_i(\tau)$, $\chi_i(\tau)$ are used with
$f(\omega_n)\equiv\int_0^\beta f(\tau) e^{i \omega \tau} d\tau$. In the limit of zero temperature, which we will primarily study, $\beta \summ_{\omega_n}$ is replaced by $\int\frac{d\omega}{2\pi}$.

\subsection{Scales, Parameters and Regimes \label{regimes}}

There are several important energy and length scales in the mesoscopic model of a chain of RSJJ's, as well as several key dimensionless parameters.

The superconducting energy scale is the Josephson coupling energy, $E_J$. Competing with it are  (twice) the charging energy of a grain,
\be
E_C=\frac{(2e)^2}{C} \ 
\ee
and the analogous   non-electrostatic portions of the energies of adding a normal electron,
\be
E_{DN}=e^2D_N\ ,
\ee
 or a Cooper pair, 
 \be
 E_{DS}=4e^2D_S \ ,
 \ee
  to a grain.
In most  situations of interest the energy scales $D_N$ and $D_S$ satisfy, 
\be
D_N\sim D_S \ll 1/C \ ,
\ee
so that  the electrostatic energy dominates.
At low energies the $D$'s drop out and only $C$ is important. 
The ratio of the Josephson to the charging energy then determines the dimensionless quantum phase-stiffness
\be
K=2\pi \sqrt{E_J/E_C} \ .
\ee

In the absence of dissipation, the low temperature behavior is controlled by the parameter $K$. For large $K$, the phase differences between neighboring grains are small and the Josephson coupling can be approximated by $E_J (\phi_{i+1}-\phi_i)^2/2$. 
This yields the conventional quadratic Hamiltonian for the superconducting degrees of freedom. With inter-grain spacing $a$, the  characteristic velocity of the phase fluctuations is
\be
\vms=\frac{a\sqrt{E_JE_C}}{\hbar},
\ee
which corresponds to the {\it Mooij-Sch\"on velocity}. \cite{Mooij,Camarota2001}   In this superconducting phase, the correlations decay as powers of distance and imaginary time with an exponent proportional to $K$. 

The dissipation can be parametrized by the dimensionless resistances,
 $R/R_Q$ and $r/R_Q$ where we used the quantum of (Cooper pair) resistance, 
\be
R_Q=\frac{2\pi\hbar}{4e^2} \ :
\ee
these dimensionless resistances are key control parameters.  When $R\ll r$,  the effects of the two resistances --- $R$'s in series and $r$'s in parallel --- becomes comparable at a  length scale
\be
\xz\approx\ax\sqrt{r/R}
\ee

which plays an important role.   In this small $R$ regime, the dimensionless measure of the dissipation is 
\be
\aeff\approx \frac{R_Q}{2\sqrt{rR}} \ .
\ee
When $R$ is comparable to or larger than $r$, the characteristic length and dissipation measure have more complicated dependences, in particular,
\be
\aeff=\frac{R_Q}{\sqrt{R^2+4rR}} \ :
\ee
as discussed in the next section, this can be understood from electro-dynamical considerations.

The {\it dissipative energy scale} is determined by the competition between the compressibilities (which did not directly affect the superconducting degrees of freedom) and the resistances. The effective resistance, 
$R^* \sim \min(r,R)$ and effective capacitance, 
 \be
 C_{SN}=\frac{1+CD_S}{D_N+D_S+C D_N D_S}\approx \frac{1}{D_N+D_S} \label{Ceff}
 \ee
 together give  the characteristic relaxation time that determines  the energy scale $T^*$ parametrizing the coupling between the normal and superconducting currents:
\be
 T^*=\frac{\hbar}{R^* C_{SN}} \ .
 \ee
The energy scale, $T^*$, can also be written in terms of the other
 energy scales. From Eq.(\ref{Ceff}), one finds that  $T^*$ is
 proportional to the escape time of normal electrons from
 the grain, i.e., to the Thouless time of the grain. It is thus very large for macroscopic grains, but its existence is
 an essential property of the mesoscopic physical content of our model.

 At temperatures higher 
than $T^*$, the  dissipation across separate junctions is effectively decoupled. In contrast, for $T\ll T^*$,  the interactions between dissipation across different junctions and within different grains are important. These interactions are crucial for the quantum dynamics.
At low energies, the existence of the compressibilities thus matters crucially. although the values of the $D$'s do not.

Before proceeding with more sophisticated analyses, it is instructive
to consider the linearized dynamics in the presence of dissipation.
As all the terms in the model action except the Josephson coupling are
quadratic, we can integrate out all but the superconducting phase,
$\phi$, and  expand the Josephson coupling energy about zero phase
difference. This results in a rather messy form of the action that is
given in Appendix \ref{appA}.
In the limit of low frequencies and long wavelengths the dissipative
effects are negligible and all that matters are the conventional
superconducting parts of the action including the suppression of
imaginary-time changes of the phase by the inverse of the total
effective charging energy. These give rise to the simple phase modes
discussed above. Nevertheless, the fact that the short wavelength
fluctuations are controlled by the dissipation, makes the terms that
appear negligible at long wavelengths also important for the quantum
dynamics and thence the phase diagram.

In the absence of dissipative effects, the mean square fluctuations of the phase difference between neighboring grains would be inversely proportional to the quantum stiffness, $K$.
But these fluctuations  are dominated by  wavelengths of order the
inter-grain spacing,  and frequencies of order the Josephson junction
plasma frequency  (proportional to  $\vms/a$). Thus, in actuality, the
modes that dominate the phase fluctuations will be affected by much of
the details of the high frequency dynamics, including the dissipation
and the cutoff frequencies of both the dissipative and the
superconducting degrees of freedom. We will often crudely approximate
these by a high-frequency cutoff $\Omega_0$. The short wavelength
processes also control the action of a quantum phase slip via
properties of its core, including its space-time  size and ``shape";
in particular, the ``fugacity" of phase slips will be exponentially
small when their core action is large.

Because of the importance of the high energy physics, it is not
obvious what the significant dimensionless parameters are, beyond the
obvious one discussed above,  nor what role these might play at low
energies. In practice, whether a chain of grains is in a ``strong" or
a ``weak" Josephson coupling regime will be determined by many
properties. Thus we will use  these terms loosely to describe various
regimes in which the behavior simplifies: small QPS ``fugacity", $\zeta$ for strong coupling, and small Cooper pair tunneling rates for weak coupling.  Care must thus be exercised in considering phase diagrams of more explicit models as changing one parameter can result in, for example, changing both the low frequency   dissipation, and the fugacity of quantum phase slips. 

To simplify discussions of phase boundaries and the behavior near them, we will generally consider tuning the shunt resistance $R$ and the strength of the Josephson coupling, either $E_J$ itself or the QPS fugacity as a proxy for this.  The phase diagrams 
in general need to be considered as functions of both $r$ and
parameters related to other high energy processes as well.


\subsection{Circuit analysis \label{ctm}}

\begin{figure}
\includegraphics[width=7cm]{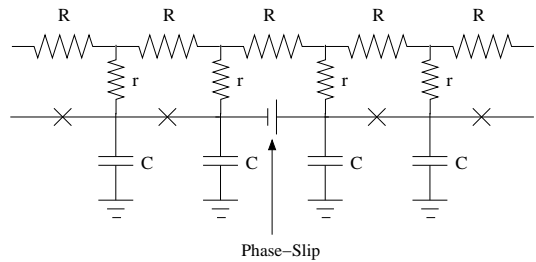}
\caption{A phase slip creates a potential drop across the junction, which propagates currents in the rest of the chain.\label{pscircuit}}
\end{figure}

In the analysis of the two junction problem in Ref. \onlinecite{Refael2003}
we showed that electrical circuit properties determined the (lowest order)
RG flows in various  limits. Before proceeding further, we analyze the linearized electrodynamics  of the chain model to gain insight into its behavior.

The basis of the circuit approach is the following simple
interpretation of the dissipation-driven transition in a single
resistively shunted Josephson junction \cite{Schoen-Zaikin, INGOLD1992}. When the junction is
in the insulating state, the Cooper pairs are localized on the
electrodes, and the phase difference across the junction is uncertain
due to proliferation of quantum phase slips (QPS).  In the opposite limit when the junction is superconducting, the phase difference between the grains is well defined and, phase slips will not occur on long time scales. A phase slip across the junction causes a voltage burst with the Josephson relation giving 
\be
\int dt V(t) =\frac{h}{2e}=\frac{2\pi\hbar}{2e}
\ee
which leads to a current flow in the shunting
resistor, $R_S$.  It turns out that the superconducting state of the junction is stable only when a single QPS would cause a charge flow, $\Delta Q$, which is more
than a Cooper pair charge $2e$. From 
Ohm's law we then obtain $\Delta Q/2e=R_Q/R_S$, so superconductivity will be observed when $R_S < R_Q$. A related analysis from the insulating state in terms of Cooper pair
tunneling events can be used to argue for the same condition: $R_S>R_Q$ stabilizes the insulating state.

\subsubsection{Cooper pair tunneling}

In the normal phase the grains are only coupled resistively, although
they can exhibit superconducting correlations within each grain. Thus
we can consider a putative Cooper pair tunneling from one grain to the
next. This is very similar to the single junction case, except that
now the total shunt resistance is $R+2r$ because of the contributions
of the conversion resistances in each grain (see
Fig. \ref{pscircuit}). We are interested in the effect of this tunneling on the SC phase difference, $\phi\equiv\phi_{i+1}-\phi_i$  across the junction. By the Josephson relation, the change in $\phi$ is given by
\begin{widetext}
\be
\Delta \phi =\int dt \frac{d\phi}{dt}=\int dt \frac{2eV(t)}{\hbar} =\frac{2e V(\omega=0)}{\hbar}=\frac{2eI(\omega=0)Z(\omega=0)}{\hbar}=\frac{2e\Delta Q Z_S}{\hbar}=2\pi \frac{Z_S}{R_Q}  
\ee
\end{widetext}
with $I$ the current, $\Delta Q=2e$ the charge transfered, and $Z$ the impedance.  Thus we see that the phase difference between the grains will change by more than or less than $2\pi$ according to whether the shunting impedance, $Z_S=R+2r$, is more or less than $R_Q$.  Since the transition from the normal phase to the FSC phase in which each grain has a well-defined SC phase is via this local process, it is not surprising that the condition for this to occur involves this combination. As we shall see, the behavior is in fact more subtle due to the effects of multiple Cooper pair tunnels on each other. 

It is also instructive to consider the classical action associated
with the tunneling of a Cooper pair.  Since the phase and the Cooper
pair number are canonically conjugate, the action will be given by
\begin{widetext}
\be
S_{pt}=\hbar \int \frac{Q}{2e} d\phi = \int dt Q(t) V(t)=\int \frac{d\omega}{2\pi} Q(\omega)V(-\omega)= \int \frac{d\omega}{2\pi} \frac{I(\omega)}{-i\omega +0}V(-\omega)
=\int \frac{d\omega}{2\pi} \frac{Z(\omega)|I(\omega)|^2}{(-i\omega+0)}  \ ;
\ee
\end{widetext}
note the integral  of the potential energy appearing.
Since in the limit of zero frequency the impedance is purely real and $I(\omega)\rightarrow 2e$,  the zero frequency part of this integral gives $\pi \hbar Z_S/R_Q$.  The finite frequency parts will be negligible if the transfer is slow; otherwise they will decrease the action so that 
\be
S_{pt}\leq \frac{\pi \hbar (R+2r)}{R_Q}.
\label{Sptr}
\ee
In imaginary time, the $1/(-i\omega+0)$ becomes $1/|\omega|$ and the integral over frequencies diverges logarithmically at low frequencies or low temperatures. It is exactly the competition  between this logarithmic action and the quantum ``entropy" --- log of the range of imaginary time $\hbar/T$ --- in which the event can occur --- that determines whether the junction is superconducting, as we shall see.  In real time, the significance of Eq. (\ref{Sptr}) is not clear, in particular, whether $\exp(iS/\hbar)=\exp(i\pi)$ is significant as far as whether or not Cooper pair tunnels can destructively interfere, and if they can,  thereby suppressing superconductivity for large shunt resistance. 

\subsubsection{Effective shunting resistance in the strong Josephson
  coupling limit \label{Rinfty}}
  
 We now turn to the superconducting phase in which the Josephson junctions are all superconducting.  
Consider a phase slip across one junction in the chain.
By analogy with the single junction case, one would guess that the relevant quantity is the low frequency limit of the impedance, $Z(\omega)$, of the circuit parallel to it. This parallel circuit involves all the other superconducting junctions and the resistors.
The total dissipation now comes from 
both the network of the Ohmic
resistors and the Josephson junction chain itself. 

\begin{figure}
\includegraphics[width=7cm]{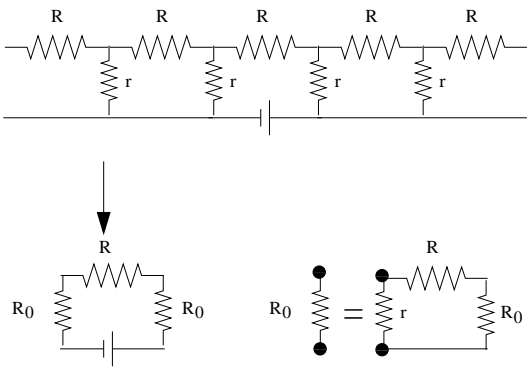}
\caption{The resistor network parallel to a single phase slip.
The shunting resistance can be
  calculated using the ``infinite resistor chain'' trick. \label{Reff}}
\end{figure}

The parallel shunting impedance of a junction in the chain
splits into two contributions: the resistance of the
network of
resistors $r$ and $R$, and the impedance of the ``telegraph line'' of
the chain. The first part is self explanatory; let us denote it as
$R^s_{eff}$. The second piece requires an explanation. When the
Josephson coupling is strong we can assume that Josephson junctions
are superconducting except for the times in which they exhibit a phase
slip. A superconducting Josephson junction has the impedance of a
solenoid with the (``kinetic") inductance $L=\hbar/(2e)^2\,E^{-1}_J$. Therefore the line
of junctions and capacitors resembles a line of solenoids and
capacitors, a simple model of a telegraph line. When a phase slip
occurs, it creates a short-lived voltage drop on the junction (see
Fig. \ref{pscircuit}). This pulse has two effects; the first is a DC
current that flows through the shunt and conversion resistors, $R$ and
$r$ and the second effect is sending {\it plasma waves} through the line
of junctions and capacitors. The latter  also behaves like a resistance as  it corresponds to energy
being carried away by plasmons. Let us denote 
the effective resistance describing this mechanism
of dissipation as $R^P_{eff}$. 
The two mechanisms of dissipation separate at low frequencies, 
 since the resistive contribution comes
from local currents which decay rapidly, whereas the
plasmon contributions arises from waves propagating at speed $\vms$.  Hence, the two mechanisms are totally out of phase with each other at low frequencies.

The effective resistance a phase slip feels
due to the plasmons is 
\begin{eqnarray}
R^P_{\rm eff}=2\sqrt{\frac{L}{C}} = 2R_Q\cdot\frac{1}{2\pi}\sqrt{E_C/E_J}=\frac{2}{\j}R_Q
\label{ReffDipole}
\end{eqnarray}
(see, for instance, Appendix 5D of Ref. \onlinecite{RefaelThesis}). Note that $R^P_{\rm eff}$ include contributions from the
telegraph lines on both sides of the given
junction.

The resistance $R^s_{eff}$ can be found easily
using the construction in Fig. \ref{Reff}. In the figure it is shown
that the shunting resistance can be broken into three resistors in
series, where the semi-infinite ladder of resistors $r$ and $R$ is replaced by
an effective resistor, $R_0=\frac{-R+\sqrt{R^2+4rR}}{2}$.
We thus have an  effective shunting resistance
of the network of Ohmic resistors
\be
R^s_{eff}=2R_0+R=\sqrt{R^2+4rR}=\frac{R_Q}{\aeff}.
\label{ecm4}
\ee

The total impedance, $Z_S=Z(\omega=0)$, shunting  a junction in the fully superconducting phase (FSC) is thus given by the two contributions in parallel:
\be
\frac{R_Q}{Z_S}= \half \j + \aeff \ :
\ee
as we shall see, this combination controls the action of QPS in the FSC phase.
From analogy with the single RSJJ case, 
we would expect that the FSC phase would become unstable to QPS when the
effective inverse shunting resistance equals $R_Q$. As above for Cooper pair tunneling, we can see that the charge transfered associated with the quantum phase slip of $2\pi$  is $V(\omega=0)/Z_S=2e R_Q/Z_S$.  Again by analogy with the Cooper pair tunneling, we can consider the action associated with the phase slip, finding that this is $\le \pi \hbar R_Q/Z_S$. Thus we would guess that the condition for suppression of QPS tunneling is 
when 
\be
\frac{R_Q}{Z_S} = \frac{\j}{2}+\aeff\ge 1.
\label{wrcond}
\ee
However, this analogy reflects 
only the {\it local} physics of phase slips and leads to the {\it
  wrong} condition: nevertheless, as we shall see, the correct
condition involves the same combination. 

In Sec. \ref{res} we discuss a region of parameter space in the
NOR phase which we
call a quasi-metallic regime. In this region the measured
resistance of the JJ chain first slowly drops as the temperature is
lowered. But then, when a crossover temperature is reached, the
resistance takes a sharp upturn (see Fig. \ref{Fig5}). This effect
occurs in the region of the NOR phase where the (wrong) local
condition for the stability of the FSC phase to
single phase slips (Eq. \ref{wrcond}) is fulfilled, but the FSC 
is unstable against dipoles.

\begin{figure}
\includegraphics[width=8.5cm]{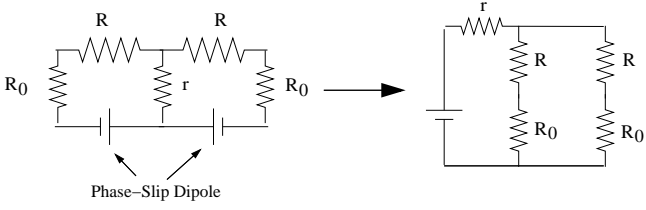}
\caption{Effective circuit for a phase-slip dipole with $s=1$. $R_0$ is the
  effective resistance of a semi-infinite resistor ladder as given
  above Eq. (\ref{ecm4}).
The dissipation seen by a dipole has no contribution
from plasma waves at low frequencies.
\label{Dcircuit}}
\end{figure}

To fully understand the effects of QPS, we also need to consider phase
slip {\it dipoles} in which a phase slip occurs on one junction
simultaneously with a  phase slip of the opposite sign on another
junction --- say $s$ grains away.  In the simplest case, $s=1$, this
has the effect of slipping the phase of one grain relative to the rest
of the system to which it is coupled.  More
generally, $s$-dipoles slip the phase of $s$ consecutive grains
relative to the rest of the chain.  Because the effects of the two
opposing phase slips cancel at long length scales, there will be no
contribution to the effective shunting impedance from plasmons: it
will be entirely dissipative.  Considering all but the two junctions
across which the slips occur, to be superconducting and thus short
circuits at low frequencies, the effective shunt conductance between a
row  of $s$ grains and the rest of the chain is found to be Fig. \ref{Dcircuit}, 
\be
\frac{R_Q}{R^S_{dipole}} =2\aeff (1-\w^s),
\ee
where 
\be
\ba{cc}
\w=1+\frac{R-\sqrt{4rR+R^2}}{2r} <1 & \aeff=\frac{1}{\sqrt{R^2+4rR}} .
\ea
\ee
Of particular importance is the resistance between one grain and the
rest of the chain that shunts the two (parallel) Josephson
junctions. This is $r+\frac{1}{4}(R+\sqrt{R^2+4rR}) =R_Q/\beff$ with
$\beff=2\aeff(1-w)$ (see Fig. \ref{Dcircuit}).

When $R\ll r$. $\aeff \approx 1/2\sqrt{rR}$, $\beff\approx R_Q/r$ and $\w\approx 1-\sqrt{R/r}$ 
so that the effective shunt resistances  of strings of $s$ grains only separates into two resistors in parallel  --- one at each end ---  for large $s>\lambda_Q/\ax\approx \sqrt{r/R}$.

\subsubsection{\scs phase and dipole proliferation}

In the FSC state individual junctions have well defined phase
differences so that the above calculations are relevant. But in
the \scs phase the superconducting
phases of the individual grains fluctuate strongly enough to decouple the superconductivity from the resistive shunts. In this case the energy of
phase slips is primarily dissipated by the plasmons and the relevant
impedance becomes just the $R^P_{eff}= 2R_Q/K$ with $K$ the
dimensionless superconducting quantum stiffness.

\begin{figure}
\includegraphics[width=8.5cm]{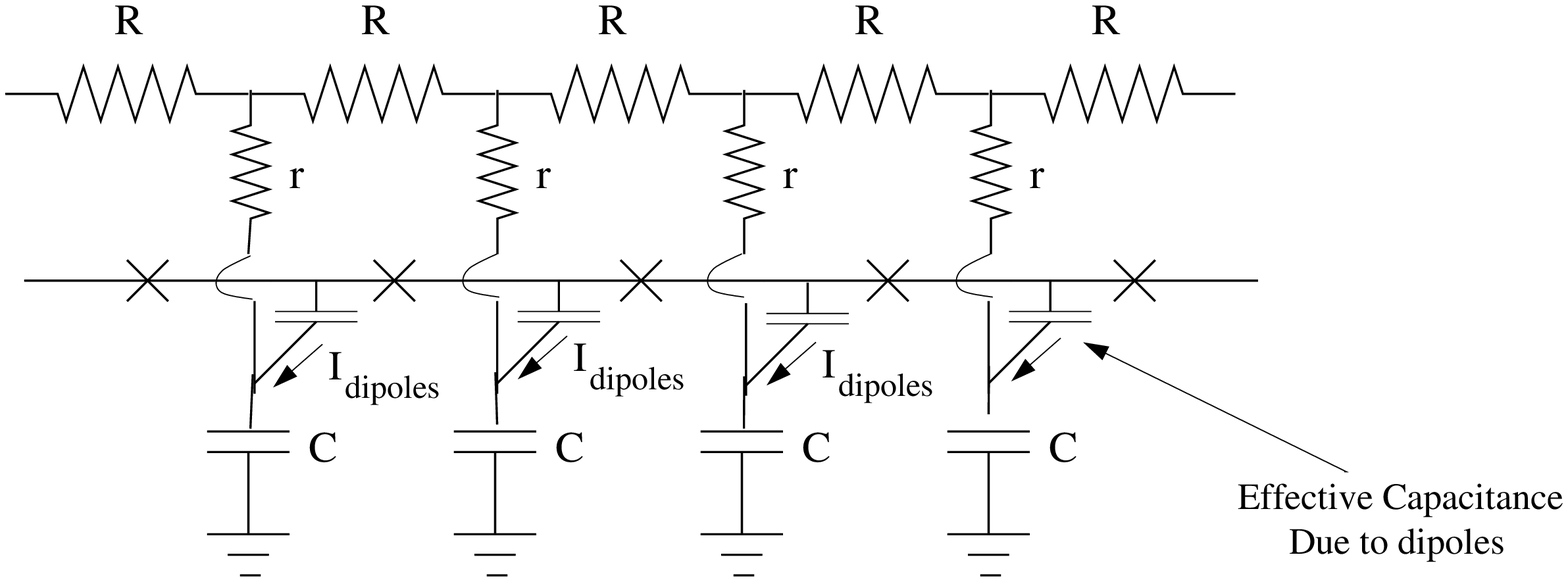}
\caption{Phase-slip dipoles create an effective capacitance which inhibits current from leaving the  line of  Josephson 
  junctions. This
  capacitance screens  the normal-to-superconductor
  conversion resistance, $r$.  But it only renormalizes the plasma-wave dissipation via the 
  change of the total effective capacitance of each
  grain. \label{dipC}}
\end{figure}

A well known result is  that a Josephson junction in its
insulating state seems like a capacitor (see, for instance, Appendix
5E of Ref. \onlinecite{RefaelThesis}).  For dipole proliferation, the relevant junction is the combined junction from one grain to the rest of the system. 
To show this we first note that the effective capacitance of a Josephson junction when phase slips
proliferate describes the electrical response of the ``plasma" of phase slips. This
effective capacitor is charged whenever current tries to cross the
junction itself. For a dipole, the relevant current is the current that tries to leave the horizontal
Josephson-junction line in Fig. \ref{dipC}. The
current leaving the junction-line into the vertically drawn wire is
the only current that interacts with phase slip dipoles, hence the
effective capacitor that appears when dipoles proliferate can get
charged only by the current associated with $I_{dipoles}$ as in the
Figure (\ref{dipC}). 
This effective capacitance thus occurs between the superconducting and
normal parts of a junction, and it is in {\it series} to the finite conversion resistance which is important at
low frequencies  --- $\omega \ll T^*/\hbar $.
When dipoles proliferate, they effectively block the low-frequency
conversion, as the effective capacitance of the dipoles dominates.

The induced capacitance due to dipoles blocks the N-S
conversion in each grain, and thus screens the dissipative interaction
between phase slips: at low frequencies, this   is essentially equivalent to setting $r\rightarrow\infty$. 
But the plasmon interaction is a dynamic effect.  As can be seen from Fig. \ref{dipC}, the
dipole-induced capacitance renormalizes the self capacitance $C$
of each grain.  Thus the plasmon interaction survives the proliferation of
dipoles, albeit with a renormalized plasma speed.  These results are formally derived
in Sec. \ref{scsscreensec} from the effective action of the chain. 

Since dipoles disconnect the N-S conversion, in the \scs phase, the
effective shunting resistance that is felt by a phase slip in the JJ chain
is just that from the renormalized plasmon impedance. Therefore,   
\be
\frac{R_Q}{R^s_{total}}=\frac{\j}{2}.
\ee
$R^s_{total}$ is the effective {\it local} dissipation in the \scs
phase. One would expect, in analogy to the single junction that when
\be
R_Q/R^s_{total}<1
\label{wrcond2}
\ee 
the \scs phase would be unstable to QPS proliferation. 
As we shall see, the relationship between this shunting impedance and
the stability of the \scs phase to QPS, is different than that for a
single junction, essentially due to the  ''entropy" of the
translational freedom of the QPS. Nevertheless, as in the case of the
FSC-NOR QPS-driven transition, the local condition in
Eq. (\ref{wrcond2}) demarcates the regime  with non-monotonic temperature
dependence of the resistivity.  In the region
of the NOR phase where condition (\ref{wrcond2}) applies, there will
be a low-energy crossover between a decreasing resistance as
temperature decreases, to insulating behavior (see Sec. \ref{res}).


\section{Strong Josephson 
Coupling Limit \label{sec2}}
\label{StrongJosephsonCouplingSection}

We now turn to the analysis of the low energy properties of the chain,
going well beyond the linear electrodynamics discussed in the
previous section. We first derive the quantum-phase-slip Coulomb-gas
representation of the action (Sec.\ref{QPSsec} and transform it
into a sine-Gordon representation (Sec. \ref{sgsec}). This enables analysis of the phases of the JJ chain
(Sec. \ref{phases1}), and  an  RG approach for  the phase diagram (Sec. \ref{sec3}). 

\subsection{Quantum Phase Slips and Representations of the Partition
  Function \label{QPSsec}}

In the limit of strong Josephson coupling, the phase differences between
neighboring grains will predominantly be localized in the vicinity of 
the minima of the Josephson potential which occur at $\phi_i - \phi_j = 2 \pi n_{ij}$,
with $n_{ij}$ an integer.  Occasionally, however, the phases
will depart from their classical superconducting form, and  tunnel between
neighboring minima with different $n_{ij}$: these events are {\it quantum phase slips}  (QPS).  Note that as discussed  in the section on parameters,  strong coupling will in practice be defined by the rareness of QPS which depends on the high energy physics as well as the dimensionless Josephson coupling, $\j^2 \propto E_J/E_C$.

Physically, QPS at zero temperature are caused by the charging energy 
(in Eq. \ref{allaction})  not commuting with the Josephson
potential. Quantum phase slips correspond to vortex-like phase configurations in the space-$\tau$ (imaginary-time) plane, while in real time they involve the launching of
one dimensional plasma-waves (plasmons) through the chain.   

At low temperatures, quantum phase slips may destroy
superconducting coherence in the JJ array, and their potential role in this way makes them the  basic excitations in terms of which the low energy physics can most readily be described when the local superconductivity is ``strong".
We can thus try to expand in the QPS fugacity in this strong Josephson coupling limit.
As we shall see, one also has to consider phase slip dipoles: bound pairs of QPS of opposite signs.

\subsubsection{Coulomb Gas Representation}

To
analyze the effects of QPS in the partition function, it is convenient
to use a Villain transformation. As usual first writing
\be
\ba{c}
\exp\left\{\int d\tau E_J\left[\cos\l(\phi_{i+1}{(\tau)}-\phi_{i}{(\tau)}\r)-1\right]\right\}\vspace{2mm}\\
\approx\summ_{\{\mu_i{(\tau)}\}}\exp\left\{-\int d\tau \frac{E_J}{2}\left[\phi_{i+1}{(\tau)}-\phi_{i}{(\tau)}+2\pi\mu_i{(\tau)}\right]^2\right\}
\ea
\label{vtrn}
\ee
The Villain transformation breaks the cosine
function of phase differences down to a sum over its troughs with $\mu_i{(\tau)}$ an integer valued function
labeling the 
troughs in the Josephson potential of the junction between the $i+1$'th and the $i$'th grains. 
A phase slip on the $j$'th junction corresponds to a sudden change of
$\mu_j$ by $\pm 1$, thus we can write it in terms of a density of discrete QPS:
\be
\rho_{QPS}(j,\tau)=\der{\mu_j{(\tau)}}{\tau} = \summ_m p_m \delta_{j,j_m} \delta(\tau - \tau_m), 
\label{LocalNeutrality}
\ee
with the $m$'th  QPS with ``charge" $p_m= \pm 1$ occurring at imaginary time $\tau_m$ and on junction  $ j_m$. 
Periodic boundary conditions on the original 
phases, $\phi_i(\tau=0) =\phi_i(\tau=\beta)$ imply  an integrated
neutrality condition for {\it each} junction:
\be
\int_0^\beta \rho_{QPS}(j,\tau) d\tau = 0 \ \ \ \ \ \ \forall j\ .
\ee

Using Eq. (\ref{vtrn}) in the partition function, Eq.
(\ref{allaction}), makes  the action quadratic in $\phi_i$ as well as
in the other fields, all of which can be integrated out to obtain the partition function solely in terms of the QPS
configurations. 
The
action of a set of $N$ QPS  with integer charges $\{p_m\}$  at space
time locations $\{(x_m=\ax j_m,\tau_m)\}$ , with (integer) $j_m$ labeling junctions,  has the form:
\be
S_N\l(\{p_m\}\r) = \frac{1}{2} \summ_{m\neq n}^N p_m p_n G{(x_m-x_n,\tau_m-\tau_n)},
\label{inter0}
\ee
%
where $G{(x,\,\tau)}$ is 
the interaction between phase slips. The QPS partition function is then:
\be
Z_{QPS}=\tilde{\summ_{\{p_m\}}} \zeta^N \exp\l[-S_N\l(\{p_m\}\r)\r],
\label{ZQPS}
\ee
with  $\zeta$ the {\it fugacity} of QPS; $\zeta$ has units of
frequency: in the absence of interactions between them,  it would be
the rate at which phase slips of each sign occur across a single
junction. Its bare value is obtained by considering the action of the
instanton which describes the short time motion of the phase variables
following a phase slip in $\mu_i$. 
The sum in Eq. (\ref{ZQPS}) is over distinguishable QPS configurations
with the restriction that {\it for each junction} the total QPS charge
is zero - a neutrality condition. Note that  strictly speaking, because of this local neutrality, with $m$ and $n$ on different junctions  $G(x_m-x_n, \tau_m-\tau_n)$ is only the finite part of the interaction: the infinite parts that would arise if the local neutrality condition were not satisfied  have been subtracted out.

The full form of the interaction between the QPS is very complicated
and has several different regimes. As discussed in the introduction, the
temperature (energy) scale that arises from the dissipation and the finite size of the grains, $T^*$, 
plays an important role. At temperatures higher
than $T^*$ dissipation on separate junctions is effectively decoupled. By contrast, at low temperatures, $T<T^*$, there is considerable interaction between QPS on
neighboring junctions: we  focus here and henceforth on this low temperature regime. 

\subsubsection{Low Temperature Limit}

In the low temperature limit, $T\ll T^*$, 
the effective interactions between QPS is  given   by 
\be
G{(x,\,\tau)}\approx \j \log\frac{\at}{\sqrt{x^2/\vms^2+\tau^2}}
+\aeff e^{-|x|/\xz}
\log\frac{\at}{|\tau|},
\label{inter11}
\ee
where the quantum stiffness
$\j =2\pi\sqrt{E_J/E_C}$,
(assuming $D_{N,\,SC} C\ll 1$),
the Mooij-Sch\"on velocity 
$\vms=\ax\frac{\sqrt{E_JE_C}}{\hbar}$, and the strength of the dissipative interactions,
\be
\aeff=\frac{R_Q}{\sqrt{R^2+4rR}}, \label{aeff0}
\ee
were all introduced earlier, and the characteristic length scale is
\be
\xz=\frac{\ax}{\log(1/\w)}
\ee
with 
\be
\w =1+\frac{R-\sqrt{4rR+R^2}}{2r}
\ee
which is approximately $1-\sqrt{R/r}$ when $R\ll r$.  The origin of
$\lambda_Q$ is the decay length of currents in the resistive network
shown in Fig. \ref{Reff}, and can also be inferred from the continuum
equations (\ref{eq1}).

The discreteness of the chain is important even at large length scales, particularly when $\xz$ is comparable to $\ax$. The strength of the dissipative interaction arises from 
\be
\aeff=\int\limits_{|k|<k_m}\frac{dk}{2\pi}\frac{R_Q\ax}{r(2-2\cos ka)+R}  \ .
\label{int2}
\ee
This yields simply  the inverse of the shunt resistance for a single
junction, as found in Sec. \ref{ctm}.  [In the continuum approximation with
a sharp momentum space cutoff at $\pi/\ax$  the QPS interaction would
yield $\aeff=\alpha\frac{2}{\pi}\arctan\l(\pi\sqrt{\frac{r}{R}}\r)$:
this is  close to Eq.(\ref{aeff0}) over the whole range but does not
correctly correspond to the shunt resistance. Similarly, in the
continuum approximation, $\xz$ would always have its asymptotic form,
$\ax\sqrt{r/R}$.]

The cutoff for the plasma interaction is given by $\at\sim a/\vms=\hbar/\sqrt{E_J
  E_C}$. The cutoff for the dissipative interaction in the strong
coupling limit, on the other hand, is $\at\sim \min\{\hbar R_Q/(R+2r)
E_J,\,\hbar/\sqrt{E_J E_C}\}$. When $R+2r\gg R_Q$ the cutoff time for the
local interaction is  lower than that for the isotropic
plasmon-related interaction. This may give rise to additional
crossovers in the temperature range $RE_J/R_Q>T>\sqrt{E_J
  E_C}$. However, we will only analyze the Coulomb-gas action, Eq. (\ref{inter11}), for
energy scales and temperatures $\Omega<\sqrt{E_J/E_C}$.   

The two logarithmic interactions between QPS in  Eq. (\ref{inter11}) have very
different physical origins. The first part, which is isotropic in space-time,
is present for $1+1$ dimensional X-Y models even in the absence
of dissipation; it gives rise to a quantum
Kosterlitz-Thouless transition, as  originally discussed 
by Bradley and Doniach \cite{Bradley1984}
(see also Ref. \cite{Efetov1980}).  The
second term in Eq. (\ref{inter11}) originates from the Ohmic dissipation.
It is logarithmic in the {\it time}
separation of phase-slips. Only the limit $r=0$ ($\xz=0$) has, to our knowledge, been analyzed 
previously \cite{Schoen,KorshunovA, KorshunovB}. In this limiting case the dissipative interaction of Eq. (\ref{inter11})  takes the form: 
\begin{eqnarray}
G_{local}{(x,\,\tau)}=a\delta{(x)}\frac{1}{R}
\log\frac{\at}{|\tau|}.
\label{locform}
\end{eqnarray}
More generally, for non-zero $r$, the  length scale, $\xz$,  is the range of {\it normal} currents that are induced when a phase slip
occurs and thus, effectively, the dissipative size of a QPS: $\xz$, controls the exponential fall-off with spatial separation of the dissipative  interactions between QPS's.  Not surprisingly, $\xz$ plays a particularly important role in the physics of 
 finite length  chains \cite{finitechain}.

\subsection{Sine-Gordon Representation \label{sgsec}}

From the Coulomb gas partition function, Eq. (\ref{inter0}), 
we derive the sine-Gordon
representation of the strong Josephson coupling limit.  The Coulomb gas representation provides an
expression for the probabilities of specific configurations of 
phase slips, whereas the sine-Gordon representation is more amenable
to a renormalization group analysis. 
We introduce two separate Hubbard-Stratonovich  
fields, $\theta_j$ and $\psi_j$ located on the {\it junctions} between the grains $(j,j+1)$,  in order to decouple the two contributions to the interactions 
between QPS of Eq. (\ref{inter11}) (for details see Appendix 5A.2 of
Ref. \onlinecite{RefaelThesis}): in terms of the physical variables, $\psi$ decouples the normal degrees of freedom, and $\theta+\psi$ the superconducting degrees of freedom. This transformation yields the dual action of the model of Eqs. (\ref{allaction}):
\be
\ba{c}
S_{\rm dual}
=\int_0^\beta d\tau\,\, \frac{1}{4\pi a^2 \jx} \summ_{i}
(\theta_{i+1} - \theta_i)^2
+ \int_0^\beta d\tau \,\, \frac{1}{4\pi \jt}
\summ_i (\partial_\tau \theta_i)^2\vspace{2mm}\\
+\frac{\beta}{2\pi} \summ_{\omega_n} \frac{|\omega_n|}{2}
\left\{
\frac{R}{R_Q} \summ_i  |\psi_{i,\omega_n}|^2
+ \frac{r}{R_Q} 
\summ_{i} |\psi_{i+1,\omega_n}-\psi_{i,\omega_n}|^2
\right\}
\vspace{2mm}\\
-\zeta \int_0^\beta d\tau \summ_i \cos (\theta_i + \psi_i)
\label{action_dual_2}
\label{action}
\ea
\ee
The effective stiffnesses are
\be
\jx \equiv \j /\vms =\frac{2\pi\hbar}{\ax E_C},
\ee
 and 
 \be
 \jt \equiv \j  \vms= \frac{\ax E_J}{2\pi\hbar} \ .
 \ee

\subsubsection{Phase-slip Dipoles}

In our recent work on pairs  of shunted
Josephson junctions \cite{2JJ}, we demonstrated the important
role of QPS dipoles:  these are instantons consisting of two opposite sign QPS that occur almost simultaneously
on nearby junctions. The simplest  phase-slip dipole, on neighboring junctions, is thus an event in
which the phase of a single grain slips by
$2\pi$ {\it relative to the chain}. 
In non-dissipative XY models such dipoles disappear when
short wavelength fluctuations are integrated out isotropically
in space and time, and the closely-spaced dipoles  act only to renormalize the stiffness $\jx$ and thereby the  interaction between other QPS.

For granular systems with dissipation, the physics on the scale of the grains plays an essential role, and 
a better approach is to perform the RG
procedure in time only. This means that dipoles  remain
as independent degrees of freedom and should be considered explicitly
along with individual QPS. [Although it is clumsy to do so, we will show later that the results for the isotropic Kosterlitz Thouless transition can be recovered from this anisotropic RG. ]
Dipoles can be included in the sine-Gordon representation by adding the
following term to the action
\be
\ba{c}
S_{\rm dipole}=\\
\summ_{s=1}^\infty \eta_s
\summ_i\int d\tau 
\cos\l[ \l(\theta_{i}{(\tau)}+\psi_{i}{(\tau)}\r)-\l(\theta_{i+s}{(\tau)}+\psi_{i+s}{(\tau)}\r)\r]\vspace{2mm} \\
=\summ_i\summ_s\int  d\tau 
\eta_s\cos\l[ \dx_s\theta_{i}{(\tau)}+\dx_s\psi_{i}{(\tau)}\r]
\ea
\label{Sdipole}
\ee
with $\eta_s$ the {\it dipole fugacity} for a dipole with separation $sa$ between its
positive and negative QPS; we will refer to $s$, loosely, as the {\it
  moment} of the dipole; $\dx_s$ is the difference operator of
distance $s$: $\dx_s\theta_i\equiv \theta_{i+s}-\theta_i$.    

The argument of the cosine function in Eq. (\ref{Sdipole}) places two phase
slips of opposite signs on junctions  a distance $s\ax$ apart.
Otherwise, this term is completely analogous to the  cosine term in
Eq. (\ref{action}).
Note that the dipole term resembles the Josephson term
in the original action for the JJ array, Eq. (\ref{allaction}), but it
is written in the dual variables. The similarity is explained by the
following description of the effect of the two terms:
$\cos(\phi_i-\phi_j)$ removes a Cooper-pair at site $j$, and creates a
Cooper pair at site $i$. Similarly,
$\cos\l[\l(\theta_{i}{(\tau)}+\psi_{i}{(\tau)}\r)-\l(\theta_{j}{(\tau)}+\psi_{j}{(\tau)}\r)\r]$,
creates a phase-slip on bond $i$, and an anti-phase-slip on bond $j$.

Initially, there are no phase slip dipoles, and $\eta_s=0$ for all $s$.  However,
upon coarse-graining in time, dipoles will form
and have to be analyzed as independent entities, on the same footing as single
phase slips.


\subsection{Phases of the JJ Array \label{phases1}}

The three phases of the JJ array can be described in terms of phase slips.  The simplest  is the normal phase (NOR), in which
phase slips proliferate, and phase coherence is lost.   
In the $SC^{\star}$ phase,  phase-slips and anti-phase-slips bind into
(neutral) dipoles and  the
chain becomes {\it globally} superconducting although the phase
fluctuations on each grain are large enough that  the local phase is
no longer well defined and the dipoles still proliferate. 
The phase
fluctuations are strong enough that there is no quasi long-range order
of the phase. Nevertheless uniform supercurrent can flow unimpeded in
the chain. 

In the fully superconducting phase (FSC), the phase slip dipoles bind into
quadrupoles and annihilate at low energy scales: the phase differences
on each
junctions are then well defined, there is quasi long-range order, and the
chain is {\it  locally}, as well as globally, superconducting.  We
first study these phases and the phase transitions between them via a
renormalization group analysis of the sine-Gordon representation of
the QPS action.

The two superconducting phases were discussed before, and were dubbed respectively \cite{Schoen,KorshunovA, KorshunovB, Fazio}.
The phase diagram of the RSJJ chain for $r=0$ was discussed
by Chakravarty {\it et al.} \cite{Chakravarty}
using the self-consistent harmonic approximation to treat
the model of Eq. (\ref{allaction}).
While this analysis is appealing due
to its simplicity, it misses an important
distinction between the two distinct superconducting phases.
A contemporary publication by Korshunov
\cite{KorshunovA, KorshunovB}, analyzed the Dyson equation of the
sine-Gordon representation Eq.(\ref{action_dual_2}) and found the two superconducting phases, dubbing them SC-1 and
SC-2 . 
Unfortunately, this approach was not very transparent
and does not give a simple physical picture of the superconductor-to-normal transition.
Fazio
{\it et al.} \cite{Fazio} obtained a similar phase diagram to Korshunov via
numerical calculations. In all these papers
only the $r=0$ case was considered. In this paper
we  analyze the model
of Eq. (\ref{allaction})  for general $r$ and provide further insight into the nature of the superconductor-to-normal transitions. For the case $r=0$, we reproduce a
phase diagram derived by Korshunov in Refs. \onlinecite{KorshunovA,
  KorshunovB}. We should emphasize that our RG analysis is the first
to allow an investigation of both phase boundaries and critical
phenomena in the entire phase diagram (with the exception of
intermediate Josephson strength, where, to date, no analytical method
applies).  Having the RG framework allows, for the first time, to
discuss both finite size and finite temperature effects of the
dissipative JJ chain.

\subsection{Renormalization Group Flows for Strong Josephson Coupling
\label{sec3}}

As we showed above, in the limit of strong Josephson coupling
the RSJJ chain of Fig. \ref{Fig1intro}(c) can be described by a spatially discrete sine-Gordon model
\be
Z=\int D[\theta]D[\psi] e^{-S_{{\rm dual}}[\theta,\psi]
- S_{\rm dipole}[\theta,\psi]}.
\label{pp2}
\ee
We now  derive and analyze the RG flow equations
in this strong Josephson coupling limit, treating the QPS fugacities as
small, but at the same time allowing any values of the shunt and conversion resistances.

The flows of the anisotropic RG we use, are  produced in the sine-Gordon representation by integrating over high frequency modes, $\theta^>$ and $\psi^>$, in a frequency shell,
$\Omega-d\Omega<|\omega|<\Omega$, of $\theta_{i}{(\tau)}$ and
$\psi_{i}{(\tau)}$, and then rescaling the time as
$\tau\rightarrow\tau'=\l(1-\frac{d\Omega}{\Omega}\r)\tau$ so that
$\Omega$ returns to its original value, and the differential flow
parameter is

\be
d\ell\equiv \frac{d\Omega}{\Omega} \ .
\ee

But, crucially, lengths are  {\it not} rescaled  and the spatial discreteness is kept. In doing this, however, there appears a characteristic length scale, $\Lambda(\Omega)$, that plays an important role because of the anisotropic RG procedure: in terms of the (renormalized) phase velocity, $\vms$, which grows under renormalization because of the rescaling of time this is
\be
\Lambda=\frac{\vms}{\ax \Omega} \ .
\ee

Appendix 5.B of Ref. \onlinecite{RefaelThesis} describes in detail the derivation of the RG-flow
equations perturbatively in powers of the fugacities; we will need these up to second order.

Integrating out the  fast modes, effectively
increases the short time cut-off scale from $\at$ to $\at+d\at=\at
(1+d\Omega/\Omega)$ so that part of the phase slip action is absorbed into its core yielding a linear
suppression of the phase-slip fugacities, 
But there  is also an increase in the fugacities from rescaling time as 
 they  represent  the rate of phase slips. 
Combining these two effects we get: 
\be \ba{c} \zeta \rightarrow
\zeta
\l[1+\frac{d\Omega}{\Omega}\l(1-\frac{1}{2}\sqrt{\j}-\aeff\r)\r],\vspace{2mm}\\
\eta_s \rightarrow 
\eta_s\l\{1+\frac{d\Omega}{\Omega}\l[1-2\aeff(1-\w^s) -\j(1-e^{-s\ax/\Lambda}) \r]\r\}
\ea
\label{fo123}
\ee
where $\j=\sqrt{\jx\jt}$.

The effective dissipation for the $s$-dipoles is
\be
2\aeff(1-\w^s) \ ,
\ee
which can be seen to be the inverse of the shunt resistance for the
pair of  (here parallel) Josephson junctions that couple a row of $s$
grains to the rest of the chain, as derived in Sec. \ref{ctm}. 
As it is necessary to treat the smallest dipoles, $s=1$, specially,
we define
\be
\beff \equiv 2\aeff(1-\w) \ .
\ee
The $\j$ dependent term in the renormalization of $\eta_s$ requires some explanation. For dipoles with large $s$, the two constituent QPS will be separately renormalized by the phase fluctuations that have been integrated out: thus the factor proportional to $K$ in this limit. In terms of the circuit analysis, this dissipation can be thought of as arising from the frequency $\Omega$ plasmons that propagate from the two opposite phase slips before annihilating far away.  In contrast, dipoles with small $s$ will couple only weakly to the low energy phase modes and thus $K$ has little effect on their renormalization.  The energy-dependent length scale $\Lambda$ separates these two regimes. 

 In principle, there should also be energy-scale dependence of the renormalization of $\zeta$, but this can be ignored as long as $\Lambda$ is substantially larger than $\ax$.  We will assume this henceforth at the cost of ignoring only initial transient renormalizations (which will in any case only be part of the other relatively high energy scale processes we have ignored).

Nonlinear contributions to the renormalization of $\zeta$ and the  $\{\eta_s\}$ come from
second order processes. One such process is the combining of two
phase-slips of opposite signs into a dipole. The two phase-slips have to
occur on at almost the same time (within $\at$ of each-other) on
Josephson-junctions separated by $s$ grains.  This process gives a renormalization of $\eta_s$ 
\be
\eta_s\rightarrow \eta_s+\zeta^2 \frac{d\Omega}{\Omega}\at\l(2\aeff\w^s +Ke^{-sa/\Lambda}\r) \ ;
\label{so1}
\ee
again note the $s$ dependence of the $K$ term: the dipoles are only formed with appreciable amplitude when the spacing between them is less than $\Lambda$; for convenience we define the factor
\be
\u(\ell)\equiv e^{-a/\Lambda}
\ee
by analogy with $\w$.

A dipole can combine with a phase-slip on one of the junctions of the dipole  to form a phase slip on the other junction.
This adds  second order contributions to $\zeta$, for each $s$:
\be
\zeta \rightarrow
\zeta+\zeta\eta_s\frac{d\Omega}{\Omega}\at \l[ 2\aeff(1-\w^s)+  K(1-u^s)\r] .
\label{so2}
\ee

Two overlapping dipoles can combine to make a single dipole with moment the sum of the constituent moments. This yields
\begin{widetext}
\be
\eta_s \rightarrow \eta_s + \frac{d\Omega}{\Omega}\frac{1}{2}\summ_{\sigma} \eta_{|\sigma|}\eta_{|s-\sigma|}\l[2\aeff(1+\w^s-\w^{|\sigma|}-\w^{|s-\sigma|})+K(1+\u^s-\u^{|\sigma|}-\u^{|s-\sigma|})\r] \ ,
\label{so4}
\ee
\end{widetext}
where the sum runs over all $\sigma$ to take into the account both relative orientations of the dipoles.

Another  important contribution comes from combining  two phase-slips
of opposite signs on the same pair of Josephson junctions with a small delay between
them [$1/\Delta\tau>\Omega(1-d\Omega/\Omega)$]. 
Their polarization gives rise to a  renormalization of the parameter $\jt $, which is related to the superconducting stiffness of the Josephson
junctions. This yields a contribution to the quadratic action:
\be
\frac{1}{4}\at^2\zeta^2\summ_i\int d\tau\frac{d\Omega}{\Omega} \at^2 \sqrt{\jt\jx}\l(\der{\theta_i{\l(\,\tau\r)}}{\tau}\r)^2.
\label{so3}
\ee
In addition, the rescaling of imaginary time --- but not space --- leads to a rescaling of
the parameters, $\jx \rightarrow \jx \l(1-d\Omega/\Omega\r)$, and $\jt
\rightarrow \jt \l(1+d\Omega/\Omega\r)$ which is reflected in the
renormalization of the velocity $\vms$ and hence $\Lambda\propto
\vms$; at linear order in the fugacities, $K=\sqrt{\jx\jt}$ is not
renormalized.

The differential RG flow equations for $\jt,\,\zeta$, and $\{\eta_s\}$ can now be
obtained by combining the terms from Eqs. (\ref{fo123},
\ref{so1}, \ref{so2}, \ref{so4}, \ref{so3})
\begin{widetext}
\be
\ba{c}
\frac{d\jt}{dl}=\jt -\frac{\pi}{2}\jt ^{5/2}\jx ^{1/2}\zeta^2\at^2
,\vspace{2mm}\\
\frac{d\jx}{dl}=-\jx ,\vspace{2mm}\\
\frac{d\zeta}{dl}=\zeta\l(1-\frac{1}{2}\sqrt{\jx \jt}-\aeff\r)+2\zeta\summ_{s>0} \eta_s\at\aeff(1-\w^s) ,\vspace{2mm}\\
\frac{d\eta_s}{dl}=\eta_s\l[1-2\aeff(1-\w^s)\r]+\zeta^2 \at\l[2\aeff\w^s  +\sqrt{\jx \jt}\r] +\frac{1}{2}\summ_{|\sigma|} \eta_\sigma\eta_{|s-\sigma|}2\aeff(1+\w^s-\w^{|\sigma|}-\w^{|s-\sigma|}),\vspace{2mm}
\ea
\label{rg232}
\ee
\end{widetext}
valid in the limit of small fugacities
$\zeta$ and $\eta_s$ and for low energy scales. As we shall see, for some purposes these flow equations are not sufficient even in this limit: the dependence of the renormalizations of and by dipoles on their moments relative to $\Lambda(\ell)$  --- whose renormalization is determined since $\Lambda \propto \sqrt{\jt/\jx}$  --- need to be included even though, formally, these disappear in the low energy limit of interest. 

For other purposes, we need to consider the case of  large
$\{\eta_s\}$ but small $\zeta$. 

\section{Phase diagram and phase transitions for strong Josephson coupling\label{sec5}}

\begin{figure*}
\includegraphics[width=13cm]{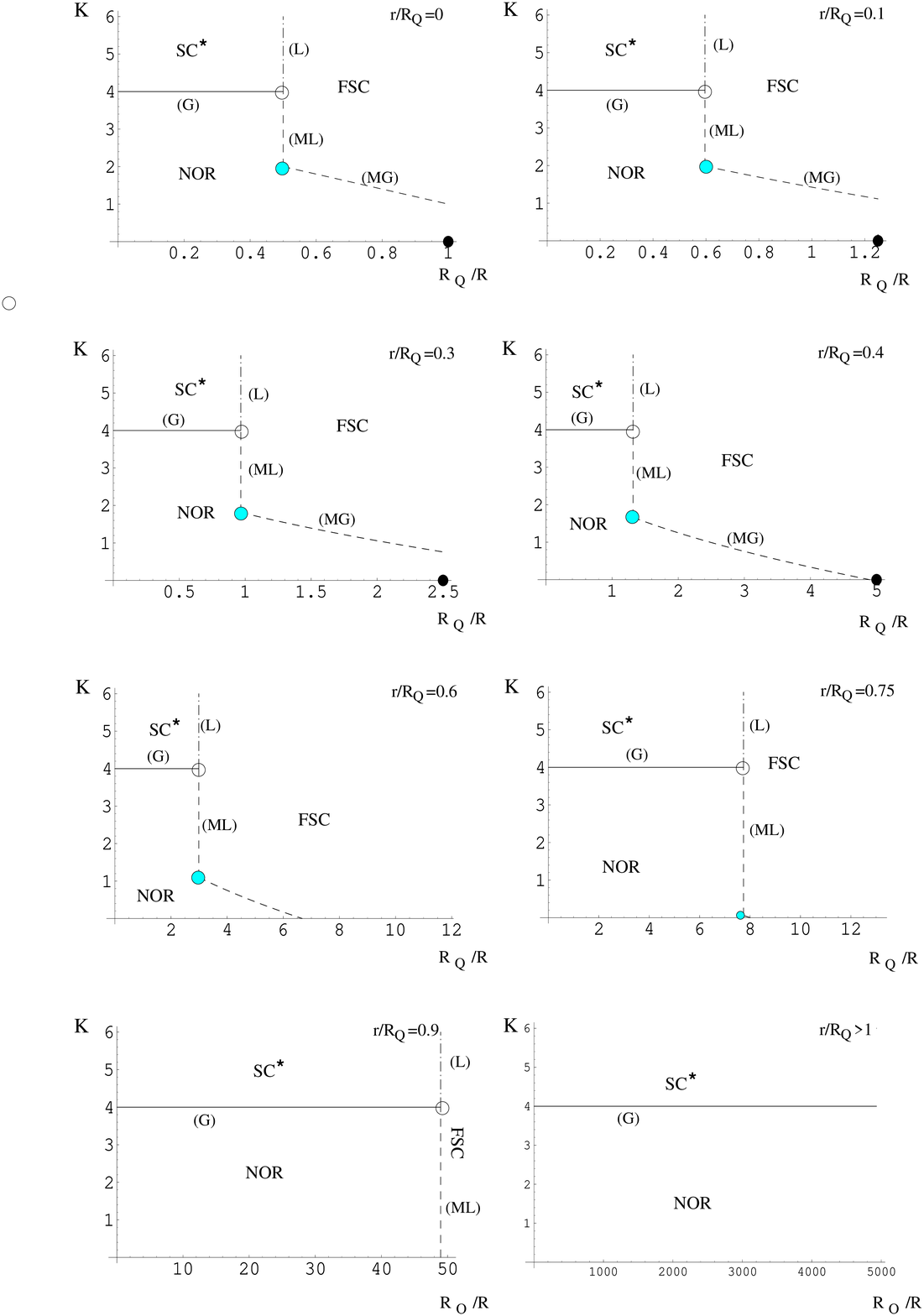}
\caption{Strong coupling phase diagram in $\j$--$R_Q/R$ plane 
for various values of $r$. 
The three phase boundaries between the FSC, \scs, and NOR
  can be local, global or mixed. They are marked as following: (G) marks global (solid line), (L) -
  local (dashed-dotted line), and (M) marks mixed (dashed lines). The
  FSC-NOR transition is partially local (ML) and global (MG) as marked
  in the figure. The meeting point where the FSC-NOR boundary changes
  its nature is a bi-critical point marked by the gray dot. The open
  dot Marks the meeting of the three phase boundaries, and is a
  multi-critical point. Black dots mark
where superconductivity breaks down in the limit of $\j \rightarrow
0$.  Note that the phase diagram in the vicinity of the sharp corners
(bicritical points) may
not be exactly correct due to large renormalizations of $\j$ which we
otherwise neglect. These regions may give rise to quite complicated
multi-critical behavior. The critical behavior in the vicinity of the
bicritical and multi-critical points is
richer than near the other phase boundaries, and is characterized
by very long crossovers. \label{iphases}}
\end{figure*}

Equation (\ref{action}), the action for phase slips, allows a
complete description of the phase diagram in the strong coupling limit and controlled expansions in the vicinity of the phase transitions for some ranges of parameters. 

When the fugacity of individual QPS,
$\zeta$, is relevant, the superconductivity should break down completely and the chain exhibit normal behavior. As $\zeta$ grows, it will also induce phase slip dipoles.  If $\zeta$ is irrelevant, on the other hand,
the chain should --- at least naively ---  be  superconducting, but which superconducting phase obtains depends on the behavior of the dipoles.

The simplest phase in the strong coupling limit is  the fully
superconducting phase (FSC). In this phase,  the fugacities $\{\eta_s\}$ and $\zeta$ are all
irrelevant about the Gaussian fixed line of the sine-Gordon model,
Eq. (\ref{action}), 
which thus controls the low energy behavior of the FSC phase.  Not
only will phase-slips occur only in tightly  bound pairs that will not
be apparent at low energy scales,  but phase slip dipoles will bind
into quadrupoles, and isolated dipoles will not be evident at low
energies. 

When both $\zeta$ and the $\eta_1$ --- the smallest dipoles that are
least costly --- are relevant, they will grow, inducing the other
$\{\eta_s\}$ to grow as well, until the small fugacity expansion
breaks down.  Below this scale, there will be free QPS's and no
superconductivity: the resulting normal phase is best studied from the
weak coupling limit as we do in the next section, Sec. \ref{wcsec}.

The behavior when $\eta_1$ is relevant but $\zeta$ is not is more subtle: either the \scs phase or the normal phase can obtain. If the 
phase-slips still bind into dipoles, but the dipoles do not bind into
quadrupoles, the \scs phase obtains. In this phase we can ignore the
ohmic dissipation and therefore also the dipoles as separate objects: the
proliferated dipoles essentially render the conversion resistance $r$
infinite. Once dipoles proliferate, the \scs phase may still be
unstable to single QPS. An instability to $\zeta$ once dipoles are
proliferated drives the \scs phase to the NOR phase, but this
stability must be considered differently.

The qualitative discussion above gives rise to the main features of
the phase diagram of Fig. \ref{iphases}: the exception being the
weak-coupling section of the mixed transition that occurs for small
$\j$ and small $R$; this we discuss in the next section, Sec. \ref{wcsec}.

The relevance of $\zeta$ or $\eta_1$ about the FSC limit triggers the
phase transitions from this phase to the normal and \scs phases as
well as controlling the bicritical point at which the mixed phase
transition changes character.  In the limit of very strong coupling,
the {\it locations} of the phase boundaries are given by the vanishing
of the linear eigenvalues of $\zeta$ and $\eta_1$.

The \scs-normal phase boundary can be understood straightforwardly
from the \scs phase (in which dipoles have proliferated) purely in
terms of individual QPS. This transition is triggered  by relevance of
$\zeta$.  Its location, however, cannot be precisely obtained in terms
of the original parameters, even in the  limit of strong coupling,
because of the effects of the proliferation of dipoles.

\subsection{FSC- \scs phase transition [line (L)] \label{scs11}} 

In both the FSC and \scs phases single phase slips are irrelevant. Therefore,
to determine the transition between the FSC and the \scs phase we need
to know when $\eta_1$ (the least costly dipole) changes from
irrelevant to relevant about the Gaussian fixed line. The primary role
of the non-linear RG flows for these purposes, is the generation of
$\{\eta_s\}$ from $\zeta^2$ in Eq. (\ref{rg232}) so that these
will appear and must be considered even when they are absent
initially. [Note, however, that the scale at
which $\eta_1$ becomes important is determined by the $\zeta^2$ term so these  second order terms can be 
crucial for understanding finite-temperature and other crossovers.]

The first order terms of Eq. (\ref{rg232}) for the flow of $\eta_1$:
\be
\frac{d\eta_1}{dl}\approx\eta_1\l(1-\beff \r)
\label{zsrg2fo}
\ee
control the behavior near this transition. In the limit of infinite
coupling they yield a boundary line between the FSC and the \scs phases:
\be
\beff^\infty_L=2\aeff\l(1-w\r)=1,
\label{scsfsc}
\ee
which is thus a condition on the critical resistances, e.g. $R_L^\infty(r)$ for this local transition. The quantitative superfluid properties drop out as is the case for individual junctions. As we shall see, for strong but finite coupling, the situation with a chain of junctions is more complicated.

When $\eta_1$ is irrelevant about the Gaussian fixed manifold, one must investigate whether there is a critical value of the fugacities above which $\eta_1$ grows.
For $\beff$ close to unity, an expansion in the fugacities is still possible.
The crucial terms are the creation of $\eta_2$ by $\eta_1$ and the feedback of $\eta_2$ on $\eta_1$.  The structure is
\be
\frac{d\eta_1}{dl}\approx\eta_1\l(1-\beff \r)+C_{112} \eta_1\eta_2
\ee
\be
\frac{d\eta_2}{dl}\approx\eta_2\l(1-2\aeff[1-w^2] \r)+C_{211} \eta_1^2
\ee
with the $C$s coefficients. Since the (linear) eigenvalue of $\eta_2$ is negative, there is a critical fixed point for $\beff>1$ at 
\be
\eta_1^*\sim \sqrt{\beff-1} \ \ \ \ {\rm and} \ \ \ \ \eta_2^* \sim \beff-1 \ .
\ee
It can be seen that the higher moment dipoles have corresponding fixed point values 
\be
\eta_s^* \sim (\beff-1)^{s/2}
\ee 
and can thus be neglected to leading order in $\beff-1$. 
Since $\eta_1$ will be of order the bare $\zeta^2$ after the brief initial transients, we see that for 
$R_L^\infty-R $ small but positive, there will be a critical value of $\zeta$ for the local transition
\be
\zeta_L \sim \l(R_L^\infty-R\r)^\frac{1}{4}
\ee
using, for convenience, $R$ as the control parameter; with $r$ varying instead, the behavior is similar only with a different coefficient.  

Just above the critical fugacity, the crossover energy scale below which the chain will no longer appear critical  goes to zero near the critical point as
\be
\E_{\times} \sim |\zeta-\zeta_L|^{\nuz} \ .
\ee
The critical exponent $\nuz$ varies  continuously with the resistive parameters. For $\beff-1$ small, 
\be
\nuz\approx \frac{1}{2(\beff-1)} \ .
\ee

Because this transition is essentially local --- with dynamic scaling exponent $z=\infty$ ---  the diverging length scales can be subtle. But there should be some  crossover in the spatial correlations at a length scale
\be
\xi \sim \vms/\E_{\times} 
\ee
beyond which the superconducting correlations behave somewhat differently.  

For this discussion, we have not been careful about where the RG flows go when the dipoles
proliferate. If they go to the \scs phase, the above obtains.  But it is also possible that they go near the \scs regime only to flow away at lower energies if  $\zeta$ is relevant. In this latter case, the transition will be from the  FSC directly to the normal phase, as we discuss below.   But we first consider the effects of dipole proliferation.

\subsection{\scs phase  and screening of 
  dissipative interactions \label{scsscreensec}}

In the region of parameter space where $\eta_1$ is relevant, the simple perturbative RG flow,
Eqs. (\ref{rg232}), breaks down, and phase-slip dipoles proliferate. But this is not in itself sufficient to
destroy superconductivity; the lead-to-lead coherence can persist as long as the quantum phase slips remain bound in pairs.
\cite{Refael2003, WernerRefael}  Dipoles do, however, allow
the phase of a single grain to jump by $2\pi$ relative to its
neighbors. This is 
the \scs in which  each grain no longer has a well defined phase. 
But the global superconductivity survives as long as $\zeta$ 
is irrelevant about the fixed line that controls the \scs phase.

In the \scs phase, the dissipation no longer plays an essential role. 
This can be most easily seen by comparing the dipole term with the quadratic parts of the action for the $\psi$ field in the Sine-Gordon action.  At low energies, the fluctuations of the field $\theta$ are small enough that one can ignore the $\theta_{j+1}-\theta_j$ in the dipole term and approximate it by
\be
\eta_1\cos[\psi_{j+1} -\psi_j] \ .
\ee
When $\eta_1$ becomes of order $\Omega$, the highest remaining frequency, the dipole term is comparable to the $R|\omega | |\psi_j(\omega)|^2$ and the $r|\omega | |\psi_{j+1}(\omega)-\psi_j(\omega)|^2$ term at frequencies of order $\Omega$. 
[If $r\ll R$ the comparison is somewhat more subtle.]  

Below this crossover frequency, the dipole term strongly suppresses fluctuations of neighboring $\psi_j$ at low frequencies: together with the $R|\omega | |\psi_j(\omega)|^2$ term, it cuts off the divergent fluctuations of $\psi$ while $\theta$ continues to fluctuate similarly to in the absence of the dipoles.   At this point, we can take the continuum limit without worry, replace the lattice with a high-momentum cutoff and expand out the dipole terms 
(analogous to treating screening in a metal as if by continuous charges --- the Debye-H\"uckel approximation) as
\begin{widetext}
\be
\eta\l(\cos\l(\ax\l[\der{\psi{(x,\,\tau)}}{x}+\der{\theta{(x,\,\tau)}}{x}\r)\r]-1\r)\approx -\frac{\eta}{2}\l(\ax\l(\der{\psi{(x,\,\tau)}}{x}+\der{\theta{(x,\,\tau)}}{x}\r)\r)^2.
\label{largezs}
\ee
An approximate action for the \scs phase is then
\be
\ba{c}
S=\int\frac{d\omega}{2\pi} \int\frac{dk}{2\pi} \l\{\l(k^2\l(\frac{1}{2\pi \jx}+\ax^2\eta\r)+\frac{\omega^2}{2\pi \jt}\r)\theta^2+\l(\ax^2\eta k^2+\frac{\xz}{2\pi \alpha}|\omega|\l(k^2+\frac{1}{\xz^2}\r)\r)\psi^2\r\}\vspace{2mm}\\
-\summ_i\int d\tau \zeta\cos\l( \theta_{i}{(\tau)}+\psi_{i}{(\tau)}\r)
\label{adipoles2}
\ea
\ee
\end{widetext}
Where we have neglected $\theta-\psi$ cross-terms proportional to
$k^2$. These do not contribute to or alter the singular interaction
between phase slips in this regime. Note that the plasmon part of the
interaction (the $\theta^2$ part) has been renormalized by the phase-slip dipoles,
but it is not screened, i.e., it still vanishes at $\omega=k=0$.
The
dissipative part, however, does
get screened reducing the dissipative interaction between
phase slips to $S_{diss}{(x,\,\tau)} \propto 1/\tau^2$ when
$|\tau|\gg\frac{\xz}{4\pi \alpha\eta \ax}$. 
This should be compared with Eq. (\ref{inter11}). Because of the finite screening time of
the dissipative  interaction between phase slips, this does not play a
role  for the asymptotic behavior of either the \scs phase, or the
\scs to normal transition. The screening of the dissipative interaction by dipoles can be understood simply by circuit methods
(Sec. \ref{ctm}) or by considering the phase-slips as a Coulomb gas. It plays
a crucial role for {\it finite} JJ chains for which the effects of
dissipation are not completely eliminated. This will be discussed in a future publication
\cite{finitechain}.

If the system flows to the \scs fixed line when the dipole fugacity
grows, then the critical behavior of the FSC-\scs transition is that
found in the previous subsection. Before returning to the trickier
question of what happens when the flows eventually go away from the
\scs regime, we briefly discuss the direct \scs-NOR transition.

\subsection{\scs --- Normal transition [line (G)]\label{brscs}}

Once dissipation has been eliminated,  we are left with a simple low-energy effective action for the \scs phase: the standard chain of grains
with self capacitance and  connected by
Josephson junctions.  In the Sine-Gordon representation, we have
\be
\ba{c}
S_{\rm SC*}\approx\\
\int\frac{d\omega}{2\pi} \int\frac{dk}{2\pi} \l(\frac{k^2}{2\pi \tilde{\jx}}+\frac{\omega^2}{2\pi \tilde{\jt}}\r)\theta^2-\summ_i\int d\tau \zeta \cos\l(\theta_{i}{(\tau)}\r)
\label{newa1}
\ea
\ee
where the tildes over the couplings are a reminder that they include renormalizations from integrating out the effects of the dipoles as in Eq. \ref{adipoles2}.

The action Eq. (\ref{newa1}) is most naturally studied by an isotropic RG 
rescaling both $x$ and $\tau$, and integrating out the modes of $\theta$ that have large $\omega$ or 
large $k$. By carrying out the RG flow in $x$ and $\tau$ directions, 
the usual K-T RG equations for $\j=\sqrt{\tilde{\jx}\tilde{\jt}}$ and $\zeta$ obtain
\cite{Minnhagen}:
\be
\ba{c}
\frac{d\j}{dl}=-\frac{\pi}{2} \j^2\at^2\zeta^2\vspace{2mm}\\
\frac{d\zeta}{dl}=\zeta\l(2-\frac{\j}{2}\r).
\ea
\label{ktrg}
\ee
The \scs-normal phase boundary --- a global transition --- thus occurs for 
$\zeta\ll 1$ at 
\be
\j_G=4+\cal{O}(\zeta ).
\label{pbscs}
\ee
When the initial $\zeta$ is  not small, the critical value of $\j$ will be increased  both by the direct effects of $\zeta$ and also via its generation of dipoles ($\{\eta_s\}$) and their later modification of $\jx$ at intermediate scales as in Eq. (\ref{adipoles2}).

In the normal phase near this global transition, the characteristic time and length scales both diverge exponentially as $\exp[1/\sqrt{\j_G-\j}]$ --- and similarly in $R-R_G$ when dissipation plays a role at intermediate scales and can control the transition.  This contrasts with  the power law divergences near the FSC-\scs local transition.

Although we have here used the conventional isotropic RG to study the
\scs - normal transition, it is important to understand whether the
anisotropic RG we use in the rest of this paper can reproduce the KT
results in the appropriate regime. We address this issue below in
Sec. \ref{revisit}.

\subsection{FSC --- Normal transition [line (M)] \label{brfsc}}

We now turn to the most subtle transition: the mixed-character transition from the 
FSC to the normal phase.  As the phase slip fugacity is increased, the
superconducting stiffness, $\j $, is reduced, or the resistances increased, a quantum phase transition between
the FSC phase and the normal phase of the chain can occur directly without an intermediate \scs phase. This mixed transition can be driven by two distinct mechanisms even in the strong coupling limit in which the Sine-Gordon description is useful.
If single QPS's become relevant when dipoles are still irrelevant about the FSC fixed line, they will drive the transition inducing proliferation of both dipoles and isolated QPS as the fugacity $\zeta$ grows.

But if small dipoles become relevant when single QPS are still irrelevant about the FSC fixed manifold, it is nevertheless 
still possible for the dipole-driven transition to be between the FSC and normal phases.
This occurs if the  flow towards the \scs phase from the proliferating dipoles is interrupted 
by the relevance of individual QPS about the \scs fixed line. We first discuss the latter case.

\subsection{Dipole driven mixed transition [line (ML)] \label{ssE}}

The RG flows for this dipole driven transition are somewhat complicated.
As shown earlier, if the QPS fugacity is irrelevant about the FSC
fixed manifold when the dipole fugacities start growing --- such as
when the bare fugacity, $\zeta$, is above a critical value --- $\zeta$
will continue to decrease until the effects of the dipoles are strong
enough to change its flow.  Crudely, this will happen when the
$\eta_s\zeta$ terms in  Eq. (\ref{rg232}) become large
enough to dominate over the $(\frac{1}{2}K+\aeff-1)\zeta$ term. As this
will happen when some of the $\etas$ are of order the energy scale
$\Omega$, say at energy scale $\E_\times$, the system will soon after
approach the \scs fixed line as discussed in Sec. \ref{scs11} (other
operators neglected perturbatively will also become important on these
scales). But by this point, if the system is close enough to
critical, $\zeta$ will have become extremely small. Thus even when it
is relevant about the \scs fixed line and hence will turn around are
grow on scales smaller than $\E_\times$, $\zeta$ will not become large
enough to make individual QPS proliferate, until a much lower energy
scale, $\E_N<\E_\times$.

Physical properties should exhibit both of the energy scales just beyond the FSC-normal
transition: At high energies, the system will appear critical ---
FSC-\scs critical --- at intermediate energies,
$\E_N\ll\E\ll\E_\times$, it will appear to be in the \scs phase  ---
albeit with a phase stiffness that is too low to sustain
superconductivity at long scales ---  and at asymptotically low
energies, $\E\ll\E_N$, normal behavior will obtain.  This behavior is
characteristic of a {\it dangerously irrelevant operator} --- in this
case $\zeta$ --- about the critical fixed point.  The energy scale
$\E_N$ will go to zero as a power of the distance from criticality,
but its exponent will be larger than that of $\E_\times$ by an amount
controlled by {\it both} the flow  of $\zeta$ towards the FSC fixed
manifold, and its flow away from the \scs fixed line, as well as the
flow of $\eta_1$ away from the critical fixed point.  The behavior of
the length scales will be similarly complicated and the dynamic
critical exponent relating $\E_N$ to the superconducting correlation
length, will be larger than unity.

\subsection{QPS driven mixed transition [line (MG)]\label{ssF}}

In the regime in which the FSC-normal transition is driven by QPS --- the fundamental topological excitations --- the phase boundary and nature of the transition can be studied
from Eqs. (\ref{rg232}) for the flow of $\zeta$. For this transition,
the dipoles play only a secondary role.

In the region near the FSC fixed manifold, the perturbative
RG equations (\ref{rg232}) are valid and, in the regime of interest, the $\etas$ are renormalized to zero and can be ignored. We thus need only
the first two RG flow Eqs. (\ref{rg232}). Working in units of the
cutoff energy scale $\Omega$ and length scale $\ax$, we have:
 \be
\ba{c}
\frac{d\jt}{dl}\approx\jt -\frac{\pi}{2}\jt ^\frac{5}{2}\jx^{\frac{1}{2}}\at^2\zeta^2 ,\vspace{2mm}\\
\frac{d\jx}{dl}\approx-\jx ,\vspace{2mm}\\
\frac{d\zeta}{dl}\approx\zeta\l(1-\frac{1}{2}\sqrt{\jx \jt}-\aeff\r),
\ea
\label{zflow1}
\ee
Equations (\ref{zflow1}) are 
similar to the Kosterlitz-Thouless (K-T) flow equations, except for  the anisotropic scaling of $\jx$ and $\jt$.  We first proceed naively and transform to variables in which the flows {\it appear} more isotropic, defining:
\be
\ba{c}
\overline{\jx} =\jx  e^l,\vspace{2mm}\\
\overline{\jt} =\jt  e^{-l},\vspace{2mm}\\
\oz=\zeta e^{\frac{1}{2}l}.
\ea
\label{rescale1}
\ee
whereby  Eqs. (\ref{zflow1}) become: 
\be
\ba{c}
\frac{d\overline{\jx}}{dl}=0,\vspace{2mm}\\
\frac{d\overline{\jt}}{dl}=-\frac{\pi}{2}\overline{\jt}^\frac{5}{2}\overline{\jx}^\frac{1}{2}\oz^2 \at^2,\vspace{2mm}\\
\frac{d\oz}{dl}=\oz\l(\frac{3}{2}-\frac{1}{2}\sqrt{\overline{\jx} \overline{\jt}}-\aeff\r).
\ea
\label{zflow2}
\ee
Shifting and rescaling the variables via
\be
\ba{c}
x=\frac{\sqrt{\overline{\jx}  \overline{\jt}}}{2}+\aeff-\frac{3}{2}\vspace{2mm}\\
z=\sqrt{\frac{\pi \overline{\jx}}{8}}  \overline{\jt}\at\oz .
\ea
\label{newxz}
\ee
so that $x$ parametrizes deviations from the special point, the flow Eqs. (\ref{zflow2}) assume the canonical K-T form 
\be
\ba{cc}
\frac{dx}{dl}\approx -z^2,       &       \frac{dz}{dl}\approx -zx.
\ea
\label{ktc}
\ee

The flows of Eq. (\ref{ktc}) suggest a  phase
boundary between the normal and FSC phases. In the FSC phase $z$ flows to zero
whereas in the normal phase $z$ diverges and $\j $ flows to zero. The critical line separating  the two phases is thus
\be
x_M\approx -z,
\label{bd1}
\ee
When $\zeta\ll 1$ the FSC-normal phase boundary in Eq. (\ref{bd1}),
$x\approx 0$, translates to 
\be
\j_M =3-2\aeff. 
\label{bd2}
\ee
Although the phase boundary of Eq. (\ref{bd2}) is correct for the {\it renormalized} $\j$, for non-zero initial QPS fugacity $\zeta$, the critical value of $\j$ will be slightly larger --- by ${\cal O}(\zeta^2)$ when the bare fugacity is small ---  because of the renormalizations  of Eqs. (\ref{zflow2}). 

On the normal side of the transition, the behavior is subtle.  The subtlety arises because the transformation between $\oz$ and $\zeta$ (Eq. \ref{newxz} and \ref{zflow2} )
involves the scale of the RG explicitly: $\zeta\propto
e^{-l/2}\oz\propto e^{-l/2} z$, so that $\zeta$
may decrease even though $z$ is formally relevant.   But as $z$ grows to the order unity, it will substantially decrease $x$, concomitantly, $\overline{\jt}$, and hence the further growth rate of $\oz$ from Eq. (\ref{rescale1}). The growth rate of $\oz$ will then become rapid enough to overcome the $ e^{-l/2}$ factor and make $\zeta$ grow as well.  Thus after  initially decreasing because  of the factor
$e^{-l/2}$, $\zeta$ eventually turns around and then becomes large (at not much lower energies) at which point the behavior will be characteristic of the normal phase. 

The dipole fugacities, $\etas$, will be small as long as  $\zeta$  is, but when $\zeta$ becomes large, they will grow (for large $s$ as well as small). This implies
that Josephson junctions are no longer independent of each other, and
phase slips become spread-out over several junctions. In
principle, at this stage we could switch to carrying out the RG isotropically, but by then other higher order processes will also come in and expansions in powers of the fugacities break down.

The critical behavior near this QPS driven mixed transition will be
similar to near a conventional Kosterlitz-Thouless transition, with,
for example, energy scale, $ E_\times$, vanishing exponentially in the
inverse square root of the distance from criticality. The
corresponding exponent is $\nuz=\infty$ in contrast to the finite but
variable $\nuz$ that characterizes the dipole driven mixed transition.
Associated with the $E_\times$ will be one or more diverging length
scales. There should be a scale that grows as $1/E_\times$ corresponding to
dynamical exponent $z=1$, but there may be a second scale associated
with the onset of substantial screening of $\jx$ and spatially (rather
than temporally) separated QPS that will occur only after the dipoles
come into play. We will not explore this issue further here.

To make the above analysis solid, It is essential to show that the
perturbative RG  equations on which it is based are valid out to
scales at which $\j(\ell)$ starts to decrease significantly from the
screening by pairs of QPS.  The rescaling from $\zeta$ to $\oz$ was
chosen to make these renormalizations, as in Eq. (\ref{zflow2}), small
if and only if $\oz$ remains small. This could better (but
equivalently to the needed order) have been done by defining instead
$\tilde{\zeta}=\sqrt{\jt}\zeta$ thereby making $d\log J/d\ell$ depend
only on the combinations $\tilde{\zeta}$ and $\j=\sqrt{\jt\jx}$: the
latter (in contrast to, e.g. $\vms=\sqrt{\jt/\jx}$) is of order unity
in the regime of interest near the transition. With this choice, it
means that the dipole fugacities, $\etas$, are induced by
$(\tilde{\zeta})^2/\jt$ which decreases rapidly after the initial
transients that create small dipoles. Thus the $\etas$ should not be
problematic as they will be decreasing and only give rise to {\it
  relative} changes: terms of order $\zeta\eta_s$ .

The primary potential problems are additional renormalizations of $\jt$ (or $\jx$) that are not small when $\tilde{\zeta}$ is small. These can arise, for example, by the combination of a {\it tripole} --- a QPS ``dressed" by a nearby small dipole --- and an opposite sign QPS. Such terms are equivalent to terms of order $\tzeta^4$ in other RG schemes, but these, and other terms higher order than quadratic, do not occur directly with the RG scheme we use: they are generated from other operators as in this case. This has the advantage of making a cataloging of all operators and how they renormalize each other linearly and quadratically, equivalent to ``all orders" arguments in other perturbative RGs.
 One can readily  show that the generation of tripoles from a dipole and a QPS, and the subsequent renormalization of the tripole fugacities by dissipation plus the stiffness $\j$, will result in effects of the tripoles back to $\tzeta$ and to $\jt$ that decay with energy scale while $\tzeta$ (and equivalently $z$) is decreasing. Further analysis along these lines shows that indeed, the result for the critical $\j_M$, Eq. (\ref{bd2}) is correct.  What is rather surprising, is that this
 arises from a basic (power-counting-like) renormalization of $\zeta$
 that is midway between the `1' for purely temporal rescaling, and the
 `2' for space-time rescaling. 

By showing that the RG remains controlled out to this stage, we have verified that the FSC-normal phase boundary is in one regime 
indeed given by Eq. (\ref{bd2}) in the limit of small fugacities;  more generally the location of the transition will be exactly $\j=3-2\aeff$  if the renormalized --- and hence measurable  --- low energy parameters, $\j$ and $\aeff$, are used.

Nevertheless,  near the phase boundary on the normal side, the fact that $\zeta$ first flows toward zero and only
at a later stage becomes large, gives rise to a
crossover in the resistance vs. temperature.  From the last of Eqs. (\ref{zflow1}) we see that this will occur
as long as 
\[
 \j >2-2\aeff. 
\]
Thus the crossover occurs in the range of parameters 
\be
2-2\aeff<\j <3-2\aeff\ ;
\label{mphase}
\ee
 the
physical consequences and origin of this region is discussed in
Secs. \ref{res} and  \ref{ctm}.

\subsection{FSC---Normal bicritical point \label{bicrit}}

As shown above in Secs. \ref{ssE} and \ref{ssF}, the FSC-normal
transition can have one of two distinct characters: either a dipole
driven transition with power-law singularities [line (ML) in Fig. \ref{iphases}], continuously variable
exponents, and non-trivial relations between length and time scales,
or driven by single QPS, with exponentially-rapidly decreasing energy
scales and at least some space and time scales scaling similarly [line
  (MG)]. In
the limit of small (bare) fugacities, the transition occurs where in the $\j,\aeff,\beff$ space the system is at intermediate
energy scales.  The two critical surfaces in this space meet along a
bicritical  line, across which the critical behavior changes, although
the phases on both sides are the same.   At non-zero fugacity,
$\zeta$, the behavior is similar with the critical surfaces (actually
now hypersurfaces) and location of the bicritical  line (actually now
a two-dimensional manifold) changed. When the critical fugacity is
small, as occurs when $\beff-1$ (see Eq. \ref{zsrg2fo}) is small (and positive), the behavior
near the bicritical  line can be treated perturbatively.  If the
multicritical manifold is approached along the dipole-driven critical
manifold, the former can be found by considering the eigenvalue of
$\zeta$ at the point on the critical fixed manifold, parametrized
perturbatively by $\eta_1^*$.  This eigenvalue will be increased by
${\cal O}(\eta_1^*)^2$ from its value at the corresponding point on the
FSC fixed manifold (all $\etas$ zero).  When the {\it critical}
eigenvalue, rather than $1-\frac{1}{2}\jx-\aeff$ passes through
$-\frac{1}{2}$, the critical fixed manifold will become unstable to
individual QPS at low energies. This condition thus characterizes the
multicritical manifold. For $\eta_1$ just  below the bicritical
manifold the flow will be to the FSC phase, but because of the initial
effects of $\eta_1$ on $\zeta$, will end up closer to critical than it
would have been.  Associated with the multicriticality will be
complicated crossover behaviors that we will not explore.

\subsection{\scs--- Normal transition revisited \label{revisit}}

In the standard treatment of isotropic 2d, or 1+1, X-Y
models, we use isotropic RG to obtain the K-T transition. This is also the way in which we analyzed
the Global (G) NOR-\scs transition in Sec. \ref{brscs} once dipoles
are proliferated. As mentioned earlier, for consistency, we should be able to analyze
this global \scs-normal transition by the anisotropic RG we use in the
rest of this paper. This also serves to make a more convincing case
for the peculiar behavior of the QPS driven FSC-normal transition.
Analyzing the global transition by the anisotropic RG is possible but
clumsy; we only outline the basics here. 

It is useful to first consider what modification there would be to the
small-fugacity  flow equations if opposite-sign QPS on {\it different}
junctions could annihilate each other under renormalization rather
than combining into dipoles.  Because of factors of
$\u^s=\exp[-\ax/\Lambda]$ --- similar to those in the generation of
$\eta_s$ by $\zeta^2$ in Eq. (\ref{rg232}) --- only QPS pairs separated by
less than or of order $\Lambda(\ell)=\sqrt{\jt/\jx}/\Omega$ will be
renormalized away.  But this is a factor of $\Lambda/\ax$ more
combinations than were included from QPS pairs on the same junction:
the renormalization of $\jt$ of order $\zeta^2$ will thus be larger by
a similar factor.  We now  recombine parameters similarly to what was
in the mixed case: to make the renormalizations of $\jt$ small when
$\tzeta$ is small and $\j$ of order unity. This now requires
$\tzeta=\jt\zeta$ (contrast with $\sqrt{\jt}\zeta$ in the mixed case)
which will change the eigenvalue of $\tzeta$ to $2-\half \j -\aeff$
while keeping $\j$ marginal.  In the absence of dissipation, this
would yield  a KT transition with critical value $\j_G=4$ as obtained
from the isotropic RG of Sec. \ref{brscs}.   What is not immediately clear,
is at what point the flows have crossed over into the normal regime
and $\zeta\cos \theta_j$ can be well approximated by $\zeta
(1-\half\theta_j^2)$ and the $\summ_j$ approximated by
$\frac{1}{\ax}\int dx$.  This can be checked  by estimating the
mean-square fluctuations of $\theta_j$ using the resulting quadratic
continuum form. As at low energy scales $1/\jx$ is large and $1/\jt$
small, this will be dominated by wave-vectors of order $1/\Lambda$ or
smaller. The  remaining frequency integral will be cut off at
$\sqrt{\zeta \jt}$ and thus dominated by  frequencies of
order $\Omega$ when $\tzeta=\zeta \jt/\Omega$ is of order unity as
would be naively guessed from the rescaling that gave the KT flows.

But this is not the whole story.  In addition to the renormalizations
of $\jt$, there should be similar (asymptotically) renormalizations of
$\jx$ which controls the spatial gradients of $\theta$. In the
anisotropic RG, these arise from replacing the dipole operators in the
absence of $\psi$,  $\cos(\theta_{j+s}-\theta_j)$, by
$1-2(\theta_{j+s}-\theta_j)^2\approx 1-2s^2\ax^2(\partial_x\theta)^2$.
This is justified if $\theta$ is sufficiently slowly varying in space.
With the anisotropic cutoff, this is justified for $s<\Lambda$ as,
because of the low frequency cutoff relative to the wave-vector
dependence, $\langle (\theta_{j+s}-\theta_j)^2\rangle \approx
2\j|s|/\Lambda$ in this regime.  Thus the renormalizations of $1/\jx$
to order $\zeta^2$ are of the correct form. By summing over the
appropriate range of $s$, their amplitude can be found to be of order
$\tzeta^2 \Lambda$ which have the additional $\Lambda$ factor that
compensates exactly for the linear growth of $1/\jx$ with the inverse
of the energy scale in the anisotropic RG. Similarly, the argument in
the previous paragraph for how $\jt$ is renormalized, can   be
justified by  the replacement of $(\partial_\tau
[\theta_{s+j}+\theta_j])^2 \cos(\theta_{j+s}-\theta_j)$ by
$4(\partial_\tau \theta_{\half s+j})^2$ renormalizing $1/\jt$.

For the \scs - normal transition when the fugacities are small, the
analysis above can only be done once the dipoles have been induced and
proliferated enough to suppress the fluctuations of $\psi$ and enable
it to be neglected as in Sec. \ref{brscs} . Again,  the intermediate regime
of the flows, where the  $\etas$ are of order unity, cannot be handled
in a controlled manner.  But this occurs over a small range of energy
scales and thus will only result in factors of order unity.

\subsection{Multicritical point \label{multi}}

The three phase boundaries, FSC---\scs, \scs---normal, and FSC---normal come together at a multicritical point as shown in the schematic phase diagram, Fig. \ref{iphases}. The behavior near this multicritical point involves an interplay between the \scs-like behavior when dipoles proliferate, and the FSC-like behavior when the dipole fugacities are small.  How these come together at the multicritical point is likely to involve flows in the intermediate coupling regime in which controlled calculations are beyond the scope of the methods of this paper.


\section{Weak Josephson coupling limit \label{wcsec}}

In this section we analyze the JJ chain when the Josephson
interaction is weak: $E_J\ll E_C$. In this limit the starting point is
to assume there is no phase coherence between grains so that the
currents between grains are predominantly normal. We can then consider
{\it Cooper-pair-tunneling} events perturbatively. In this limit the
Josephson coupling is sufficiently weak that superconductivity can
only be established if the dissipation is sufficiently strong to drive
the system into the FSC phase: the \scs phase cannot occur as its
stiffness $K=2\pi\sqrt{E_J/E_C}$ is too small. We  thus can explore
only the dissipation-dominated transition directly from the normal to
the FSC phase: as both local and global superconducting coherence are
established by this transition, it has  {\it mixed} character.  As
discussed in Ref. \onlinecite{Fisher-zwerger}, pair tunneling events are dual to quantum phase slips. 
Nevertheless, the treatment is substantially simpler than that of the mixed transition in the strong coupling limit analyzed in the previous section.

\subsection{Pair-tunnel events \label{ptesec}}

The basic objects that need to be considered are Cooper pair-tunneling
events between grains. Their action can be formulated in the
Coulomb-gas representation. This allows the RG flows
when pair-tunneling is rare to be worked out perturbatively.

Consider first the Hamiltonian of a
single capacitively shunted Josephson junction:
\be
H_{JJ}=\frac{1}{2C}\hat{Q}^2-E_J \cos\phi.
\label{pt0}
\ee
In the weak coupling limit, the charge fluctuations will be small relative to the phase fluctuations, and the junction can be described in terms of charge states with wave functions:
\be
\Psi{(\phi)}=e^{i\frac{q}{2e}\phi},
\label{pt1}
\ee
where $q$ is the charge imbalance across the 
 junction. The cosine term in
the Hamiltonian thus  induces hopping of Cooper-pairs, since $e^{i\phi}$ is
a translation operator that changes the charge imbalance by
2e. We can describe the normal-to-superconductor
phase transition of a {\it resistively} shunted Josephson junction in
term of pair-tunnel events: when these are suppressed at low energies,  the
junction is well described by the charge state, Eq. (\ref{pt1}), and is in the normal state.
But when pair-tunnels proliferate, superconductivity
is established and the phase becomes the good quantum number.  In this
dual description, in opposition to that in terms of QPS, the quantum fluctuations are responsible for the superconductivity rather than destroying it.

To  explore the pair-tunnel physics, we can start with   the action in
terms of the superconducting phases, $\{\phi_j\}$,  of the grains in the chain.  This
is given by Eq. (\ref{allaction}). Since we are interested in the low energy dynamics on small length scales, 
we take the limit of small $\omega$ of the quadratic parts of the action, keeping the spatial discreteness but dropping the ${\cal O} (\omega^2 \phi^2)$ charging energy. The partition function in this limit can be written in the form:
\begin{widetext}
\be
Z=\int D[\phi_{i}{(\tau)}]exp\l[-\frac{1}{2} \int \frac{d\omega}{2\pi}\int \frac{dk}{2\pi}
\frac{|\omega|R_Q}{2\pi \ax \l(R+r\l(2-e^{ik\ax}-e^{-ik\ax}\r)\r)} \l|\Delta{(k,\,\omega)}\r|^2
+\summ_i\int d\tau E_J \cos\l(\Delta_{i}{(\tau)}\r)\r]
\label{Taction23}
\ee
\begin{figure}
\includegraphics[width=8.5cm]{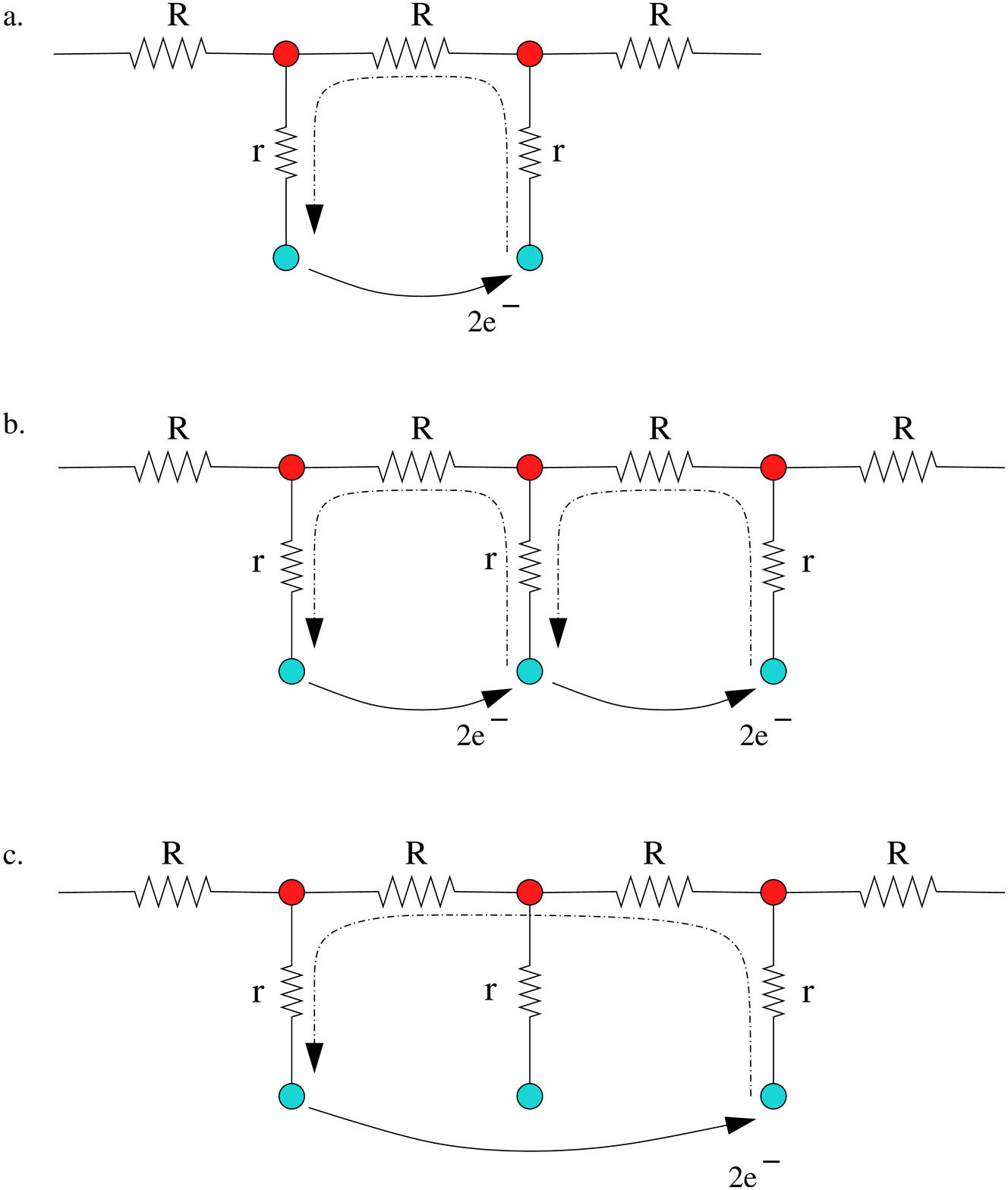}
\caption{(a) Cooper pair tunneling event,  between two neighboring
  grains (solid line): the basic process for weak Josephson
  coupling. Charge relaxation is via the resistance $2r+R$ (dotted
  line). 
(b) Two nearly simultaneous pair-tunnels involving a common grain: the current through the conversion resistance $r$ of that grain cancels giving rise to an effective
interaction between the two pair-tunnels.  These two pair tunnels 
can renormalize into a pair-tunnel over two junctions (c). \label{pair1}}
\end{figure}
or equivalently with a local part on each grain, and an exponentially decaying  --- with $\w^s$ --- interaction of opposite sign between phases on grains $s$ apart that exactly cancels for the total interaction (i.e., vanishes at $k=0$). This is because 
it is the phase {\it differences} across the junctions,
\be
\Delta_{i}{(\tau)}=\phi_{i+1}{(\tau)}-\phi_{i}{(\tau)}.
\ee
that enter here. Note that in the normal phase --- the starting point
here -- it would be more proper to write the resistive part of the
action in terms of $d\Delta_j/d\tau$ which is proportional to the
electrochemical potential differences across the junctions as the
phase variables are not defined modulo $2\pi$. The absence of
superconducting aspects other than the cosine terms is because the
quadratic action explicitly does not include the Josephson coupling.
At low energies and when the capacitative interactions dominate, the
densities of states parameters, $D_N$ and $D_S$, drop out.  Note that
if there were no normal conduction across the junctions, i.e. $R=0$,
the quadratic action would separate into a sum over non-interacting
grains.

The role of pair-tunneling events, and hence the superconducting
behavior of the JJ chain, can be seen by carrying out an expansion of Eq. (\ref{Taction23}) in powers of $E_J$. The 
$n$'th order term corresponds to 
the probability weight of $n$ pair-tunnels occurring
in  space-time. If the weight of the partition
function (\ref{Taction23}) is concentrated in
the term of $0$'th order in $E_J$, it means that no
superconductivity is taking place at $T=0$, since pair-tunnel events
are suppressed. On the other hand, if the weight is concentrated on
high order terms, pair-tunnel events proliferate ---
i.e., they happen frequently --- and the chain is superconducting.  

The expansion in $E_J$ leads to a Coulomb gas representation
of the action of pair-tunnel events and the partition function:
\be
Z=
\summ_N \l(\frac{E_J}{\Omega}\r)^N \tilde{\summ_{\{\sigma,\,x,\,\tau\}}}\exp\l[-1/2 \summ_{m=1}^N  \summ_{n=1}^N \sigma_m\sigma_n G(x_m-x_n,\tau_m-\tau_n)\r],
\label{pfpt2}
\ee
where $\{\sigma=\pm 1\}$ are the ``charges" --- i.e. signs --- of the tunnels, 
the sum is over all distinguishable neutral configurations with
$\summ_n \sigma_n=0$, and the interaction between tunnel
events is 
\be
G{(x,\,\tau)}=-2q_1q_2\log\l(\frac{|\tau|}{\at}\r)\l(\frac{R+2r}{R_Q}\delta_{x,\,0}-\frac{r}{R_Q}\l(\delta_{x,\,\ax}+\delta_{x,\,-\ax}\r)\r),
\label{ptinter}
\ee
\end{widetext}
with $\delta_{x_1,\,x_2}$ the  Kronecker $\delta$. The short time cutoff, $\at\sim(R+2r)C/\hbar$ is now the time duration of a phase slip, 
essentially the RC relaxation time of a charge imbalance between two
neighboring grains. This and other cutoffs from the high energy physics, we approximate by   a UV cutoff $\Omega_0=1/\at$. The {\it tunneling fugacity} we denote
\be
\xi_1 \propto E_J,
\ee
and it has units of frequency.  
  
The distance dependence of the  interaction Eq. (\ref{ptinter}) is easy to understand. If two pair-tunnel events happen
through the same Josephson junction, the strength of the logarithmic interaction
between them is $2\frac{R+2r}{R_Q}$  the 
total resistance
that the Cooper-pair has
to go through in order to relax back to an equilibrium charge distribution (Fig. \ref{pair1}(a)) as found for 
 a single shunted Josephson junction
(see e.g. Ref. \onlinecite{Refael2003}). If two pair-tunnels happen on neighboring junctions, the Cooper-pair
relaxation current overlaps on the common grain and thus involves the normal-to-superfluid resistance within the grain, $r$ (Fig. \ref{pair1}(b)).
This reduces  the logarithmic interaction strength of
$2\frac{r}{R_Q}$ between neighboring tunnel-events. Events separated by more than one junction have independent relaxations and hence no interaction as in Eq. (\ref{ptinter}).

If two tunneling events that involve a common grain occur close together in time, the charge changes of the common grain can cancel and the combination is equivalent to a single pair-tunneling
between further separated grains (see Fig. \ref{pair1}(c)).  We denote by $\xi_n$ the fugacity of a
pair-tunnel between grains separated by $s \ax$. Although in the original
partition function, Eq. (\ref{pfpt2}), $\xi_n=0$ for $n\geq 2$, the
long range pair tunnels will be produced during the RG flow.

The strength of the mutual and self interaction of pair-tunnel events will determine
whether they proliferate and induce superconductivity in the chain, or whether pair-tunnels
form bound pairs and annihilate each other, keeping the chain normal. For
weak-Josephson-coupling, this can be analyzed via a perturbative RG analysis in powers of the $\{\xi_s\}$.

\subsection{RG Flow of the pair-tunnel fugacities \label{wrg}}

To study the RG flows, it is easiest to go back to the initial action in terms of the superconducting phase differences, $\Delta_j$,  between the junctions:
\begin{widetext}
\be
Z=\int D[\phi{(\tau)}]exp\l[-\frac{1}{2} \int \frac{d\omega}{2\pi} \int \frac{dk}{2\pi}
\frac{|\omega|\ax R_Q}{2\pi \l(R+r\l(2-e^{ik\ax}-e^{-ik\ax}\r)\r)} \l|\Delta{(k,\,\omega_n)}\r|^2
+ \summ_{i}\int d\tau \summ_s \xi_s  \cos\l(\summ_{n=0}^{s-1}\Delta_{i+n}{(\tau)}\r)\r].
\label{Taction24}
\ee
\end{widetext}
As explained in the previous section, this is equivalent to a Coulomb gas of  pair-tunnel events
because 
\be
\xi_s\exp\l(\summ_{n=0}^{s-1}\Delta_j\r)=\xi_s\exp(\phi_{j+s} -\phi_j)
\ee
increases the charge on grain $j+s$ by 2e, and reduces the charge on
grain $j$ by 2e at rate $\xi_s$. We have explicitly included  longer
range tunneling as this will, in any case, be produced under renormalization from the basic nearest neighbor tunneling which has rate $\xi_1=E_J$. 
The dissipative interaction between a  pair-tunnel of range $s$ and a tunnel in the reverse direction a time $\tau$ later, is 
\be
\frac{2r+sR}{R_Q} \log(\tau/\at)
\ee
proportional to the 
total resistance through which normal relaxation must occur to compensate for  a Cooper-pair tunneling between grains separated by $s$.

The RG flows in powers of the fugacities of pair-tunnels are obtained similarly to those in the strong coupling regime in terms of QPS (see Sec. VIIB of Chapter 5 of
Ref. \onlinecite{RefaelThesis}): indeed they have a similar structure
to the RG for the various QPS dipoles in the absence of individual QPS
(Sec. \ref{revisit}).

Including the combining of pair tunnel pairs --- such as one from
$j-s$ to $j$ together with one from $j$ to $j+m$ into a single one
across $m+s$ junctions --- the RG flows are
\be
\ba{c}
\frac{d\xi_s}{dl}=\xi_s \l(1-\frac{2r+sR}{R_Q}\r)\\
+\at\half \summ_{m\neq s}\frac{2r+mR+|s-m| R-sR}{R_Q}\xi_m\xi_{|s-m|}.
\label{xirg}
\ea
\ee
Note that we have  neglected the renormalization of two
overlapping pair-tunnel events which do not share
any grains. Such events renormalize into a complicated non-contiguous
compound four-tunnel events, and are more strongly suppressed.

\subsection{Phase diagram and Normal-FSC transition for weak coupling \label{wpd}}

 \begin{figure*}
\includegraphics[width=15cm]{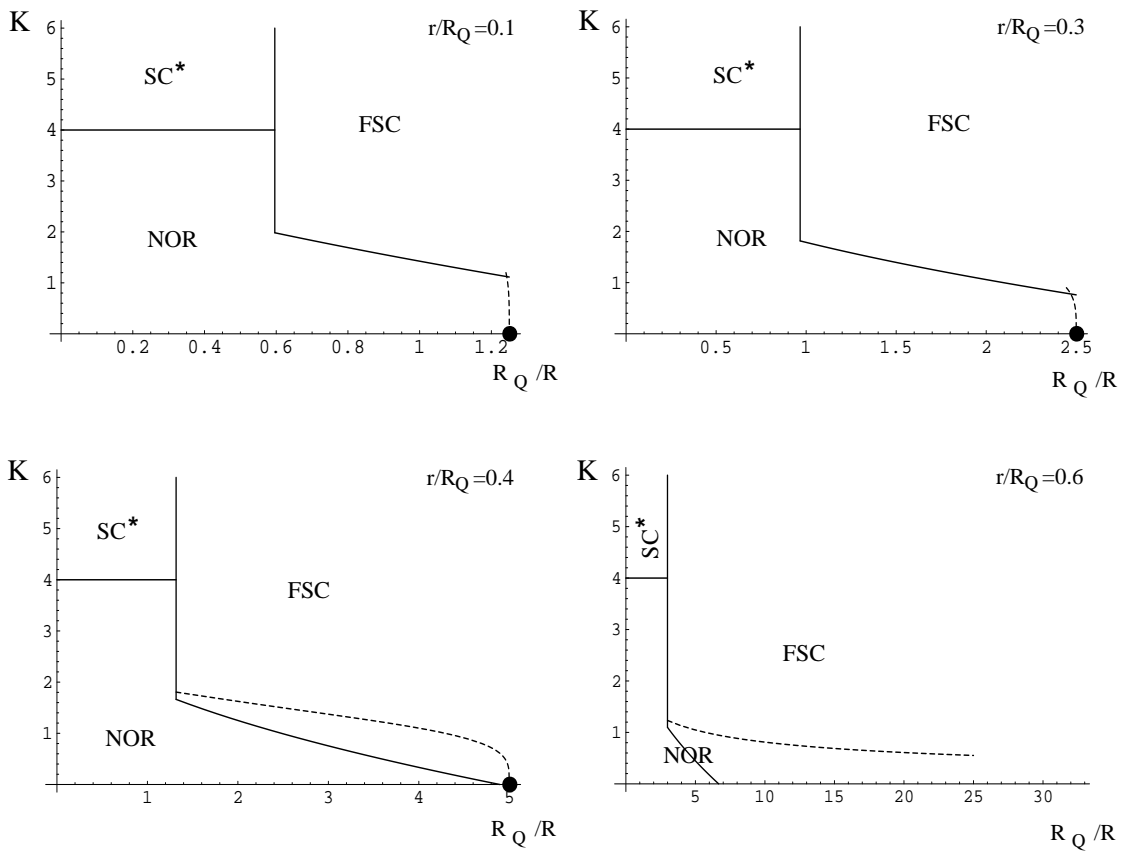}
\caption{Phase diagram with the weak Josephson
coupling results. The solid lines represent the
transition lines obtained with the strong coupling phase-slip approach (Sec. \ref{sec5}). The
dotted line is obtained from extending Eq. (\ref{xifpt}) by solving
the flow equations (\ref{xirg}):
  $\j\sim\frac{1}{\sqrt{r}}\sqrt{\delta}$. The two approaches seem to
  cross into each other continuously. The meeting point of the weak
  and strong-coupling boundaries is likely to be a bi-critical point
  where the transition changes its nature, but our methods are limited
  in that regime.\label{wphases}}
\end{figure*}

The trivial fixed point of the RG equations for the pair-tunnel fugacities is the gaussian line with no pair-tunnel events:
\be
\xi_s=0\ \ \ \ \forall s \ :
\ee
this controls the low energy behavior of the normal phase of the chain.  The normal fixed point is stable as long as all the $\xis$ are irrelevant: since $\xi_1$ is the least irrelevant, this obtains  when:
$\frac{2r+R}{R_Q}>1$ ,
so that with $R$ as the tuning parameter, the critical point in the limit of zero coupling 
is given by
\be
R_M^0=R_Q-2r \ .
\label{norcon}
\ee
When $R<R_M^0$, the normal fixed point becomes unstable: Cooper pair tunneling events proliferate and give rise to the superconducting phase. Since all the non-linear terms in the RG flows of the $\etas$ are positive, these can only accelerate this flow so that the condition $R<R_M^0$ is certainly {\it sufficient} for superconductivity
Formally, the flow is then  toward  large $\xi$ and eventually to  the FSC fixed manifold where the QPS fugacity, 
$\zeta$ is zero. 

Because the non-linear contributions to the flows are positive, superconductivity can still occur when $R+2r>R_Q$ and the $\xis$ are irrelevant about the Gaussian fixed point.
If 
\be
\delta\equiv \frac{R+2r}{R_Q} -1
\ee
is small and positive, $\xi_1$ will be weakly irrelevant while the other $\xis$ are still strongly irrelevant.  Nevertheless (as in Sec. \ref{revisit}) the feedback of these to $\xi_1$ important.  For small $\xi_1$ this is dominated by the creation of $\xi_2$ at order $\xi_1^2$, and the feedback of this into $d\xi_1/d\ell$ via the $\xi_1\xi_2$ term.  There is thus a critical fixed point with
\be 
\xi_1^*\sim \sqrt{\delta}
\label{xifpt}
\ee
and $\xi_2 \sim \delta \ll \xi_1$.  It can be seen that the longer range $\xis$ attain fixed point values $\xi_s^* \sim \delta^{s/2}$. [Indeed, with the simple structure of the flows to quadratic order, an exact fixed point can be found: $\xi^s \propto (B)^s$ with $B$ an $r$ and $R$ dependent factor that is of order $\sqrt{\delta}$ for small $\delta$. But this is {\it not} a controlled expansion for the fixed point except when $\delta$ is in any case small, or perhaps in some special part of parameter space.]
Equation (\ref{xifpt}) corresponds to  a critical fixed line that controls the critical surface of the mixed character normal---FSC transition. For $\delta$ small, the critical    $E^c_J \sim \xi_1^c\approx \xi^*_1(R,r)$: when
$E_J<E_J^c$ the chain is in the NOR phase, while when 
$E_J>E_J^c$ the chain is superconducting.   Note that equivalently, one could pass from normal to superconducting phase at fixed Josephson coupling by decreasing $R$ through its critical value $R_M$, which, in the weak coupling limit is 
\be
R_M \approx R_Q -2r + {\cal O}(\E_J^2) \ .
\ee
Thus the Josephson coupling decreases the amount of dissipation that
is needed to cause superconductivity.  This is the same underlying
physics as that that makes the control parameter in the strong
coupling limit $\half \j+\aeff$ which is the sum of a Josephson part,
and a dissipative part since $\aeff\propto 1/\sqrt{R^2+2rR}$. But, not
surprisingly given the basic electrical properties of the chain
discussed in Sec. \ref{ctm}, the combination of parameters that enter in the strong and weak coupling limits are quite different.  In terms of the superconducting stiffness parameter, $\j=2\pi \sqrt{E_J/E_C}$, the critical value of $K$ for the mixed transition is
\be
K_M\sim (R+2r-R_Q)^{\frac{1}{4}}
\ee
when this is small. 

The critical behavior  near the FSC-NOR transition  will be
characterized by the exponent $\nuz$  with the crossover energy scale
going to zero as $E_\times\sim |E_J-E_J^c|^{\nuz}$.  The critical exponent, $\nuz$   is a continuously varying function of the
resistances. For small $\delta$, the above analysis yields
\be
\nuz \approx \frac{1}{2\delta} \ .
\ee 
The rapid fall-off with distance, $s$, of  the fixed point values,  $\xi_s^*$, of the pair tunneling amplitudes suggests that the fixed point does not have a long length scale associated with it,  at least coming from the normal side.   Whether there is a more subtle diverging length scale --- e.g. that characterizes the decay with distance of the superconducting correlations, we leave for future investigation.

The important features of the weak-coupling phase diagram are
shown in Fig. \ref{wphases}.  
When $2r+R>R_Q$, the normal phase can exist; otherwise the chain will be fully superconducting.  But even for this high resistance regime, strong enough Josephson coupling can cause superconductivity. Along the phase boundary --- actually a critical hypersurface in the parameters of our model --- the critical behavior will vary continuously.

This analysis of the weak coupling regime thus almost completes the phase diagrams of Fig. \ref{wphases}.  The one exception is the nature of the bicritical point at which the weak coupling section of the mixed-transition phase boundary meets the small-intermediate coupling KT section. As mentioned in the introduction, this does not appear to be  amenable to perturbative analysis. A related question is whether, for intermediate coupling in terms of the QPS fugacities and pair tunneling amplitudes as well as $\j$, the weak coupling segment of the mixed transition boundary can meet the large-intermediate coupling dipole driven segment without an intervening KT section.  And, if so, do they come together smoothly or with another type of bicritical point. Again, this is not analyzable by our methods.

\section{Nanowires
  \label{ContinuumLimitSection}}

The discussion so far has concentrated on the phase diagram of a
discrete chain of shunted Josephson junctions. One of the motivations for
this work, however, came from the
observation of superconducting-normal transitions
in nanowires reported in Refs. 
\cite{Bezryadin,Tinkham,Lau,BollingerUP,Bezryadin05}
(for earlier experiments
see also \cite{GiordanoA,GiordanoB,GiordanoC,GiordanoD}). 
In this section we will show how one can describe
the superconductor-to-normal transition in superconducting nanowires by carefully taking the continuum
limit $\ax\rightarrow 0$ of the discrete model
discussed before. 
Thermally activated phase slips in superconducting wires have been 
considered
by Langer and Ambegaokar \cite{Langer1967} and
by McCumber and Halperin \cite{McCumber1979}. 
Quantum phase slips
in superconducting wires have been discussed
previously in Refs.
\onlinecite{GiordanoA,Saito1989,Chang1996,Duan1995,ZAIKIN1997,Golubev-Zaikin,Buechler}. 

\begin{figure}
\includegraphics[width=7.5cm]{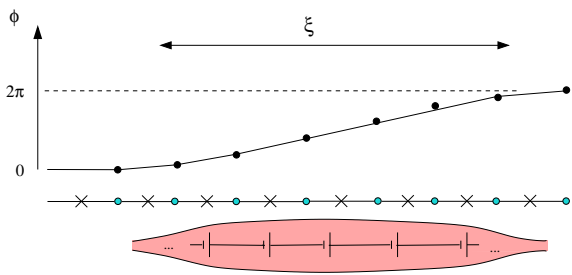}
\caption{Schematic of a QPS in a continuous wire. The size of
the core of a QPS --- in which the superconductivity is suppressed ---   is of order the superconducting coherence
length $\xi$. The wire can be  considered as the limit of a chain of grains whose separation, $\ax$,  is much smaller than $\xi$.
\label{fig17_1}}
\end{figure}

The fundamental difference between a continuous wire and a discrete JJ
chain is the existence of the lattice constant in the chain.
The interplay between the lattice constant, $\ax$, and the charge
relaxation length, $\xz$, played an important role in  JJ chains.  In a wire, the characteristic dissipative length becomes
\be 
\xz=\ax\sqrt{r/R}=\sqrt{\frac{\ax
r}{R/\ax}}=1/\sqrt{\gamma\rho} 
\ee 
where $\gamma=\frac{1}{r\ax}$ is the phenomenological parameter
describing the conversion conductance-per-length from the normal to
superconducting fluids, and $\rho=R/a$ is the resistance-per-length of the
wire. Nanowires are often modeled as discrete chains of Josephson
junctions with a lattice constant of the order of the superconducting
coherence length,  $\ax=\xi$. \cite{ZAIKIN1997} The motivation for this is that a typical
size of a QPS should be of the order of $\xi$. But from our discussion of JJ chains,  it is clear that another important length is the smallest possible distance
between QPS. In the case of JJ chains it is $\ax$, but for
continuous wires it may be arbitrarily small.  So the correct
description of the wire can be obtained from JJ models only by
carrying out the $\ax\rightarrow 0$ limit with the coherence length, $\xi$,
fixed (see Fig \ref{fig17_1}).

\subsection{Dipoles in nanowires}

Physically, the continuity of the wire and the limit $\ax\rightarrow
0$ allows the formation of dipoles of phase slips that are arbitrarily
close to each other --- up to charge discreteness effects which are discussed below.  The closer together the phase-slips are, the better
is their screening from the dissipative interaction, which is the only low energy
effect that suppresses the proliferation of dipoles in a discrete JJ
chain. This decreased dissipative interaction between them means that small dipoles  in a wire will always proliferate. At zero temperature, a wire will thus
always be either in the \scs phase or in the normal phase.

\begin{figure}
\includegraphics[width=7.5cm]{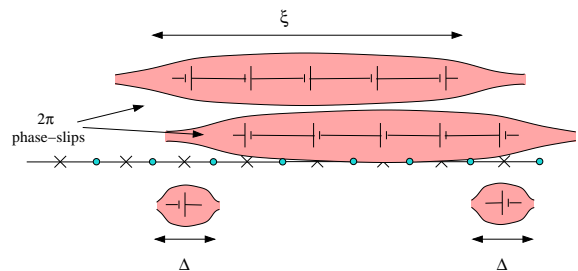}
\caption{Dipoles formed of a phase slip and anti phase slip that
are  a distance $\Delta\ll\xi$ apart. Their net effect is two partial phase slips of size $\Delta$, a distance $\xi$ apart.  \label{fig5}}
\end{figure}

This point can be made clearer by specifying how the continuum
limit of a wire is taken. One method of taking this
limit is spreading a phase slip over about $\xi/\ax\gg 1$ junctions, since a phase slip in a thin (diameter less than $\xi$) wire suppresses the
superconducting order parameter over a length $\xi$. This can be taken into account in the sine-Gordon action 
 by changing the cosine term in Eq. (\ref{action}):
\be
\ba{c}
\cos\l(\psi_{i}{(\tau)}+\theta_{i}{(\tau)}\r)\rightarrow\\
\cos \l[\ax\sum\limits_n h(\ax n)\l( \psi_{i+n}(\tau)+\theta_{i+n}(\tau)\r)\r]
\label{cont1}
\ea
\ee
where $h(x)$ is a weighting function concentrated in a region of  length of order $\xi$ with the constraint
\be
\int dx h(x) \approx\ax\summ_n h(\ax n) =1 
\ee
This constraint on $h$ ensures that the total phase involved  corresponds to one phase slip: it is forced by the global symmetry $\theta \rightarrow \theta+2\pi$. 
The UV cutoff will no longer be set by the shortest lengths, but be of order   $\Omega=\vms/\xi$ or the Ginzburg-Landau time, which is
related to the gap (in a way that depends on whether the superconductivity is  in the dirty or clean limit). Although the operator in
Eq. (\ref{cont1}) breaks a phase slip into parts that are
fractions of $2\pi$,  it does not lead to
independent fractional phase-slips; because of the $2\pi$ periodicity of the cosine, the fractional phase slips always
appear together as a part of a complete phase slip. The structure of a smeared
phase slip  is shown in Fig. \ref{fig17_1}.  The smearing 
means that the induced electrochemical  potential drop caused by the phase slip is split into fractional potential drops that occur over roughly 
$\xi/\ax$ junctions adding up to the quantized  drop of a single
phase slip.  The continuum limit $\ax\rightarrow 0$ can now be taken.

Because of the dissipation, an individual QPS in the wire will have action that diverges logarithmically at low temperatures, as in the chain.  The effective dissipation parameter is given by
\be
\aeff = \int dx\, h(x) \int dy\, h(y) \half \sqrt{\frac{\gamma}{\rho}} e^{-|x-y|/\xz}.
\ee
There are now two limits of interest. If $\xz\ll\xi$ (but still $\xz\gg \ax$ so that the discreteness  does not matter) the parameter is 
\be
\aeff \sim \frac{R_Q}{\rho\xi}
\ee
with  the coefficient depending on the form of $h(x)$ as should be expected: there is no unique way of defining the core size of a phase slip.  This result involves the normal resistance over a distance $\xi$: thus in the wire $\rho\xi$ (rather than $\rho \ax$) plays the role of $R$ in the JJ chain. But in general,  $\rho\xi$ is {\it not} the relevant resistance, in contrast to what would have been guessed from this analogy. 

If $\xz \gg \xi$, $\xi$ does not play a role in the dissipation and the behavior is similar to the JJ chain for $\xz\gg \ax$ so that
\be
\aeff \approx \half R_Q \sqrt{\frac{\gamma}{\rho}} 
\ee
analogous to $1/2\sqrt{rR}$ for the chain. In this limit, $\aeff$ will be of order unity when the normal resistance of a length $\xi$ of the wire is still small.  

By analogy with the JJ chain, one would expect that the condition for
breakdown of superconductivity would be controlled by the coefficient
of the total  interaction between QPS at the same position but
different times, $ \j +2\aeff$.  From the arguments of Sec. \ref{ssF},  one would expect the critical value of this coefficient to be three.   Indeed, there should be a change in the behavior near when this condition is met, but it will not be a true transition because of the role of dipoles. 

A pair of opposite sign phase slips separated by a distance $\Delta$ --- a dipole with moment $\Delta$ --- will have infinite action at zero temperature because of the dissipation.  But the 
dissipative interactions between the QPS that extend out to $\Delta$ of order the {\it greater} of $\xz$ and $\xi$ will reduce the coefficient, $\beff(\Delta)$  of the logarithmic action of the dipole. 

As in the chain, a pair of equal and opposite dipoles with the same location but separated in time  will have action 
\be
2\beff(\Delta) \ln (\tau/\at)
\ee
with the dissipative parameter a function of the resistances and the dipole moment, $\Delta$, relative to $\xi$ and $\xz$.   In the limit that $\Delta \ll  \min(\xi,\xz)$, for smooth weighting function $h$, the action can be expanded in powers of $\Delta$ yielding
\be
\beff \sim \frac{R_Q}{\rho \xi} \frac{\Delta^2}{[\max(\lambda,\xi)]^2}.
\ee
For $\xi\ll\xz$, there will a change in behavior from quadratic to linear dependence on $\Delta$ for $\Delta \sim \xi$: the larger moment behavior then being like the JJ chain with $\xz\ll\ax$ so that $\beff$ saturates to $\approx 2\aeff$  at $\Delta\sim\xz$. For $\xi\gg\xz$, $\beff$ will crossover from quadratic to saturating at $\Delta\sim\xi$.

Because $\beff$ decreases to zero as $\Delta \rightarrow 0$, independent of the dissipative parameters or of $\xi$, for sufficiently small $\Delta$, $\beff$ will be less than unity and thus dipoles will proliferate --- as long as very small dipoles can be modeled as we have done here.  The characteristic moment $\Delta$ below which dipole proliferation occurs is 
\be
\Delta_\times \sim \max(\xz,\xi)\sqrt{\rho\xi/R_Q}
\ee
provided this is less than $\xi$, otherwise a different dependence obtains.
Nevertheless, we see that independent of the relative magnitudes of $\xi$ and $\xz$, dipoles of moment $\xi$ will {\it not} proliferate  as long as  $1/\sqrt{\rho/\gamma} < C_h R_Q$ with $C_h$ an order-unity coefficient that depends on details of the structure on scale $\xi$.  

If dipoles of size of order $\xi$ can proliferate, the behavior will be similar to the JJ chain when $s=1$ dipoles, the smallest size, proliferate. The dipole proliferation will, at low energies, lead to screening of the dissipative interactions between individual QPS and thus to either the \scs phase or the normal phase depending on whether the renormalized $K$ is greater than or less than four. 

But if only dipoles of size $\Delta\ll \xi$ can proliferate on their own, the behavior will be quantitatively different.  Although it is not clear what pairs of QPS whose separation is much less than $\xi$ represent, they should correspond to large amplitude localized fluctuations of the superconductivity that act to decouple the superconducting from the normal degrees of freedom.   The effects of these will be to renormalize the dissipative interactions between larger moment dipoles and individual QPS by decreasing the rate of the processes, parametrized by $\gamma$, that relax imbalances between the  superconducting 
and normal voltages.   Again, if the dipole moments can be arbitrarily small, this will eventually lead to either the \scs or the normal phase.  But if the bare rate of the small-dipole-like processes, $\eta(\Delta)$, is small, there should be interesting crossovers as temperature is lowered associated with the growth at low energies of these dipole fugacities before they fully screen the dissipative interactions.
 
In principle, even in a chain of grains linked by Josephson junctions
QPSs need not take place only across the junctions but could happen
inside a grain --- albeit with much higher action and hence
exponentially lower fugacity.  But in the limit of extremely low
temperatures, such processes need to be considered and the chain will be more like a nanowire.  In particular,
the effects of QPS dipoles with very small size within one grain will,
by the arguments above, be relevant.  Eventually,
these will screen the superconducting fluctuations from the
dissipative interactions and drive the system into the \scs or normal
phases: again, if arbitrarily small dipoles can occur,  the FSC phase will not exist in the limit of
asymptotically low temperatures.

\subsection{Minimum size QPS dipoles?}

In all the above, we have assumed that arbitrarily small QPS dipoles can exist: in the continuum model we have used, even with short distance cutoffs, there seems to be no reason these cannot occur. But as the QPS are quantum objects, we must be careful of the possibility of complex --- rather than purely real --- actions.

In a QPS dipole, the superconducting phase, $\phi(x)$,  {\it
  inside} the dipole winds by $2\pi$ as $\tau$ goes from $-\infty$ to
$+\infty$ (see Sec. \ref{sgsec}).  As long as $\phi(x, \tau)$ is periodic in $\tau \in (0,\beta)$, a QPS dipole of the opposite sign is needed to satisfy this condition. But in reality, the true condition is that $\phi$ is periodic in $0$ to $\beta$ modulo $2\pi$. In the limit of zero temperature, this will have no effect in the action we have studied.  But in general, there is an additional term in the action, analogous to a Berry phase,  of the form:
\be
S_w= i \int dx \int_0^\beta d\tau\,  \tilde{n}_S \frac{\partial \phi}{\partial\tau} \ , 
\ee 
with $\tilde{n}_S$ having units of number density. In a Galilean invariant system, at zero temperature $\tilde{n}_S$  would be the density of Cooper pairs, equal to half the electron number density.  But more generally, it can have any value, positive, negative, or zero:  $\tilde{n}_S$  is some effective density for the superfluid that has no known simple interpretation.  It will depend on particle-hole asymmetry and other factors that are not
understood \cite{geshkenbein1,geshkenbein2}. For vortices in a 2d film, $\tilde{n}_S$ corresponds to a
  dual magnetic field, and it gives rise to a vortex hall effect \cite{AoThouless}.
[Note that in Ginzburg-Landau theory, only the combination $n_S/m^*$ is physical: $n_S$, in contrast to the phase stiffness, $\Upsilon$ --- often called $\rho_S$ --- has no physical meaning. Quantum mechanically, this is no longer true.]

The action of a dipole of moment $\Delta$ has an imaginary part of its action from the Berry phase term: 
\be
S_w^{dipole}=2\pi i \tilde{n}_S \Delta \ .
\ee
This suggests that dipoles for which $\tilde{n}_S\Delta$ is not an
integer will undergo destructive interference implying that $\Delta$
should be quantized in units of  $1/\tilde{n}_S$.  The QPS dipoles can
be considered as a vortex which first crosses the wire in one
dircetion, and then returns and cuts the wire in the opposite
direction at a distance $\Delta$ away from the first crossing, before
returning to its origin.  Interpreting
$\tilde{n}_S$  as Cooper pair density, the quantization of $\Delta$ means that the vortex had to ``go around" integer numbers of Cooper pairs\cite{FisherLee}.

If there is indeed a minimum size QPS dipole
with 
\be
\Delta_{min}=\frac{1}{\tilde{n}_S} \ ,
\ee
and the low frequency coupling to the ``normal" degrees of freedom is still effectively dissipative for such small dipoles, then it is possible that the dissipation could be sufficiently strong to suppress these smallest dipoles. [Although, the effective coupling, $\beta(\Delta_{min})$, would be determined by the shorter length scale physics which will not be well parametrized by $\gamma$ and $\rho$.]  If the smallest dipoles are indeed suppressed, then an FSC phase --- and a transition from it to an \scs or a normal phase --- would be possible in nanowires. 

 It is worth noting that  effects of a small local disturbance {\it
   can} be  long range in time and  give rise to actions that diverge
 logarithmically as  temperature is lowered.  An example is  an
 impurity in a metal that is moved by a small amount: the divergent
 action is associated with the ``orthogonality catastrophe" caused by
 the changes in the electron wave-functions induced by the altered
 potential. \cite{hamann} Better known, but closely related, is the
   X-ray edge singularity \cite{nozieres}.  Thus it is not 
unreasonable to think that a local change in the superconducting degrees of freedom could result in a logarithmically infinite action associated with the response of the quasiparticles and other low energy degrees of freedom to the change.

 Whether or not the FSC phase  can occur, even in principle,  we leave as an intriguing open question.

\section{Resistivity \label{res}}
\label{ResistivitySection}

In the previous sections we concentrated on the zero temperature
phase diagram of the RSJJ chain shown in Fig. \ref{Fig1intro}(c).
We now extend this analysis to discuss the temperature
dependence of the resistivity of the system using
the RG flow equations obtained in Section \ref{sec3}.
One of the most surprising results is that in the normal phase close to the superconductor-to-normal transition
the system exhibits a non-monotonic $R(T)$ dependence.
In this regime one finds that the resistivity
first decreases with decreasing temperature,
then saturates and stays nearly constant at a value $R(T)\ll R$,
i.e., much smaller than the normal resistance of the chain, down to very
low temperatures, and only then finally starts
to increase. During this long crossover, the Josephson junctions
themselves form a resistive channel, parallel to the shunt resistors. Since experiments have a lower limit of  temperature, such behavior may appear to support the existence
of the ``metallic phase'' \cite{Kapitulnik2000}
separating the superconducting and insulating ones.

We start by showing how one can obtain the resistance vs. temperature
curves from the RG flow for the QPS fugacity. We then proceed to
discuss the finite temperature resistivity in the FSC and \scs
regions, and explain the origin of the quasi-metallic behavior
in some of the NOR region.

\begin{figure}
\includegraphics[width=7.5cm]{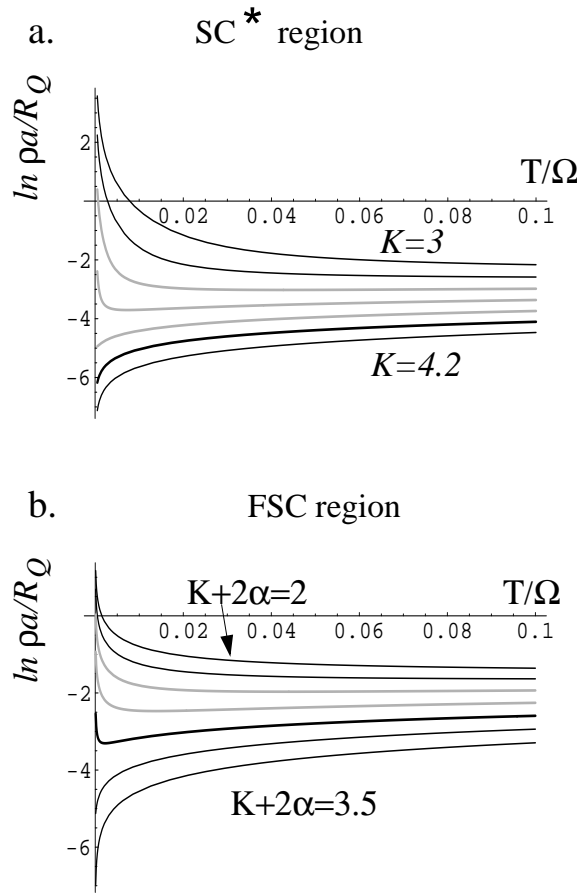}
\caption{Semi-log plot of the resistance vs. temperature for various values of $\j$. 
(a) Resistivity in the \scs phase. $\j=3,\,3.2,\,3.4,\,3.6,\,3.8,\,4,\,4.2$.
(b) Resistivity in the FSC phase. $\j+2\aeff=2,\,2.25,\,2.5,\,2.75,\,3,\,3.25,\,3.5$.
In both plots the resistivity at the bare critical value of the
stiffness appears as a thick black line ($\j=4$, and $\j+2\alpha=3$). The non-zero phase-slip
fugacity shifts these lines slightly from criticality.
Resistivities in the non-monotonic pseudo-metallic region are shown as
thick gray lines.  
Temperature is normalized by the plasma frequency of an individual
junction. The
contribution of the parallel channel of normal electrons
to the total conductivity has been subtracted.
We use the approximation
  $\zeta\sim e^{-\sqrt{8}\j/\pi}$ as initial condition, which
corresponds to the bare action of the QPS $S_0=8\sqrt{E_J/2E_C}$ 
\cite{likharev,paalanen}. 
 \label{Fig5}}
\end{figure}

\subsection{Resistance of a single junction \label{1J}}

The RG analysis presented in this paper allows one to calculate the scaling
behavior --- including crossovers --- of the temperature dependence of the resistance of a JJ chain. The simplest way to do this  is to carry out
the RG flow until the UV cutoff becomes of order the temperature. 
At lower energy scales  the renormalized phase slips are only weakly
interacting and their effects independent. 

To obtain physical quantities, we must distinguish between the ``bare" parameters --- those that appear in the action with the initial high energy cutoff --- and the renormalized parameters.  In situations in which there might be confusion, we denote the renormalized parameters with  subscripts, $\ell$, corresponding to energy scale $\Omega=\Omega_0 e^{-\ell}$. 

We first consider  a single resistively shunted
Josephson junction and estimate the (superconducting) voltage fluctuations across it at a low temperature, $T$.  At low frequencies, the resistance of the superconducting part controls the low frequency fluctuations via 
\be 
\langle V(\omega)V(\omega') \rangle = R T \cdot 2 \pi \delta(\omega+\omega') \ .
\label{VVR}
\ee
  The voltage fluctuations caused by QPS are tricky to estimate because of the interactions between them. But if we renormalize until an energy scale at which the QPS are weakly interacting, $T(\ell) \sim \Omega$, then the fluctuations can be estimated simply. 

The phase-slip fugacity $\zeta_\ell$ and the temperature of a single RSJJ obey the flow equations \cite{Fisher-zwerger}:
\be
\ba{c}
\frac{d\zeta_\ell}{dl}=\zeta_\ell\l(1-\frac{R_Q}{R}\r),\vspace{2mm}\\
\frac{dT_\ell}{dl}=T_\ell
\label{r1}
\ea
\ee
with the high energy cutoff, $\Omega_0$, initially of order the  plasma frequency of the circuit. When the RG flow reaches the stage in which
$T_l\sim\Omega$, only one pair of  phase slips (they must come in pairs because of the periodic boundary conditions in $\tau$) is likely in the
rescaled imaginary time duration $1/T_\ell$ and these are, in any case, weakly interacting. At this scale, the renormalized fugacity is 
\be
 \zeta_T=\zeta(\ell=\ln[\Omega_0/T]) \ .
\ee
The phase slip pairs will then occur with probability of order $\zeta_T^2/\Omega^2$ in time of order $\hbar/T_\ell \sim \Omega$. A phase slip contributes to $V(\omega=0)$ --- a quantity that is scale independent --- an amount $h/2e$. The rescaling of $T$ and of $\delta(\omega+\omega')$ in Eq. (\ref{VVR}) cancel, and the renormalization of $\zeta$ to $\zeta_T$ then yields
\be
R_{JJ}\sim \frac{h}{(2e)^2}\l(\zeta_T/\Omega\r)^2\sim R_Q\l(\zeta\Omega_0\r)^2\ \l(\frac{\Omega_0}{T}\r)^{2-2\aeff} \ .
\label{r4}
\ee
The resistance increases with decreasing temperature in the
normal phase which obtains when $R_Q/R<1$, and goes to zero at zero
temperature in the superconducting phase which obtains when
$R_Q/R>1$. 
This result, Eq. (\ref{r4}), is valid only for 
$T<\Omega_0$.  For higher temperatures, 
  $T>\Omega_0$ activated scaling of the form
 $R_{JJ}\sim R_Q \exp\l(-C_JE_J/T\r)$ with $C_J={\cal O}(1)$ obtains.  How these match together can be understood in terms of the fugacity $\zeta$. This is of order $\Omega_0 \exp(s_0/\hbar)$ with $s_0$ the action of a phase slip $s_0 \sim E_J\tau_0\sim E_J/\Omega_0$ since the time scale of a QPS is $\tau_0$. If the imaginary time $\hbar/T$ is smaller than $\tau_0$, the action of the phase slip will be reduced by a factor of order $\hbar/T\tau_0$. This gives rise to the activated resistance in terms of $E_J/T$.
   
The measured resistance of the shunted junction is
\be
R_m^{-1}\approx \l(R_{JJ}\r)^{-1}+R^{-1}.
\label{e9}
\ee 
If in Eq. (\ref{r4}), $\zeta\sim \Omega_0\frac{R}{2\pi R_Q}$ or larger, the Josephson junction becomes essentially insulating, and
most of  the current flows through the shunt resistor,
$R$. 

Note that at any non-zero current, some Cooper-pair tunneling will occur and non-linear resistance $V/I$, will no longer be given by  Eq. (\ref{e9}). But these 
  corrections vanish in the limit of zero-temperature and zero-current.

\subsection{Resistivity in the FSC region \label{RFSC}}

We now turn to the JJ chain. In the FSC phase, each junction behaves
similarly to a single junction in its superconducting phase and thus
contributes roughly independently to the total resistance. Phase slip
dipoles, with fugacities $\etas$, are the explicit manifestation of
interaction between phase slips on different junctions. Thus when all
the $\etas$   are irrelevant about the FSC fixed line,  the phase
slips in each junction will be almost independent --- although their
dissipation involves overlapping resistors. Thus each Josephson
junction will act like the a single junction with  resistance given by
Eq. (\ref{e9}), with the appropriate $\zeta_T$ --- the renormalized
fugacity at scale $T\sim\Omega$  --- of individual QPS in the chain:
\be
 R_{JJ}
\sim \frac{h}{(2e)^2}\at^2\zeta_T^2.
\label{r5}
\ee
 At intermediate temperatures (what we mean by
intermediate is discussed in Sec. \ref{ssF}), Eq. (\ref{r5})  predicts a measured resistance per unit length,
\be
\rho_m \approx R_{JJ}/\ax \sim T^\Gamma
\ee
with the (positive) exponent $\Gamma$ depending on the normal resistances and the superconducting stiffness, $\j$.

In the FSC phase, the exponent $\Gamma$ is given by minus twice  the eigenvalue $\zeta$   [from Eqs. (\ref{zflow1})]
\be
\Gamma = K+2\aeff-2 \ .
\ee
Since the boundary between the FSC and normal phase when the $\etas$ are irrelevant --- the mixed transition --- is given by $K_M=3-2\aeff$, we see that  on the critical line, 
\be
\rho_m^M \sim \frac{R_Q T}{\ax \hbar \Omega_0}
\ee
in contrast to the single junction critical point at which the measure resistance will be roughly temperature independent. 
The origin of the factor of $T$  can be seen from the transformation that related the 
flow
of $\zeta_\ell$  to the flow of the parameters $z \propto \zeta_\ell \exp(-\ell/2)$ and $x=\frac{\j}{2}+\aeff-\frac{3}{2}$
which obey the Kosterlitz Thouless flows Eqs. (\ref{ktc}). When $\zeta$ is small, $x=0$ --- at which $z$ is weakly scale dependent --- marks the FSC-NOR transition. 
The scale dependence of the relationship between $z$ and $\zeta_\ell$  transforms Eq. (\ref{r5}) to 
$R_{JJ}\sim R_Q
\at^2\zeta_T^2\sim R_Q \frac{T}{\Omega}z_T^2$ giving rise to the factor of $T$ in terms 
 the ``natural" variable $z$.  As we shall see, this really is natural for the KT-like global transition from the \scs phase to the normal phase.
  
The behavior of $R_{JJ}(T)$ near the FSC-NOR transition is shown in
Fig. \ref{Fig5}(b). The results in  Eqs. (\ref{zflow1}-\ref{zflow2}) lead to a regime of parameters in which a crossover between quasi-superconducting and normal  behavior
takes place: in particular, the resistance exhibits a minimum at a low temperature.  This behavior arises on the normal side of the transition, but still close to it: when
 \be
1<\frac{\j}{2}+\aeff<\frac{3}{2},
\label{e14}
\ee
so that $\zeta_\ell$ initially decreases with scale even though $z$ is increasing. If the bare fugacity is small, this gives rise to an intermediate temperature regime with a power law decreasing resistance with exponent $0<\Gamma<1$.  But at low enough energies, the rapid increase of $z$ will give rise to proliferation of dipoles and individual QPS and the $\zeta_\ell$ will start to grow.  The junction then crosses over to insulating behavior as the temperature is lowered further. 
Near the transition to the FSC phase, the crossover temperature, $T_{min}$, at which the resistive minimum occurs becomes very low: 
\be
T_{min}\sim \Omega_0 \exp\l(-b/ \sqrt{K_M-K}\r)
\label{ktco}
\ee
with $b$ a coefficient proportional to the bare QPS fugacity, $\zeta$, when this is small. 
Below the crossover temperature, the measured resistivity, $\rho_m$,  of the chain will increase as the temperature
is lowered, until it reaches the resistance of the shunt resistors, $R/\ax$.

\subsection{Resistivity in the \scs region \label{rscssec}}

The low temperature behavior in the \scs phase can be understood similarly.  When the renormalized dipole fugacities are  large, dissipation will be screened and we need only consider individual QPS and the interactions between them mediated by the superconducting degrees of freedom. 

It is now simplest to use the isotropic RG and rescale space and time. 
 When the RG flow reaches
the stage where $T(\ell)=\Omega$, the spatial size of a QPS is
$\Omega_0/T$ times its original size. Therefore, when resistances per unit length are found from renormalized quantities, a factor of the length rescaling is needed to obtain the physical resistivities.   The resistivity of the Josephson junctions, will thus be 
\be
\rho_{JJ} \sim \frac{T}{\Omega_0} \frac{R_Q}{\ax} (\zeta_T/\Omega)^2 \sim \frac{R_Q}{\ax} (\zeta/\Omega)^2\l(\frac{\Omega_0}{T}\r)^\Gamma
 \label{e10}
\ee 
with 
\be
\Gamma= 1+(K-4)
\ee
the $K-4$ being minus twice the eigenvalue of $\zeta$ in the isotropic RG. We have assumed that the basic length scale that enters is the grain spacing, $\ax$, and used the bare QPS fugacity $\zeta$: more generally these will depend on the higher energy processes as discussed below.

The resistivity as a function of temperature is plotted for
several values of $\j$ in Fig. \ref{Fig5}(a)
Since for $\j<K_G=4$ the chain will be normal at low temperatures and the resistivity increase to $R/\ax$, for the intermediate range 
 \be
3<\j<4
\label{e11.5}
\ee 
the resistivity will decrease with decreasing $T$ and have a minimum at a crossover temperature $T_{min}$ given by the energy scale at which the QPS proliferate enough to screen the longer-range interactions between them driving the junctions normal. 
This range is analogous to the intermediate regime of 
Eq. (\ref{e14}) near the FSC phase: In both cases the resistivity seems
to show signs of superconductivity before exhibiting insulating
behavior. For $\j\approx 3$ the flat resistivity curves over a substantial temperature range could be confused for a 
quasi-metallic phase.

\subsection{Roles of dipoles \label{RDip}}

In the JJ chain, if the QPS fugacities are small and the dissipative
interactions substantial, the QPS fugacity can be irrelevant, but the
dipoles may be relevant about the FSC fixed line. In this case,  the \scs
phase behavior  can obtain only at low enough energies below which the QPS
dipoles proliferate: this makes the temperature dependence of the
resistivity much richer than that discussed so far. 
If the \scs phase is not stable, the resistivity near  the dipole
driven FSC-NOR transition, will involve a crossover from dipole to
individual QPS dominated resistivity.

The temperature behavior of the resistivity in both these cases splits into two
regimes. Since the $\etas$ are zero initially, and get generated at order $\zeta^2$, at temperatures in the range $T_D<T<\Omega_0$ we expect the JJ chain
to behave as though it were in the FSC region. The crossover  temperature $T_D$, away from the FSC regime is the
energy scale at which $\eta_1$ becomes of order $\Omega$: at lower energies, the dissipative interactions get screened (see discussion above
Eq. (\ref{largezs})). 
At energies higher than $T_D$, the scaling is similar to that of individual junctions, while at lower temperatures, it will become isotropic as in the \scs regime discussed above, and exhibit normal behavior at sufficiently low energies if $\j$ is too small to stabilize the \scs phase.

We now estimate  $T_D$, and discuss its consequences for the low
temperature behavior.  The flows of $\eta_1$ when it is relevant can
be well approximated by an initial rapid regime in which $\eta_1$
becomes of order $\zeta^2/\Omega_0$, followed by a scaling regime in
which it grows with exponent $1-\beff$ with $\beff=2\aeff(1-w)$ as in
Sec. \ref{revisit}. Thus it will become of order $\Omega$ at an energy scale
\be
T_D \sim \Omega_0  \l(\frac{\Omega_0}{\zeta}\r)^{2/(1-\beff)}
\ee
for $\beff$ below its critical value of unity for either the FSC-\scs or the FSC-NOR dipole-driven transitions.  This is valid as long as $\zeta_\ell^2$ grows less rapidly than $\eta_1$: i.e. if $2-2\aeff-K<1-\beff$. This is always the case if $K>1$ and sometimes also at smaller $K$  if $w$ is substantial.   If $\zeta_\ell^2$ grows more rapidly than $\exp[(1-\beff)\ell]$, then  $\eta_1$ follows $\zeta_\ell^2$ and individual QPS will drive the crossover to normal behavior.

For $T>T_D$, the individual-junction controlled behavior found above for the resistivity in the  FSC phase (and for crossover away from FSC) will obtain: 
\be
R\sim T^{\j+2\aeff-2} \ .
\ee
But below $T_D$ the temperature dependence of the resistivity will be like that in the \scs phase (or the flow away from that)  with $\rho_m\sim T^{K-3}$.

As  before, if $\j<4$, at low enough temperatures the chain will crossover to normal behavior and the resistivity saturate at $R/\ax$, while if $\j>4$ the resistivity  will  go to zero more rapidly than linearly.

We have seen in this section that the interplay between local and long-length scale physics gives rise to interesting behavior of the resistivities.  Perhaps surprisingly, the temperature dependence near the critical point of the QPS driven mixed transition, for which the critical value of the parameter $\half \jx +\aeff$ is $\frac{3}{2}$, midway between the values of $1$ and $2$ expected for local and global transitions, nevertheless is similar to that of the global transition with $\rho_m \sim T$, with the $T$ arising from spatial rescaling. 

Unfortunately, the form of the various crossovers of the resistivity as a function of temperature, which would yield firmer testable predictions, are not readily  accessible to our analysis
since these involve analyzing a sine-Gordon model in the intermediate
coupling-strength regime. In addition, we have been rather cavalier about which resistivity we are considering: in general, even in simple scaling regimes, the resistivity at $\omega \sim T$ and the DC resistivity can be quite different.  The latter certainly requires proper real-time analysis.

\section{Discussion \label{discuss}}

In this final section we discuss some implications of the results presented in this paper, and issues associated with these.

\begin{figure}
\includegraphics[width=8cm]{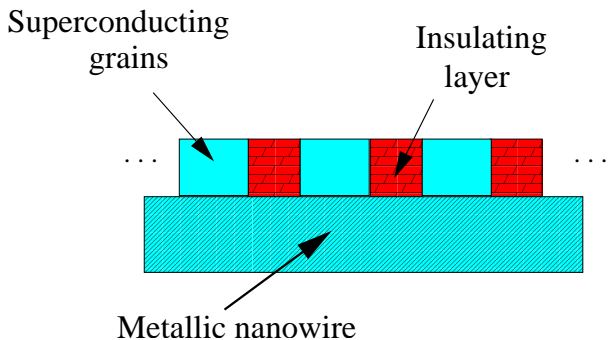}
\caption{Possible realization of a two-fluid JJ
  chain: mesoscopic superconducting
  grains deposited on top of a metallic nanowire and separated by  a thin
  insulating layer, to produce an SIS junction. \label{fig18}}
\end{figure}

{\bf Chains of Mesoscopic Josephson Junctions}.  Several groups have
recently reported experimental realizations of mesoscopic
superconducting grains \cite{ralph} and  arrays of low-capacitance
Josephson junctions 
\cite{HavilandA,HavilandB,HavilandC,HavilandD,HavilandE,paalanen,Haviland2003,W-H2003}. 
The possibility of
introducing ohmic dissipation using shunt resistors has been
demonstrated for a single quantum Josephson junction by Pentill\"a {\it et al.}
\cite{paalanen}, and Watanabe and Haviland \cite{W-H2003}, and for arrays of Josephson junctions by Takahide {\it
et al.} \cite{japan} and Miyazaki {\it et al.}
\cite{Miyazaki2002}.   A potential realization of a chain of mesoscopic
Josephson junctions with ohmic dissipation is shown in Fig. \ref{fig18}. It consists of equally
spaced small superconducting grains on top of a metallic
nanowire with thin insulating layers separating the two 
materials. This system is a realization of the model that we presented in
Sec. \ref{sec2}, with the possibility of controlling separately the parameters
$E_J$, $E_C$, $R$, and $r$ by changing the width of the metallic
wire, the size and separation  of the superconducting grains,
and the strength of the tunnel barriers between the grains and between
the grains and normal wire. 
A schematic phase diagram of this system is
presented in Figs. \ref{iphases}, and
\ref{wphases} (we did not consider odd-even effects in
the grains, which must be irrelevant if $r<R_Q$, as discussed below). Both the phase diagram and the character of the
transitions are affected by the superconducting-normal relaxation
parametrized by $r$. \footnote{For the clean \N wire and perfect \S-\N contact the SC-normal
transition may arise due to a different physical mechanism, in which
superconducting condensation energy on the grains competes with the
energy cost of inducing pairing correlations in the \N wire, see
e.g. Refs. \onlinecite{Spivak,Feigelman1998}.}.  

If the metallic wire is highly resistive  and/or the tunnel
barrier between the \S (superconducting) and \N  (normal) parts weak, the dissipation will not be important, the FSC phase will not occur,  and the superconductor-to-normal transition will be determined primarily by the competition between
Josephson coupling and Coulomb charging energy of the grains. This will be a quantum Kosterlitz-Thouless transition, isotropic in space and time, with the energy scale decreasing exponentially as the transition is approached.

If the wire is less resistive and/or the SN tunnel junction stronger, the FSC phase can occur. The FSC to normal phase
boundary and the FSC to \scs phase boundary will be dominated by dissipative effects. These change, in particular,  the  nature of the superconductor-to-normal transition which will have a mixed character with both local and global physics involved.  This mixed transition can either be KT-like --- although modified from the conventional KT transition --- or have strongly anisotropic  power law scaling with continuously variable exponents.
                                                                                                                                                                           
The renormalization group analysis presented in this paper allows us to
obtain the scaling form of the chain's resistivity as a function of temperature in the vicinity of the various 
superconductor-to-normal transitions; the results are presented in Fig. \ref{Fig5}.  Near the
\scs to normal phase boundary, our results agree with the results in
Refs. \cite{ZAIKIN1997,Duan1995}, but 
the analysis of $R(T)$ in the vicinity of the FSC-to-normal phase
boundary is new. We find that in both cases the system has extended
crossover regions on the normal side of the transition, in which superconducting tendencies give rise to a weakly  temperature dependent resistivity that decreases as  a small power of $T$  over a wide temperature range before turning up at the lowest temperatures as it becomes asymptotically normal.  Near the superconductor-to-normal phase boundary this upturn in
$R(T)$ occurs at  temperatures
that are are either power-law or exponentially small in the inverse of the deviation from the zero-temperature transition, depending on the regime.

Our RG approach to the problem is also new, and allowed us to expose
a myriad of crossover effects occurring near phase
boundaries. Particularly interesting is the interplay between global
and local mechanisms for the breakdown of superconductivity. The mixed
FSC - normal transition, for example, can arise from the local resistive
environment, or from a combination of the long-wavelength plasmon
dissipation and the local resistive parts. The phase
boundaries for which  these two
mechanisms obtain meet at a bicritical point, which has a rich crossover
behavior yet to be analyzed in full. The \scs - normal transition is
shown to be global, i.e., determined by the long-wavelength excitations, and thus, for an infinite chain, the  low energy critical behavior is 
almost independent of the resistive shunting or the
two-fluid nature of the grains, except for their effects in modifying the effective capacitances and Josephson couplings.  

Other aspects of the phases and phase transitions could also be
investigated for a chain of grains. Of particular interest would
be tunneling into pairs of grains, studying the behavior as a
function of their separation and of temperature, and how these change
near phase transitions.

\begin{figure}
\includegraphics[width=8cm]{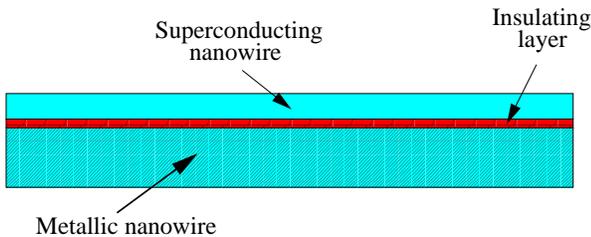}
\caption{Schematic of a type of superconducting nanowire with dissipative
  degrees of freedom: a superconducting layer is deposited on top of a
  metallic nanowire with an insulating layer in between. 
\label{fig19}}
\end{figure}                                                                                                                                                                        
                                                                                                                                                                            
{\bf Superconducting Nanowires}. Various types of
superconducting nanowires can have dissipation down to zero temperature.
The simplest is perhaps a wire with regions that remain
normal, so that it has a positive density of states at zero energy. 
Another possibility is a sandwich of superconducting and
metallic nanowires separated by a thin insulating  barrier (see Fig.
\ref{fig19}). But  even nanowires made of a conventional gapped
superconductor may have low energy degrees of freedom that give rise
to dissipation  when they are close to a superconductor-to-normal
transition. The conventional BCS theory of superconductivity, which
leads to an energy gap, is intrinsically based on a mean-field
analysis and thus unlikely to be reliable in the vicinity of a quantum
phase transition when virtual quantum phase slips occur. A roughly
analogous effect is seen in superfluid $^{4}$He in a disordered porous
medium (e.g. Vycor): an apparent normal fluid component has been
observed down to the lowest temperatures \cite{Putterman}.  In general, one would expect dissipation of some
sort --- but not necessarily ohmic --- near superconductor-to-normal
quantum transitions as long as some of the Cooper-pairs break apart near the
transition. This may not occur if the Cooper pairs form tightly bound
bosons that still exist in the non-superconducting phase, but more
generally should occur.  Furthermore, if the normal phase is a Bose or
Fermi glass, there will be a constant density of states of low energy
excitations and concomitant dissipation arising from 
local two level systems. That said, the approach we have taken 
 in this paper is phenomenological:  we assume that dissipation
is present in at least some superconducting nanowires and then study its
consequences.

For superconducting nanowires,  our  analysis of the simplest model
suggests that  nanowires will always be either in the \scs or the
normal phase with the FSC phase unstable to small QPS dipoles. The
superconductor-normal transition would then generally be of global KT
character. But if the fugacity of phase slips is low, and the
dissipative effects large, there will be interesting crossovers in the
temperature and length-scale dependent properties in both the
superconducting phase, and near the transition.

Nevertheless,  as discussed in the Sec. \ref{ContinuumLimitSection},  these
conclusions may not be correct. In particular, charge discreteness
effects  can lead to interference between QPS dipoles which would set
a minimum size for dipoles in the limit of zero temperature. If the
dissipation is strong enough --- a condition that will depend on
short-length scale physics in addition to the mesoscopic resistive
parameters --- then it is possible that the FSC could be stabilized.
These effects clearly need further exploration.

{\bf Finite length wires}. In the \scs phase at zero temperature, the only low energy effects of
dissipation in our nanowire model arise from the global $k=0$ mode and
thus depend solely on the total resistance of the wire. In finite
wires this would mean that at very low temperatures, when the wire
acts like a single Josephson junction, its {\it total} resistance will
determine whether it is superconducting or resistive.  At higher
temperatures, $T>\vms/L$, the behavior of the wire could exhibit a K-T
crossover which is tuned by the stiffness, or thickness, of the wire.

We note, however, that in finite wires the resistivity may determine
other parameters that are important for the superconductor-to-normal
transition, such as the fugacity of the QPS \cite{Tinkham} and the
strength of quantum fluctuations \cite{Golubev-Zaikin}.  So the
superconductor-to-normal transition in finite wires is determined by
both the resistance and the resistivity. A more
detailed discussion will be given elsewhere \cite{finitechain}. We
should note that other effects may also be important for the case of
superconducting nanowires, such as the suppressed $T_c$ due to Coulomb
interactions as in Refs. \onlinecite{yuval1,yuval2}.

The picture arising from our analysis conforms with and enhances that in
Ref. \onlinecite{Buechler}. There it is assumed that a finite nanowire
is shunted externally by a resistor. This resistor gives rise to a 
zero-temperature SC-NOR transition, but the plasmon degrees of freedom
in the wire give rise to sharp crossovers. Our picture puts the
picture of B\"uchler et al. \cite{Buechler} on firm microscopic
footing by showing that, quite generically, the resistivity of the
wire needs to be taken into account as a single resistor that shunts
the entire length of the wire.

{\bf Experiments on nanowires}. The hypothesis that dissipation plays an important role in
superconducting nanowires is supported by the recent experiments of
Bezryadin et al. \cite{Bezryadin,Tinkham,Lau}, 
in which  the superconductor-to-normal transition was
observed to depend on the normal state resistance of the wire. Later
experiments \cite{Tinkham} on longer wires showed a transition that
depended on the {\it resistivity} of the wire, rather than its total
resistance. Bollinger et al. \cite{BollingerUP} recently carried an
extensive study of shorter wires ($\sim 200nm$), and observed results
consistent with both previous experiments. At this point it is still
not clear whether the zero-temperature SC-metal transition in
finite-length amorphous nanowires depends on their total resistance,
or primarily on the resistivity.

{\bf Charge discreteness effects}. Before turning to broader implications, we briefly raise various
issues.  The most serious caveat to our results arise from aspects
of charge discreteness that we have ignored. In the  formalism we use,
the charge discreteness not only gives the basic periodicity  of  the
superconducting phase, $\phi$, of the Josephson coupling, but also
implies periodic boundary conditions modulo $2\pi$  in the imaginary time
path integral on $(0,\beta)$, e.g., in Eq. (\ref{allaction}).  We have ignored the effects of non-trivial windings of the phase and  thus the linear $\frac{d\phi}{d\tau}$  term that only matters when $\phi(\tau+\beta)\neq\phi(\tau)$. These can cause Berry-phase-like interference effects among QPS that we have ignored. 

For single Josephson junctions, an approach to dissipation
that respects charge quantization is to explicitly analyze 
quasiparticle tunneling as carried out, for example, in
Ref. \onlinecite{Schoen-Zaikin}. It is found that
the quasiparticle tunneling has very similar effect to Ohmic
dissipation when the (continuum) electrostatic
equilibrium corresponds to an integer number of Cooper pairs on each
electrode. The generalization of this to JJ chains in the absence of
dissipation has been analyzed in Ref. \onlinecite{FWGF} and others
\cite{AKPR}: the simple model we use is only directly applicable when there is an {\it integer number} of Cooper pairs per grain (or other commensurate filling for which  the grains can act in groups). At incommensurate 
filling factors the Cooper pairs are always delocalized in the absence of randomness. Hence, naively, one would expect
that if the ohmic shunt resistors are replaced by 
quasiparticle tunneling and motion of discrete electrons, and the conversion resistance by quasiparticle creation and annihilation processes caused by the dynamics of the superconducting phase, $\phi$, the conclusions of our paper would only be valid for integer number of Cooper pairs per grain. 
This additional effect of charge discreteness, which would  severely
restrict the applicability of our results, is most likely to be
problematic when the shunt and conversion resistances are high:
this will increase phase fluctuations and result, in any case, in
charging effects destroying the superconductivity as we have found.
When the resistances are low --- the regime in which the FSC phase is
predicted --- it is less clear what the effects of charge discreteness
are as the number of Cooper pairs on a grain is no longer a good
quantum number.  We leave for future work this issue, as well as the question of
whether near to, but on the normal side of
transitions, there is an intermediate --- or fuller --- temperature range
over which our results apply.

 The case of effectively commensurate filling, corresponding to the coefficient of the $\partial/\partial\tau$,  $\tilde{n}_S=0$ is the case in which  a transition is possible even in the
 absence of dissipation.\cite{FWGF}  From the  
superconducting side, this should be driven by proliferation of
individual QPS as the absence of resistive interactions,  dipoles
have finite action, and thus will always occur as fluctuations.

{\bf Randomness}. For another issue, the effects of {\it randomness},  the above discussion is also relevant.
Giamarchi and Schulz analyzed  a strictly one-dimensional boson system with a random potential.  \cite{Giamarchi-Schulz1988}  They
discussed the important role of the random Berry's phase, which
controls the equilibrium local density of bosons, and found a transition between the superfluid and the insulating
Bose glass phases that is driven by pinning of the density
fluctuations of the superconducting phase by the random potential. 
How wires with spatially random properties would behave in the presence of ``normal" carriers and
dissipation is unclear. A similar question arises for the KT-like
transition predicted by Altman et al. \cite{AKPR} for a commensurately
filled chain of grain with random self capacitance and nearest neighbor
Josephson interactions.

 In the strong randomness limit of a JJ chain with  dissipation, the superconducting transition is effectively local as we have found here, but with subtle interactions between the local phase slips. With strong randomness, the transition should thus be 
dominated by weak links on many length and energy scales, and may be controlled by 
an infinite randomness quantum critical point as found in other random quantum systems. 
\cite{DSF94,DSF95,Doty-Fisher,Bhatt-Lee}

The effects of randomness on normal carriers need also be
considered. Localization effects in a nanowire typically become
important below the Thouless temperature which is roughly the spacing
of energy levels in a segment of length such that its total resistance
is of order $R_Q$. Simple estimates of this for a JJ chain suggest
this temperature is of the same order as $T^*$ up to factors of
$R/R_Q$ --- which make the Thouless temperature much lower when this
factor is small, i.e., strong dissipation.  But with a system like
that of Fig. \ref{fig18}, the tunneling barriers between the grains
and the normal nanowire both play roles: in this situation, the conversion resistance, $r$ will be large, and  the Thouless temperature 
is likely to be well below $T^*$.  Near transitions to
superconductivity, the superconducting
fluctuations will affect  the normal electron transport and any possible
localization.  As these decrease the overall resistance, it is likely
that they also lower the Thouless temperature --- perhaps all the way
to zero at the transition.  This argument is supported by the results
of Giamarchi and Schulz: the \scs phase is stable to randomness if its
stiffness parameter, $K$, is sufficiently large.  These, and other issues, surely merit exploring: the interplay
between superconductivity and randomness is still largely an open
field.

{\bf Higher Dimensions}. Finally, we briefly discuss the implications
of our results for thin films and bulk materials. The percolation
picture of the superconducting to normal transition in two and three
dimensional systems of grains, \cite{MPAFisher1987,Shimshoni1998}
suggests that close to the transition the conductivity will be
determined primarily by quasi-one-dimensional percolation paths. This,
and the local effects of dissipation on each junction, suggest that
many of the effects that we study here for one-dimensional systems
should apply to transitions in granular systems more generally.  These
effects may lead to various interesting crossovers, including the
possibility of non-monotonic temperature dependence of the resistivity
(see also Ref. \cite{Refael2003}) and quasi-metallic behavior near to
putative superconducting-insulator transitions as observed in a
variety of systems. \cite{Kapitulnik1, Kapitulnik2000, mason2}  One of
the most intriguing (and poorly understood) features of such
transitions in thin films is the existence of a ``supermetallic
phase'' characterized by a small but apparently  weakly temperature
resistivity down to very low temperatures ---  extrapolating, it
appears, to zero-temperature. \cite{MASON2002} This behavior is in
striking contrast with the theoretical picture of the superconductor
to  {\it insulator} transition in two dimensional systems presented in
Refs. \cite{Efetov1980,Doniach1981,FWGF,Cha1991,vanOtterlo1993,
Kopec1999}. In this, the resistivity in the limit of low temperatures
should go to  either zero or infinity.  The origin of the observed
metallic behavior in thin films is still unclear, although several
groups have recently addressed this problem
\cite{Dalidovich2001,Dalidovich2000,Das1999,Vishwanath2002,
GalitskiRefael}. A possible origin of such a `supermetallic phase' in
thin films are the  percolation aspects of the transition combined
with effects analogous to those  that we have discussed in this paper.
In particular, these should be significant when, near the transition,
there are many low energy excitations that give rise to dissipation.

We note that after this paper was complete, Ref. \onlinecite{goswami},
which analyzes a similar problem with similar methods appeared online.

\noindent
{\bf Acknowledgments}
It is a pleasure to thank A. Bezryadin, M.P.A. Fisher, J. Free, B. I. Halperin,
A. Kapitulnik, S. Kivelson, W. Neils,
M. Tinkham, S. Sachdev, D. Shahar, and G. Zarand for useful
discussions. This work has been supported in part by the NSF via grants
DMR-0229243(DSF) and DMR-0132874 (ED), and by the Israel-U.S. BSF and an Alona grant (YO).


\appendix
\begin{widetext}
\section{Derivation of low energy  action and justification of approximations \label{appA}}

In this appendix we outline derivations of  various low-energy forms of the action that are used in the text, from the basic action Eq. (\ref{allaction}) of the chain  in terms of the superconducting phase and the
electrochemical potential of each grain:
\be
\ba{c}
S_{chain}=S_Q+S_{dis}^r+S_{dis}^R+S_J\vspace{2mm}\\
=\int d\tau \summ_i \frac{1}{(D_N+D_S+C D_N D_S)}
(\frac{1}{2}\l(V_{SC}{(x_i,\,\tau)}-V_{N}{(x_i,\,\tau)}\r)^2+\frac{1}{2}CD_SV_{N}{(x_i,\,\tau)}^2+\frac{1}{2}C D_{N}V_{SC}{(x_i,\,\tau)}^2)\vspace{2mm}\\
+\beta \summ_{\omega_n}
\l(\frac{1}{|\omega_n|r}|V_{N}{(\omega_n)}-V_{SC}{(\omega_n)}|^2\r)\frac{1}{\hbar}\vspace{2mm}\\
+\beta \summ_{\omega_n} \int \frac{dk}{2\pi}\l(\frac{1}{|\omega_n|R}|k V_{N}{(k,\,\omega_n)}\cdot \ax|^2
\r)\frac{1}{\hbar}\vspace{2mm}\\
-\int_0^\beta d \tau\summ_i\l(E_J \cos\l(\phi_{i+1}{(\tau)}-\phi_{i}{(\tau)}\r)\r)\vspace{2mm}\ ;
\ea
\label{allaction1}
\ee
at zero temperature, the sums over Matsubara frequencies can be replaced by integrals:
$\beta\sum_{\omega_n}\rightarrow \int \frac{d\omega}{2\pi}$ which we will use herein.

From Eq. (\ref{allaction1}), one can integrate out the normal voltages, $V_{N}{(x,\,\tau)}$, to obtain:
\be
\ba{c}
S_{chain}=S_J+\frac{1}{2}\int \frac{d\omega}{2\pi}\int \frac{dk}{2\pi}\l(\l[\tilde{C}-F(k,\omega)\r]\l|V_{SC}{(k,\,\omega)}\r|^2+\frac{2-2\cos(k\ax)}{\ax R|\omega|} \l|V_{SC}{(k,\,\omega)}\r|^2\r) 
\ea\label{Taction}
\ee
where
\be
F= \frac{\l(\B+\frac{\ax^2 k^2}{R|\omega|}\r)^2}{\l(C_{NS}+\frac{\frac{1}{r}+\frac{\ax^2 k^2}{R}}{|\omega|}\r)}
\ee
with
\be
\ba{c}
\tilde{C}=\frac{C\l(D_N+D_S\r)}{D_N+D_S+CD_ND_S}\approx C \vspace{2mm} \\
C_{NS}=\frac{1+D_SC}{D_N+D_S+CD_ND_S}\vspace{2mm}\\
\B=-\frac{D_SC}{D_N+D_S+CD_ND_S} \ .
\ea
\ee

In expression (\ref{Taction}), the last term and the $\tilde{C}$ term (which is a slightly modified capacitance in the limit $CD_{N,S}\ll 1$) comprise the usual action for a dissipatively shunted JJ.  Together with  these two terms, the third, complicated looking, term, with coefficient  $F(k,\omega)$, 
gives rise to the two fluid behavior. It is from the frequency dependence of $F$  that the basic energy scale
\be
T^*=\frac{\hbar}{C_{NS}}\l(\frac{1}{r}+\frac{1}{R}\r)
\label{Tstarapp}
\ee
arises.  The low energy regime of interest is $\omega\ll T^*$. 
Note that because of the local nature of some of the important
physics, we must take the low frequency limit at fixed $k$; if the
limit is taken in the opposite order, the results can be very
misleading.

At low energies,  the action becomes
\be
S_{chain}=S_J+\frac{1}{2}\int \frac{d\omega}{2\pi}\int \frac{dk}{2\pi}\l(\tilde{C} \l|V_{SC}{(k,\,\omega)}\r|^2+\frac{\ax}{|\omega|}\frac{2-2\cos(k\ax)}{R+r[2-2\cos(k\ax)]} \l|V_{SC}{(k,\,\omega)}\r|^2\r) \ ,
\label{Tactionp1}
\ee
which, on approximating $\tilde{C}$ by $C$, is the form of the action we use in both the strong and weak coupling limits.

In the weak coupling limit, we drop the capacitative term as this is unimportant at low frequencies.

In the strong coupling limit, the capacitative term is important and
the Villain approximation for the Josephson coupling and ensuing
transformations can be carried out straightforwardly as discussed in
Sec. \ref{QPSsec}. In general, this is only appropriate for low energies, in particular for $\omega \ll \Omega_{LR}$, the inductive relaxation rate
\be
\Omega_{LR}= \l(\frac{2e}{\hbar}\r)^2 E_J R,
\label{localco}
\ee
since $E_J$ is inversely proportional to the ``kinetic inductance", $L_J=\l(\frac{\hbar}{2e}\r)^2\frac{1}{E_J}$ of the junction.  Note that $\Omega_{LR}$ is the same order as the plasma frequency of the junctions, reduced from this by a factor of $\j R/R_Q$ which is of order unity in most of the  regimes of interest.  Thus the Villain approximation is valid wherever we have used it.

The action in terms of phase slips, the interactions between them, and the equivalent sine-Gordon representation in terms of the dual fields $\theta$ and $\psi$ can be derived straightforwardly. Note that we could have gone more directly to the sine-Gordon  representation by decoupling the inter-grain terms in Eq. (\ref{allaction1}) with the fields $\psi$ and $\theta$, then integrating out $V_N$ and $\phi$. The periodicity of $\phi$ then gives rise to an integer constraint on $(\theta+\psi)/2\pi$ which after integrating out the high energy fluctuations, becomes the $\cos(\theta+\psi)$ in the sine-Gordon action.  The appropriate low frequency approximations appear naturally in the intermediate representation. In particular, the quadratic coupling between $\theta$ and $\psi$ that appears if this route is followed, is unimportant at low frequencies.
\end{widetext}

\bibliography{thebib}

\end{document}